\documentclass[a4paper,fleqn]{cas-sc}

\usepackage[numbers,longnamesfirst]{natbib}
\usepackage[export]{adjustbox}
\usepackage{makecell}
\usepackage{bm}
\usepackage{nomencl}
\makenomenclature
\usepackage{framed}
\usepackage{multicol}
\renewcommand{\nompreamble}{\begin{multicols}{3}}
\renewcommand{\nompostamble}{\end{multicols}}
\usepackage{rotating}
\usepackage{pdflscape}
\usepackage{afterpage}
\usepackage{wrapfig}
\usepackage{subfigure}
\usepackage{caption}

\usepackage{tikz}
\usepackage{amsmath}
\usepackage{pgfplots}
\pgfplotsset{compat=newest}

\usepackage[title]{appendix}
\usepackage[capitalise]{cleveref}

\crefname{appendix}{appendix}{appendices}
\Crefname{appendix}{Appendix}{Appendices}

\renewcommand*\nompreamble{\begin{multicols}{3}}
\renewcommand*\nompostamble{\end{multicols}}

\definecolor{TableGray}{RGB}{240, 240, 240}
\definecolor{TableWhite}{RGB}{253, 253, 253}

\definecolor{v1}{HTML}{fde725}
\definecolor{v2}{HTML}{f8e621}
\definecolor{v3}{HTML}{f1e51d}
\definecolor{v4}{HTML}{ece51b}
\definecolor{v5}{HTML}{e5e419}
\definecolor{v6}{HTML}{dfe318}
\definecolor{v7}{HTML}{d8e219}
\definecolor{v8}{HTML}{d0e11c}
\definecolor{v9}{HTML}{cae11f}
\definecolor{v10}{HTML}{c2df23}
\definecolor{v11}{HTML}{bddf26}
\definecolor{v12}{HTML}{b5de2b}
\definecolor{v13}{HTML}{addc30}
\definecolor{v14}{HTML}{a8db34}
\definecolor{v15}{HTML}{a0da39}
\definecolor{v16}{HTML}{9bd93c}
\definecolor{v17}{HTML}{93d741}
\definecolor{v18}{HTML}{8ed645}
\definecolor{v19}{HTML}{86d549}
\definecolor{v20}{HTML}{7fd34e}
\definecolor{v21}{HTML}{7ad151}
\definecolor{v22}{HTML}{73d056}
\definecolor{v23}{HTML}{6ece58}
\definecolor{v24}{HTML}{67cc5c}
\definecolor{v25}{HTML}{60ca60}
\definecolor{v26}{HTML}{5cc863}
\definecolor{v27}{HTML}{56c667}
\definecolor{v28}{HTML}{52c569}
\definecolor{v29}{HTML}{4cc26c}
\definecolor{v30}{HTML}{48c16e}
\definecolor{v31}{HTML}{42be71}
\definecolor{v32}{HTML}{3dbc74}
\definecolor{v33}{HTML}{3aba76}
\definecolor{v34}{HTML}{35b779}
\definecolor{v35}{HTML}{32b67a}
\definecolor{v36}{HTML}{2eb37c}
\definecolor{v37}{HTML}{2ab07f}
\definecolor{v38}{HTML}{28ae80}
\definecolor{v39}{HTML}{25ac82}
\definecolor{v40}{HTML}{24aa83}
\definecolor{v41}{HTML}{22a785}
\definecolor{v42}{HTML}{20a486}
\definecolor{v43}{HTML}{1fa287}
\definecolor{v44}{HTML}{1fa088}
\definecolor{v45}{HTML}{1f9e89}
\definecolor{v46}{HTML}{1e9b8a}
\definecolor{v47}{HTML}{1f998a}
\definecolor{v48}{HTML}{1f968b}
\definecolor{v49}{HTML}{20938c}
\definecolor{v50}{HTML}{20928c}
\definecolor{v51}{HTML}{218f8d}
\definecolor{v52}{HTML}{228d8d}
\definecolor{v53}{HTML}{238a8d}
\definecolor{v54}{HTML}{24878e}
\definecolor{v55}{HTML}{25858e}
\definecolor{v56}{HTML}{26828e}
\definecolor{v57}{HTML}{26818e}
\definecolor{v58}{HTML}{277e8e}
\definecolor{v59}{HTML}{287c8e}
\definecolor{v60}{HTML}{29798e}
\definecolor{v61}{HTML}{2a768e}
\definecolor{v62}{HTML}{2b748e}
\definecolor{v63}{HTML}{2c718e}
\definecolor{v64}{HTML}{2d708e}
\definecolor{v65}{HTML}{2e6d8e}
\definecolor{v66}{HTML}{306a8e}
\definecolor{v67}{HTML}{31688e}
\definecolor{v68}{HTML}{32658e}
\definecolor{v69}{HTML}{33638d}
\definecolor{v70}{HTML}{34608d}
\definecolor{v71}{HTML}{365d8d}
\definecolor{v72}{HTML}{375b8d}
\definecolor{v73}{HTML}{38588c}
\definecolor{v74}{HTML}{39558c}
\definecolor{v75}{HTML}{3b528b}
\definecolor{v76}{HTML}{3c508b}
\definecolor{v77}{HTML}{3d4d8a}
\definecolor{v78}{HTML}{3e4989}
\definecolor{v79}{HTML}{3f4788}
\definecolor{v80}{HTML}{414487}
\definecolor{v81}{HTML}{424186}
\definecolor{v82}{HTML}{433e85}
\definecolor{v83}{HTML}{443a83}
\definecolor{v84}{HTML}{453882}
\definecolor{v85}{HTML}{463480}
\definecolor{v86}{HTML}{46327e}
\definecolor{v87}{HTML}{472e7c}
\definecolor{v88}{HTML}{472c7a}
\definecolor{v89}{HTML}{482878}
\definecolor{v90}{HTML}{482475}
\definecolor{v91}{HTML}{482173}
\definecolor{v92}{HTML}{481d6f}
\definecolor{v93}{HTML}{481b6d}
\definecolor{v94}{HTML}{481769}
\definecolor{v95}{HTML}{471365}
\definecolor{v96}{HTML}{471063}
\definecolor{v97}{HTML}{460b5e}
\definecolor{v98}{HTML}{46085c}
\definecolor{v99}{HTML}{450457}
\definecolor{v100}{HTML}{440154}

\newcommand{\gc}[1]{%
    \ifdim#1 pt < 0.01 pt v100%
    \else\ifdim#1 pt < 0.02 pt v99%
    \else\ifdim#1 pt < 0.03 pt v98%
    \else\ifdim#1 pt < 0.04 pt v97%
    \else\ifdim#1 pt < 0.05 pt v96%
    \else\ifdim#1 pt < 0.06 pt v95%
    \else\ifdim#1 pt < 0.07 pt v94%
    \else\ifdim#1 pt < 0.08 pt v93%
    \else\ifdim#1 pt < 0.09 pt v92%
    \else\ifdim#1 pt < 0.10 pt v91%
    \else\ifdim#1 pt < 0.11 pt v90%
    \else\ifdim#1 pt < 0.12 pt v89%
    \else\ifdim#1 pt < 0.13 pt v88%
    \else\ifdim#1 pt < 0.14 pt v87%
    \else\ifdim#1 pt < 0.15 pt v86%
    \else\ifdim#1 pt < 0.16 pt v85%
    \else\ifdim#1 pt < 0.17 pt v84%
    \else\ifdim#1 pt < 0.18 pt v83%
    \else\ifdim#1 pt < 0.19 pt v82%
    \else\ifdim#1 pt < 0.20 pt v81%
    \else\ifdim#1 pt < 0.21 pt v80%
    \else\ifdim#1 pt < 0.22 pt v79%
    \else\ifdim#1 pt < 0.23 pt v78%
    \else\ifdim#1 pt < 0.24 pt v77%
    \else\ifdim#1 pt < 0.25 pt v76%
    \else\ifdim#1 pt < 0.26 pt v75%
    \else\ifdim#1 pt < 0.27 pt v74%
    \else\ifdim#1 pt < 0.28 pt v73%
    \else\ifdim#1 pt < 0.29 pt v72%
    \else\ifdim#1 pt < 0.30 pt v71%
    \else\ifdim#1 pt < 0.31 pt v70%
    \else\ifdim#1 pt < 0.32 pt v69%
    \else\ifdim#1 pt < 0.33 pt v68%
    \else\ifdim#1 pt < 0.34 pt v67%
    \else\ifdim#1 pt < 0.35 pt v66%
    \else\ifdim#1 pt < 0.36 pt v65%
    \else\ifdim#1 pt < 0.37 pt v64%
    \else\ifdim#1 pt < 0.38 pt v63%
    \else\ifdim#1 pt < 0.39 pt v62%
    \else\ifdim#1 pt < 0.40 pt v61%
    \else\ifdim#1 pt < 0.41 pt v60%
    \else\ifdim#1 pt < 0.42 pt v59%
    \else\ifdim#1 pt < 0.43 pt v58%
    \else\ifdim#1 pt < 0.44 pt v57%
    \else\ifdim#1 pt < 0.45 pt v56%
    \else\ifdim#1 pt < 0.46 pt v55%
    \else\ifdim#1 pt < 0.47 pt v54%
    \else\ifdim#1 pt < 0.48 pt v53%
    \else\ifdim#1 pt < 0.49 pt v52%
    \else\ifdim#1 pt < 0.50 pt v51%
    \else\ifdim#1 pt < 0.51 pt v50%
    \else\ifdim#1 pt < 0.52 pt v49%
    \else\ifdim#1 pt < 0.53 pt v48%
    \else\ifdim#1 pt < 0.54 pt v47%
    \else\ifdim#1 pt < 0.55 pt v46%
    \else\ifdim#1 pt < 0.56 pt v45%
    \else\ifdim#1 pt < 0.57 pt v44%
    \else\ifdim#1 pt < 0.58 pt v43%
    \else\ifdim#1 pt < 0.59 pt v42%
    \else\ifdim#1 pt < 0.60 pt v41%
    \else\ifdim#1 pt < 0.61 pt v40%
    \else\ifdim#1 pt < 0.62 pt v39%
    \else\ifdim#1 pt < 0.63 pt v38%
    \else\ifdim#1 pt < 0.64 pt v37%
    \else\ifdim#1 pt < 0.65 pt v36%
    \else\ifdim#1 pt < 0.66 pt v35%
    \else\ifdim#1 pt < 0.67 pt v34%
    \else\ifdim#1 pt < 0.68 pt v33%
    \else\ifdim#1 pt < 0.69 pt v32%
    \else\ifdim#1 pt < 0.70 pt v31%
    \else\ifdim#1 pt < 0.71 pt v30%
    \else\ifdim#1 pt < 0.72 pt v29%
    \else\ifdim#1 pt < 0.73 pt v28%
    \else\ifdim#1 pt < 0.74 pt v27%
    \else\ifdim#1 pt < 0.75 pt v26%
    \else\ifdim#1 pt < 0.76 pt v25%
    \else\ifdim#1 pt < 0.77 pt v24%
    \else\ifdim#1 pt < 0.78 pt v23%
    \else\ifdim#1 pt < 0.79 pt v22%
    \else\ifdim#1 pt < 0.80 pt v21%
    \else\ifdim#1 pt < 0.81 pt v20%
    \else\ifdim#1 pt < 0.82 pt v19%
    \else\ifdim#1 pt < 0.83 pt v18%
    \else\ifdim#1 pt < 0.84 pt v17%
    \else\ifdim#1 pt < 0.85 pt v16%
    \else\ifdim#1 pt < 0.86 pt v15%
    \else\ifdim#1 pt < 0.87 pt v14%
    \else\ifdim#1 pt < 0.88 pt v13%
    \else\ifdim#1 pt < 0.89 pt v12%
    \else\ifdim#1 pt < 0.90 pt v11%
    \else\ifdim#1 pt < 0.91 pt v10%
    \else\ifdim#1 pt < 0.92 pt v9%
    \else\ifdim#1 pt < 0.93 pt v8%
    \else\ifdim#1 pt < 0.94 pt v7%
    \else\ifdim#1 pt < 0.95 pt v6%
    \else\ifdim#1 pt < 0.96 pt v5%
    \else\ifdim#1 pt < 0.97 pt v4%
    \else\ifdim#1 pt < 0.98 pt v3%
    \else\ifdim#1 pt < 0.99 pt v2%
    \else v1%
    \fi\fi\fi\fi\fi\fi\fi\fi\fi\fi\fi\fi\fi\fi\fi\fi\fi\fi\fi\fi\fi\fi\fi\fi\fi\fi\fi\fi\fi\fi\fi\fi\fi\fi\fi\fi\fi\fi\fi\fi\fi\fi\fi\fi\fi\fi\fi\fi\fi\fi\fi\fi\fi\fi\fi\fi\fi\fi\fi\fi\fi\fi\fi\fi\fi\fi\fi\fi\fi\fi\fi\fi\fi\fi\fi\fi\fi\fi\fi\fi\fi\fi\fi\fi\fi\fi\fi\fi\fi\fi\fi\fi\fi\fi\fi\fi\fi\fi\fi
}

\newcommand{\mc}[1]{
    \cellcolor{\gc{#1}!50} #1
}

\begin{document}

\shorttitle{Comparison of Clustering Approaches for Smart Meter Time Series Data}

\shortauthors{L Yerbury et~al.}

\title[mode = title]{Comparing Clustering Approaches for Smart Meter Time Series: Investigating the Influence of Dataset Properties on Performance}

  \author[1,3]{Luke W. Yerbury}[type=editor,
  orcid=0009-0005-5894-2281
  ]
  \ead{Luke.Yerbury@uon.edu.au} 
  \cormark[1]
  \credit{Conceptualisation of this study, Methodology, Coding, Conducting Experiments, Analysis, Writing, Plots}

  \author[2]{Ricardo J.G.B. Campello}[
  orcid=0000-0003-0266-3492
  ]

  \author[1]{G. C. Livingston Jr}

  \author[3]{Mark Goldsworthy}[
  orcid=0000-0001-8718-1139
  ]

  \author[4]{Lachlan O'Neil}[
  orcid=0000-0002-3981-5990
  ]

  \address[1]{School of Information and Physical Sciences, University of Newcastle, Callaghan, NSW, 2308, Australia}
  \address[2]{Department of Mathematics and Computer Science, University of Southern Denmark, Odense, Denmark}
  \address[3]{The Commonwealth Scientific and Industrial Research Organisation (CSIRO), Energy Centre, Mayfield West, NSW 2304, Australia}
  \address[4]{Independent Technical Expert, Lachlan.P.ONeil@gmail.com}

  \cortext[cor1]{Corresponding author}

 \begin{abstract}[S U M M A R Y]
The widespread adoption of smart meters for monitoring energy consumption has generated vast quantities of high-resolution time series data which remains underutilised. While clustering has emerged as a fundamental tool for mining smart meter time series (SMTS) data, selecting appropriate clustering methods remains challenging despite numerous comparative studies. These studies often rely on problematic methodologies and consider a limited scope of methods, frequently overlooking compelling methods from the broader time series clustering literature. Consequently, they struggle to provide dependable guidance for practitioners designing their own clustering approaches. 

This paper presents a comprehensive comparative framework for SMTS clustering methods using expert-informed synthetic datasets that emphasise peak consumption behaviours as fundamental cluster concepts. Using a phased methodology, we first evaluated 31 distance measures and 8 representation methods using leave-one-out classification, then examined the better-suited methods in combination with 11 clustering algorithms. We further assessed the robustness of these combinations to systematic changes in key dataset properties that affect clustering performance on real-world datasets, including cluster balance, noise, and the presence of outliers. 

Our results revealed that methods accommodating local temporal shifts while maintaining amplitude sensitivity, particularly Dynamic Time Warping and $k$-sliding distance, consistently outperformed traditional approaches. Among other key findings, we identified that when combined with $k$-medoids or hierarchical clustering using Ward's linkage, these methods exhibited consistent robustness across varying dataset characteristics without careful parameter tuning. These and other findings inform actionable recommendations for practitioners, and validation with real-world data demonstrates that our findings translate effectively to practical SMTS clustering tasks. Finally, our datasets and code are publicly available to support the development, evaluation, and comparison of both novel and overlooked methods.

\end{abstract}
 

 \maketitle

	\nomenclature[Vp]{RVI}{Relative Validity Index}
\nomenclature[Vp]{SMTS}{Smart Meter Time Series}
\nomenclature[Vp]{DLP}{Daily Load Profile} 
\nomenclature[Vp]{RLP}{Representative Load Profile}
\nomenclature[Vp]{ARI}{Adjusted Rand Index}
\nomenclature[Vp]{PSI}{Pair Sets Index}
\nomenclature[Vp]{EVI}{External Validity Index}
\nomenclature[Vp]{AMI}{Adjusted Mutual Information}
\nomenclature[Vp]{NVD}{Normalised Van Dongen}

        \renewcommand{\nomname}{\normalsize Nomenclature}
        \begin{footnotesize}
        \begin{framed}
            \vspace{-0.5cm}
            \printnomenclature
            \vspace{-0.4cm}
        \end{framed}
        \end{footnotesize}

	\section{Introduction}
\label{sec:Introduction}

Smart meters capture high-resolution time series data on energy consumption, extending the scope of electricity metering far beyond the basic function of billing. 
Their deployment has rapidly expanded across the globe, with a 2021 survey indicating that 1.494 billion households have been targeted by installation programs across 47 countries \cite{Sovacool2021GlobalTransitions}. As new technologies like distributed renewable energy generation \citep{Li2020c} and energy storage \citep{Keck2019TheAustralia} are progressively integrated into existing energy networks, the resulting increase in grid complexity makes insights from smart meters more and more crucial for effective management \cite{Banales2021}.
In this context, Smart Meter Time Series (SMTS) data have been used to improve energy forecasts \cite{Auder2018}, identify candidates for demand response initiatives \cite{Ahir2022AData}, recognise \cite{Donaldson2020} and guide \cite{AlKhafaf2022} the deployment of sustainable energy infrastructure, optimise time-of-use tariff offerings \cite{Li2016a}, encourage conscious and sustainable energy consumption \cite{Volker2021}, identify anomalous consumption \cite{Hurst2020}, and more \cite{Wang2020}. The data mining technique known as \textit{clustering} consistently serves as a fundamental data processing step within these myriad applications, facilitating the detection of patterns essential for traditional analytical approaches that would otherwise be intractable due to the enormous scale of the data.

Clustering is an unsupervised learning technique aimed at partitioning objects from datasets into groups, such that objects within groups share a greater notion of similarity or homogeneity than objects in different groups \cite{Gan2007DataApplications,Aggarwal2014}. Such a grouping or partition is obtained by applying a \textit{clustering approach} to a dataset. 
In the context of time series data, clustering approaches can be understood through five fundamental \textit{components}: a data normalisation procedure, a data representation method, a distance measure, a clustering algorithm, and a prototype definition \cite{Aghabozorgi2015,Yerbury2024}. 
Though not all clustering approaches atomise completely into \textit{all} five components, each component is known to significantly influence clustering outcomes when present. Furthermore, no single combination can be optimal across all domains and datasets \cite{Wolpert1997}, and there is no single ``cluster'' definition which universally applies in all domains and applications. The result is an enormous catalogue of components and approaches advanced in the literature --- including a host of time series specific methods \cite{Aghabozorgi2015}. Practitioners seeking to employ clustering are thus confronted with an overwhelming number of component options and combinations, with no established consensus regarding the best way to evaluate or compare their suitability for distinct scenarios \cite{Aghabozorgi2015,Yerbury2024}. A common form of recourse is to select methods based purely on precedence in the relevant domain's literature. Whilst there is a strong case to be made for the continued use of enduring methods with demonstrated success, precedence is a self-perpetuating rationale that is not guaranteed to provide optimal outcomes. Comparative or benchmark studies thus have the potential to be indispensable guides for practitioners navigating this complex decision space \cite{VanMechelen2023,Costa2023}. Whether they succeed at this depends on the comparative methodologies employed therein.

Due to the exploratory nature of clustering, the comparison of clustering approaches, components, and partitions remains a difficult open problem \cite{Aghabozorgi2015,Yerbury2024}. Meanwhile, two classes of quantitative measures have endured as the most common tools for addressing this challenge, namely Relative Validity Indices (RVIs) and External Validity Indices (EVIs). RVIs, such as the Silhouette Width Criterion \cite{Rousseeuw1987Silhouettes:Analysis} and Davies-Bouldin Index \cite{Davies1979AMeasure}, attempt to quantify the extent to which a partition exemplifies one or more generally desirable \textit{cluster concepts} \cite{Hennig2015a}, such as small within-group dissimilarities, large between group dissimilarities or effective representation by cluster prototypes. The quality of a set of partitions can be ranked by an RVI, and these ranks are frequently used in the SMTS and wider clustering literature to make various inferences, such as identifying the best number of clusters or best clustering approach for a dataset. There are two major problems with the use of RVIs for comparisons of clustering methods. Firstly, comparisons based on RVIs will be biased towards those partitions and methods that align with the primarily domain-agnostic cluster concepts inherently preferred by different RVIs \cite{Kang2015}. This was recognised in the context of SMTS clustering in \cite{Li2022} when optimal performance according to RVIs didn't translate to optimal performance in downstream applications. Furthermore, \cite{Toussaint2019} and \cite{Rajabi2020} noted that RVIs show a tendency to insufficiently penalise large, noisy clusters, and favour partitions that isolate outliers, suggesting the cluster concepts in popular RVIs are not well-suited to SMTS data.
Secondly, the use of RVIs for any comparisons of normalisation procedures, representation methods and distance measures are highly dubious due to the influence of these components on pairwise dissimilarities \cite{Yerbury2024}.

On the other hand, EVIs such as the Adjusted Rand Index (ARI) \cite{Hubert1985} and Pair Sets Index (PSI) \cite{Rezaei2016}, can only be used with labelled datasets, as they quantify the extent to which a clustering approach has recovered a dataset's ground-truth labels.  Classification datasets have often been used as the foundation for comparative studies of clustering methods employing EVIs \cite{Javed2020,Paparrizos2015}. However, the associated \textit{class labels} do not necessarily correspond to recoverable, natural clusters \cite{Farber2010}. This lack of correspondence has been observed in the context of SMTS clustering when considering the association between surveyed household characteristics and cluster membership \cite{McLoughlin2015,Roberts2019}. Furthermore, although high EVI scores reliably indicate good label recovery, low EVI scores do not necessarily indicate poor clustering structure in a partition, given the possibility of legitimate alternatives \cite{VanCraenendonck2015}. 

These concerns can be mitigated by instead utilising datasets where \textit{cluster labels} have been curated to encompass those formal and informal cluster concepts deemed important for the problem and domain according to the relevant experts. Compared to RVIs, external validation against such datasets offers a more transparent selection bias towards methods that will align with the domain-specific cluster concepts specified by experts. Importantly, external validation additionally allows for comparisons of those components which affect pairwise dissimilarities. However, this calibre of labelled dataset is typically not readily available. Labelling real world datasets --- whether fully by hand or by propagating partial labels --- can be a complicated and expensive process which still does not guarantee uniquely correct cluster labels as multiple valid clusterings may still exist for the same data. Synthetic datasets incorporating fundamental patterns from real-world data, where the intended clustering structure is precisely defined by construction, are thus a more natural alternative \cite{Hennig2018}.
As exemplified in recent comparative studies from other domains \cite{Watson2022,Costa2023,Preudhomme2021,Rodriguez2019}, labelled synthetic datasets provide the additional opportunity to systematically investigate the robustness of clustering methods to changes in key dataset properties. Comparative studies utilising labelled datasets are also uniquely positioned to compare the suitability of distance measures and representations for clustering SMTS data without interference from any particular clustering algorithm by utilising classification accuracy \cite{Paparrizos2020,Alonso2020,Paparrizos2015,Montero2014TSclust:Clustering,Wang2013b,CHEN2004,Keogh2002}.

\paragraph{} \noindent 
This methodological context reveals the primary limitations of previous comparative studies of short SMTS clustering methods \cite{Li2022,Rajabi2020,Ruiz2020AData,Toussaint2019,Yilmaz2019,Jin,McLoughlin2015,Iglesias2013,Chicco2012a,Chicco2006} and has informed key choices in the current study. 

Most significantly, the exclusive reliance on RVIs in previous studies for comparing methods compromises the reliability of their analyses, as RVIs not only demonstrate biases towards certain domain-agnostic cluster concepts, but are also unsuitable for comparing methods that affect pairwise dissimilarities (including distance measures and representation methods). Moreover, their exclusive use of single real-world datasets limits the generalisability of their findings and prevents any systematic investigation into how different methods perform across varying dataset characteristics. These studies have also typically restricted their scope to comparing a small number of recurring methods, while largely overlooking compelling methods from the general time series and SMTS literatures. These related studies are discussed in greater depth in \Cref{sec:RelatedWork}.

As Hennig \cite{Hennig2015a} argues, theory, real datasets and simulation studies all have their place when assessing the quality of clustering methods, as all of them offer something that cannot be replaced by the other two approaches. Given that previous comparative studies have relied exclusively on RVIs with real datasets, our timely simulation-based investigation offers novel perspectives and insights for practitioners, complementing existing work while addressing its key limitations. Central to our approach are expert-curated synthetic datasets designed around a core organising principle which contends that the timing and distribution of peak energy consumption events are fundamental to grid operation, and thus for the separation of short SMTS data into meaningful clusters \cite{Chai2024,Pelekis2023,Satre-meloy2020,Liu2019,Liu2013}. This principled foundation enables us to make several key contributions to both the theoretical understanding and practical application of short SMTS clustering methods.

\begin{itemize}
    \setlength\itemsep{0em}
    \item[(i)] We present an extensive comparison of methods for clustering of daily residential SMTS data, evaluating 31 distance measures, 8 representations, and 11 clustering algorithms.
    \item[(ii)] We introduce a more reliable and transparent comparative methodology based on the use of EVIs with expert-informed synthetic datasets, avoiding the biases and limitations inherent in previous RVI-based comparative studies. 
    \item[(iii)] We systematically investigate method robustness across key dataset characteristics including cluster balance, noise levels, number of clusters, number of time series and the presence of outliers --- enabling insights into method performance under specific conditions relevant to real-world datasets.
    \item[(iv)] We have made our synthetic datasets and code freely available, allowing practitioners to readily evaluate new or overlooked methods without needing to replicate the entirety of our experiments. 
    \item[(v)] We present a comprehensive review and analysis of previous comparative studies in short SMTS clustering, providing valuable context for this and future work.
    \item[(vi)] We provide guidance for practitioners designing short SMTS clustering approaches, drawing on our empirical findings to recommend robust method combinations and targeted solutions for common clustering challenges.
    \item[(vii)] We validate our findings by comparing method performance on synthetic data with results on expert-classified real-world data (made publicly available), confirming a strong relationship between synthetic and real-world efficacy, demonstrating the practical utility of our methodology.
\end{itemize}

The remainder of the paper is organised as follows. \Cref{sec:RelatedWork} will thoroughly review other comparative studies of SMTS clustering. \Cref{sec:SyntheticData} will present the particulars of our synthetic data. \Cref{sec:Methods} will detail the phased experimental methodology and the investigated clustering methods. \Cref{sec:Results} will present and discuss the experimental results, including practical, actionable insights for practitioners. \Cref{sec:RealData} will demonstrate the real-world applicability of our findings through validation with expert-classified smart meter data. Finally, \Cref{sec:Conclusion} will conclude the paper. 

        \section{Related Work}
\label{sec:RelatedWork}

Several studies have performed comparisons of clustering approaches and components specifically for SMTS data. In the following section, we have comprehensively surveyed studies where this has explicitly been the primary goal \cite{Rajabi2020,Toussaint2019,Yilmaz2019,Jin,Chicco2006}. We have also chosen to include selected studies where such comparisons comprised a substantial component of the publication, albeit alongside other objectives \cite{Li2022,McLoughlin2015,Iglesias2013,Chicco2012a}. We will analyse and discuss these nine studies in relation to their comparative scope, comparative methodologies and data. Relevant details for each study have been documented in \Cref{Tab:ComparativeStudies}, which will be elaborated upon and referenced throughout the following discussion.

\begin{table}[b]
    \scriptsize
    \rowcolors{1}{TableGray}{TableWhite}
    \begin{adjustbox}{center}
        \begin{tabular}{p{0.35cm} p{0.39cm} p{1.38cm} p{1cm} p{1.6cm} p{1.4cm} p{1.1cm} p{3.9cm} p{1.3cm}} \toprule \hiderowcolors 
            \multicolumn{1}{l}{} & \multicolumn{1}{l}{} & \multicolumn{4}{c}{\textbf{Compared Clustering Components}} & \multicolumn{1}{l}{} & \multicolumn{1}{l}{} & \multicolumn{1}{l}{} \\ \cmidrule(l){3-6}
            \textbf{Ref.} & \textbf{Year} & \textbf{Norm.} & \textbf{Rep.}\textsuperscript{a} & \textbf{Dist.}\textsuperscript{b} & \textbf{Alg.}\textsuperscript{c} & \textbf{Datasets} & \textbf{What Was Clustered?} & \textbf{Metrics}\textsuperscript{d}\\ \midrule \showrowcolors

            \scriptsize{\cite{Li2022}}\textsuperscript{$\star$} & \scriptsize{2022} & \scriptsize{Zero-One} & \scriptsize{Raw} & \scriptsize{CoD, ChD, DTW, ED, ERP, HD, MD, MAH, PC} & \scriptsize{KMn} & \makecell[tl]{\scriptsize{1 Real} \\ \scriptsize{Commercial} \\ \scriptsize{(1 hr)}} & \makecell[tl]{- DLPs (463 days from 1 consumer)} & \scriptsize{DI, S\_Dbw} \\[0.15cm]

            \scriptsize{\cite{Rajabi2020}} & \scriptsize{2020} & \scriptsize{Zero-One} & \scriptsize{Raw} & \scriptsize{ED} & \scriptsize{FCM, HAC (W), KMn, SOM} & \makecell[tl]{\scriptsize{1 Real} \\ \scriptsize{Residential} \\ \scriptsize{(30 min)}} & \makecell[tl]{- DLPs (356 days from 1 consumer) \\ - RLPs (4141 consumers, derived \\ from 1 year and grouped by \\ weekdays and weekends)} & \scriptsize{DBI, DI, MIA, MSE, SWC, WCBCR} \\[0.15cm]

            \scriptsize{\cite{Toussaint2019}} & \scriptsize{2019} & \scriptsize{Zero-One, SA-Norm, De-Minning, Unit-Norm} & \scriptsize{Raw} & \scriptsize{ED} & \scriptsize{KMn, SOM, SOM$+$KMn (with/out pre-binning)} & \makecell[tl]{\scriptsize{1 Real} \\ \scriptsize{Residential} \\ \scriptsize{(1 hr)}} & \makecell[tl]{- DLPs (3,295,848 from various days \\ and 12,945 consumers)} & \scriptsize{Combined Index (DBI, MIA, SWC)} \\[0.15cm]

            \scriptsize{\cite{Yilmaz2019}} & \scriptsize{2019} & \scriptsize{Integral} & \makecell[tl]{Raw,\\ Statistical \\ Feature \\ Vector} & \scriptsize{ED} & \scriptsize{KMn} & \makecell[tl]{\scriptsize{1 Real} \\ \scriptsize{Residential} \\ \scriptsize{(15 min)}} & \makecell[tl]{- DLPs (239,440 from 365 days and \\ 656 consumers) \\ - RLPs (656 consumers, derived from \\ 365 days, plus grouped by weekday \\ and further by season) \\ - Other (365 average daily loads)} & \scriptsize{SWC} \\[0.15cm]

            \scriptsize{\cite{Jin}} & \scriptsize{2017} & \scriptsize{De-minned Integral} & \scriptsize{Raw} & \scriptsize{BD, CoD, ChD, ED, MD, PC, sED, sqED} & \scriptsize{GMM, HAC\textsuperscript{$\dagger$} (A, C, W), } & \makecell[tl]{\scriptsize{1 Real} \\ \scriptsize{Residential} \\ \scriptsize{(1 hr)}} & \makecell[tl]{- DLPs (32,611,421 from 365 days \\ and $\sim100,000$ consumers)} & \scriptsize{CCC\textsuperscript{$\dagger$}, CDI, DBI, MIA, SMI, SWC, VRSE} \\[0.15cm] 

            \scriptsize{\cite{McLoughlin2015}}\textsuperscript{$\star$} & \scriptsize{2015} & \scriptsize{Unknown} & \scriptsize{Raw} & \scriptsize{ED} & \scriptsize{KMd, KMn, SOM} & \makecell[tl]{\scriptsize{1 Real} \\ \scriptsize{Residential} \\ \scriptsize{(30 min)}} & \makecell[tl]{- DLPs (3941 consumers on 3 \\ random days separately, then from \\ 184 days)} & \scriptsize{DBI} \\[0.15cm] 

            \scriptsize{\cite{Iglesias2013}}\textsuperscript{$\star$} & \scriptsize{2013} & \scriptsize{$Z$-norm} & \scriptsize{Raw} & \scriptsize{DTW, ED, MAH, PC} & \scriptsize{FCM} & \makecell[tl]{\scriptsize{1 Real} \\ \scriptsize{Commercial} \\ \scriptsize{(1 hr)}} & \makecell[tl]{- DLPs (124 days for each of 5 \\ buildings} & \scriptsize{CB, CVB, DBI, DI} \\[0.15cm]

            \scriptsize{\cite{Chicco2012a}}\textsuperscript{$\star$} & \scriptsize{2012} & \scriptsize{Zero-One} & \scriptsize{Raw} & \makecell[tl]{\scriptsize{Minkowski} \\ \scriptsize{Metric with} \\ \scriptsize{$p=1,2,3,4,5$}} & \scriptsize{FCM, FL, KMn, HAC (A, S)} & \makecell[tl]{\scriptsize{1 Real} \\ \scriptsize{Commercial} \\ \scriptsize{(15 min)}} & \makecell[tl]{- RLPs (400 consumers, derived from \\ ``a representative weekday of the \\ intermediate season'')} & \scriptsize{CDI, DBI, MDI, MIA, SI, SMI, VRC, WCBCR} \\[0.15cm]

            \scriptsize{\cite{Chicco2006}} & \scriptsize{2006} & \scriptsize{Zero-One} & \makecell[tl]{CCA, \\ PCA, \\ SM} & \scriptsize{ED} & \scriptsize{FCM, HAC (A, C, W), KMn, MFL, SOM} & \makecell[tl]{\scriptsize{1 Real} \\ \scriptsize{Commercial} \\ \scriptsize{(15 min)}} & \makecell[tl]{- RLPs (234 consumers, derived by \\ averaging ``spring weekdays'')} & \scriptsize{CDI, DI, SI, DBI} \\[0.15cm]
            
            \bottomrule \hiderowcolors
            \multicolumn{9}{p{\dimexpr\textwidth-2\tabcolsep-2\arrayrulewidth-0.2cm}}{\textsuperscript{a}\tiny Curvilinear Component Analysis (CCA), Principal Component Analysis (PCA), Sammon Map (SM)} \\
            \multicolumn{9}{p{\dimexpr\textwidth-2\tabcolsep-2\arrayrulewidth-0.2cm}}{\textsuperscript{b}\tiny Braycurtis Distance (BD), Cosine Distance (CoD), Chebyshev Distance (ChD), Dynamic Time Warping (DTW), Euclidean Distance (ED), Edit Distance with Real Penalty (ERP), Hausdorff Distance (HD), Manhattan Distance (MD), Mahalanobis Distance (MAH), Pearson Correlation (PC), standardised ED (sED), squared ED (sqED)} \\ 
            \multicolumn{9}{p{\dimexpr\textwidth-2\tabcolsep-2\arrayrulewidth-0.2cm}}{\textsuperscript{c}\tiny [Modified] Follow the Leader ([M]FL), Fuzzy C-Means (FCM), Gaussian Mixture Models (GMM), Hierarchical Agglomerative Clustering (HAC) --- Linkages: Average (A), Complete (C), Single (S), Ward (W), $k$-medoids (KMd), $k$-means (KMn), Self-Organising Maps (SOM)} \\
            \multicolumn{9}{p{\dimexpr\textwidth-2\tabcolsep-2\arrayrulewidth-0.2cm}}{\textsuperscript{d}\tiny Cluster Balance (CB), Cophenetic Correlation Coefficient (CCC), Cluster Dispersion Indicator (CDI), Clustered-Vector Balance (CVB), Davies-Bouldin Index (DBI), [Modified] Dunn Index ([M]DI), Mean Index Adequacy (MIA), Mean Squared Error (MSE), Scatter Index (SI), Similarity Matrix Indicator (SMI), Silhouette Width Criterion (SWC), Scatter and Density between and within clusters (S\_Dbw), Variance Ratio Criterion (VRC), Violation of Relative Standard Error (VRSE), Ratio of Within-Cluster sum of squares to Between-Cluster variation (WCBCR)} \\
            \multicolumn{9}{p{\dimexpr\textwidth-2\tabcolsep-2\arrayrulewidth-0.2cm}}{\textsuperscript{$\star$}\tiny Comparative studies by nature, but not by name.  \textsuperscript{$\dagger$}\tiny Distances were compared using this algorithm/metric only.} \\
        \end{tabular}
    \end{adjustbox}
    \caption{Comparative studies of clustering approaches for SMTS data. The columns include the reference, year, clustering components used or compared (Normalisation, Representation, Distance and Algorithm), a description of the data, a description of the format of the clustered data, and the quantitative metrics comparisons were based on.}
    \label{Tab:ComparativeStudies}
\end{table}

The collective scope of previous comparative studies has been limited in terms of both the number and variety of included methods, as well as the range of components considered. \Cref{Tab:ComparativeStudies} catalogues the methods compared in each study by component. Whilst a majority of studies included at least two clustering algorithms, a minority of studies included multiple distance measures. Furthermore, only two of the nine studies incorporated different representation methods, and only one study compared various normalisation procedures. Given that the choice of normalisation, representation and distance also significantly influence clustering outcomes \cite{Toussaint2019,Aghabozorgi2015,Javed2020}, it is notable to see the emphasis fixed primarily on clustering algorithms. Despite this emphasis, studies tend to consider only a small subset of available clustering algorithms, with $k$-means, Hierarchical Agglomerative Clustering, Self-Organising Maps and Fuzzy $c$-means frequently constituting the majority of the pool of candidates, with few alternatives. It is also notable that many compelling components and approaches from the time series clustering literature have also been absent from these studies, including various elastic distance measures \cite{Holder2023AClustering}, $k$-shape \cite{Paparrizos2015} and more. This suggests a disconnect between the respective literature. 

It should also be noted that of the surveyed studies, six chose to compare methods from a single clustering component, fixing all others. None of the other three studies compared methods from more than two components. This means that potentially fruitful interactions between components, and method robustness in combinations cannot be observed. However this is understandable given the typical volumes of real data and the capacity for combinations to grow explosively, further exacerbated by the necessity to explore the parameter spaces of many methods. Given these observations, we have taken steps to ensure that this current study considers a wide range of methods including representations, distances \textit{and} clustering algorithms --- taken from both the SMTS and broader time series clustering literature --- along with their combinations and parameter spaces. 

These previous comparative studies are all subject to methodological limitations regarding their universal reliance on real world datasets and RVIs, which restricts the utility of their results. As indicated in \Cref{Tab:ComparativeStudies} all of the studies utilised a \textit{single} dataset for their comparisons, even in those studies noted as comparative or benchmark studies. These have also exclusively been real world datasets, which don't allow for systematically investigating how methods respond to changes in dataset properties --- something that would be possible with synthetic data. Reliance on real world datasets also begets a reliance on RVIs for establishing quantitative comparisons (the RVIs used by each study have also been listed in \Cref{Tab:ComparativeStudies}), and reliance on RVIs is a problem for two main reasons. 

Firstly, inferences based on RVIs for comparing distance measures are unreliable due to the dependence of RVI computations upon pairwise dissimilarities. A recent study \cite{Yerbury2024} demonstrated that when an RVI has been computed with a particular fixed distance measure, say the frequent default Euclidean distance, a bias can be observed towards clustering output from the same and similar distance measures. This study also demonstrated that when computing an RVI using the distance measure which matches that used for clustering, the resulting RVI values cannot be meaningfully compared across different distance measures, as each produces its own distinct statistical distribution. The bias observed in \cite{Yerbury2024} was theorised in \cite{Iglesias2013} and later referred to in \cite{Li2022}. In the latter two papers, the authors decided to instead score the same partition using multiple versions of each RVI --- specifically one version for each distance measure being compared. However this innovative ``cross-test'' concept is conditional upon the pool of candidate methods, and multiple scores must also be resolved in some way to obtain a final ranking. As normalisation procedures and representation methods also directly affect pairwise dissimilarities between objects, the same observations also apply to comparisons of these components. 

The second issue with the reliance on RVIs relates to the domain-agnostic cluster concepts upon which they are generally formulated. In \cite{Kang2015} it is suggested that regular RVIs ``are not proficient in specific applications such as electricity load profile clustering [because] they do not consider domain knowledge and only focus on the internal structure of nodes.'' This view is supported more broadly by various authors \cite{Hennig2015a,VonLuxburg2012}, who argue that the search for the
``true clusters'' in cluster analysis cannot be disentangled from the domain context and clustering aims. In \cite{Hennig2015a} it is suggested that suitable clustering approaches can only be decided upon once researchers can indicate the \textit{data analytic} characteristics their clusters should have, referred to as the \textit{cluster concept}. For example, the $k$-means algorithm is a heuristic to minimise the \textit{within cluster sum of squares}, which is equivalent to maximising the Calinksi-Harabasz index (CHI) \cite{Calinski1974AAnalysis} for fixed $k$. Thus the CHI is more likely to favour partitions produced by $k$-means than from other clustering algorithms. Hence, comparative studies based on generic RVIs will tend to recommend methods that align with the domain-agnostic cluster concepts in those selected RVIs, which may not always be optimal for SMTS clustering.

\paragraph{} \noindent
There is significant variation in the minutiae of the data clustered in these studies, spanning three axes that will be discussed in turn: clustered objects, sampling rate and source. It is important for comparative studies to clearly specify these particulars so that their conclusions can be understood within context. 

The SMTS clustering literature provides many examples of different objects being clustered depending on the objectives of the study, and these objects are usually delineated by unique time-scale configurations. For instance, clustering of 24-hour, or diurnal, subsequences referred to as \textit{Daily Load Profiles} (DLPs) is quite prevalent \cite{Abreu2012,Kwac2013,Iglesias2013,Kwac2014,Hsiao2015,McLoughlin2015,Ozawa2016,Jin,Xu2017,Darby2019VictorianYear,Yilmaz2019,Nordahl2019,Toussaint2019,Toussaint2020,Rajabi2020,Choksi2020a,Bourdeau2021,Ahir2022AData}. Depending on the application, the pool of DLPs to be clustered can be taken from different days for an individual consumer, different consumers on a single day, or from different consumers over different days. This is in contrast with the clustering of \textit{Representative Load Profiles} (RLPs), which are also diurnal subsequences used to represent the energy consumption of a single consumer by averaging many DLPs together, usually over similar loading conditions (e.g. day of week or season) \cite{Chicco2006,Chicco2012a,Flath2012}. Of the nine reported studies in \Cref{Tab:ComparativeStudies}, seven analysed DLPs, four analysed RLPs and two considered both. While they share some similarities, RLPs are much smoother than their noisy, peaky DLP derivatives \cite{Yilmaz2019}. Several authors \cite{McLoughlin2015,Jin,Yilmaz2019,Luo2020,Rajabi2020} have suggested that clustering of averaged consumer load curves is likely to miss significant intra-consumer variations between days, \cite{Jin} adding that consideration of such variability can indicate consumer suitability to various demand side management strategies. For these reasons we have decided to focus the current study on clustering of DLPs. Whilst clustering of long \textit{Multi-Day Load Profiles} is also performed \cite{Motlagh2019,Ruiz2020AData,Alonso2020}, it falls outside the scope of this particular study as longer time series require their own dedicated dimensionality reduction techniques to avoid the curse of dimensionality \cite{Aggarwal2001}.

Smart meter sampling rates also vary significantly across studies (see ``Datasets'' column in \Cref{Tab:ComparativeStudies}). In a 2017 systematic literature review of consumption classification studies using SMTS data, the authors identified that datasets with 15, 30 and 60 min rates were most commonly studied \cite{Tureczek2017}. The effect of temporal resolution on raw SMTS clustering outcomes was analysed in \cite{Granell2015ImpactsProfiles}, where it was concluded that 8 to 30 minutes provided the optimal utility for energy providers attempting to discern differences between consumers. It was further observed in \cite{Trittenbach2019} that aggregating from 30 seconds to 1 minute had a much greater impact on their metrics than aggregation at coarser resolutions, such as from 5 minutes to 10 minutes. However, conclusions from both of these studies should be treated with caution due to their comparisons of essentially different data representations using RVIs \cite{Yerbury2024}. As a middle ground between common sampling rates, we have chosen to generate synthetic data emulating a sampling rate of 30 mins. 

Finally, in the aforementioned systematic literature review, $73\%$ of surveyed studies clustered residential data, compared to $44\%$ for commercial \cite{Tureczek2017}. This predilection for residential data could simply be due to the fact that more residential data is available, or because it is more difficult to cluster. The variability within residential SMTS data is attributable to many hidden factors such as mutable human behaviours, appliances, building characteristics, etc. Commercial consumption on the other hand can be more predictable with variability somewhat attributable to the operational category. This has the effect of making privacy more difficult to preserve and clustering relatively less complicated or necessary. For these reasons, we have designed our synthetic data to emulate residential consumption patterns. That is not to suggest that our findings are irrelevant for the clustering of commercial DLPs, as there is still substantial cross-over between these types of SMTS data. 

        \section{Synthetic Data}
\label{sec:SyntheticData}

In the following section, we introduce the philosophy and design strategy that guided the development of our synthetic data. We then describe how the synthetic data were generated and explore the relevance and utility of the data through some qualitative validation.

\subsection{Design Philosophy} \label{subsec:SyntheticData-Philosophy}

Simulation studies offer a unique perspective on the comparison of clustering methods, one that cannot be fully substituted by theory or analyses of real datasets \cite{Hennig2018}. However, Hennig cautions that the value of such comparisons depends on how accurately the simulated true classes model the actual clusters that practitioners expect to identify in new, unlabelled datasets \cite{Hennig2015a}. A recent study \cite{VanMechelen2023} advocating good benchmarking practices for the comparison of clustering methods concluded by urging practitioners ``to go for a deep reflection of what constitutes, within a particular context, and given particular research questions and aims, a good or desirable clustering.'' They highlighted that the result of such reflection ``may have far-reaching consequences for the choice of relevant methods, benchmark data sets, and evaluation criteria.'' To maximise the relevance and utility of method comparisons, it is crucial for synthetic datasets to embody cluster concepts that are closely aligned with the requirements of the application context.

In \cite{VanMechelen2023} the authors suggest that practitioners consider the ``intension'' of the sought-after clusters, which refers to the organising principles that define clusters and their associated features. This can be viewed in terms of the within- and between-cluster organisation. One must determine the unifying or common ground required for elements to belong to the same cluster, and the discriminating ground for elements to belong to different clusters. These organising principles could be formal data analytic cluster concepts \cite{Hennig2015a}, such as small within-cluster dissimilarities, large between-cluster dissimilarities, being densely connected regions in the data space, being characterised by a small number of variables, or being well fitted by a particular probability model.
However in \cite{Hennig2018}, Hennig suggests that if formal cluster concepts are too rigid to capture practical nuances, then even informal descriptions of the clusters of interest should be preferred rather than providing no definitions at all, or making appeals to the reader's intuition.

In this study, we aimed to imbue our synthetic data with informal cluster concepts considered fundamental for various applications of SMTS clustering. Consideration of literature and consultation with domain experts resulted in the following central organising principle, which regulates the cluster labels in our various synthetic datasets: 
\begin{quote}
    \textbf{\textit{Differences in the timing, number, shape and relative magnitudes of peak energy consumption events establish suitable grounds for the separation of residential daily load profiles into clusters.}} 
\end{quote}

\noindent This principle was informed by the frequent focus in the literature on coincident peak usage \cite{Liu2019,Liu2013}, and its consequences for demand response initiatives and grid operation \cite{Chai2024,Pelekis2023,Satre-meloy2020}. 

Additional consideration was given to the relevance and desired extent of time series ``invariances'' for clustering DLPs. Distance measures and representations can be invariant to a range of time series transformations, such that two time series that are distorted versions of one another can still be considered as similar. These invariances have typically been used to motivate novel distance measures \cite{Batista2014,Paparrizos2015}, and are well understood for selected distance measures and their parameters. However there are many viable candidate methods for which invariances are much less transparent, for instance, autoencoder representations. Commonly required time series invariances are presented in \cref{Tab:Invariances} alongside considerations of their utility specifically for clustering of DLPs. For a detailed discussion of invariances more generally, see \cite{Batista2014}. With synthetic data informed by this organising principle, our study is equipped to reveal and compare the underlying cluster concepts and invariances inherent to each of the methods themselves.

\begin{table}[!htb]
    \scriptsize
    \rowcolors{1}{TableGray}{TableWhite}
    \begin{adjustbox}{center}
        \begin{tabular}{p{1.15cm} | p{7cm} p{7cm}} \toprule \hiderowcolors 
            \textbf{Invariance} & \textbf{General Description} & \textbf{Relevance to Clustering of Daily Load Profiles} \\ \midrule \showrowcolors

            Amplitude and Offset & Invariance to affine transformations, i.e. the similarity between $X_t$ and $Y_t$ shouldn't be different to the similarity between $a X_t+b$ and $Y_t$. & This invariance is generally required, as it ensures that clustering is informed by load shapes rather than load magnitudes. It can be facilitated by appropriately normalising the data prior to clustering. \\[0.7cm] 

            Local Scaling & Some subsequences in two series may not be perfectly aligned in time, while the rest of the series are still aligned --- we may want to consider the two series as similar anyway. & Local energy consumption events in DLPs can occur slightly out of step with one another. Depending on the application, we may want to treat these events as similar despite small differences in their timing.   \\[0.7cm]

            Uniform Scaling & Similar time series may not be globally aligned due to differences in sampling rates or the speed of observed phenomena, often yielding series of differing lengths. & DLPs from the same dataset almost always share a sampling rate, so uniform scaling invariance is not required for clustering. This invariance can be accommodated in a pre-processing step by stretching time series by a constant factor. \\[0.7cm]

            Phase & Similar time series which lack a definitive starting point (e.g. periodic time series such as heartbeats) may not be appropriately aligned in time due to global circular shifts. & The hours of the day certainly provide a definitive alignment for DLPs, but we may wish to allow $X_t$ and $Y_t$ to be assessed as similar even if $Y_t = X_{t-k}$ for some small constant $k$. \\[0.7cm]

            Occlusion & If a subsequence is missing from one time series, we may wish to still consider it similar to another if the rest matches well. & Data can be missing for various reasons, and occlusion invariance may be preferable in this case --- though imputation is more common.  \\[0.7cm]

            Complexity & Two complex time series with the same shape may be judged as more similar to simple time series than to each other, because their complexity creates more opportunities for small differences to compound. Complexity invariance remedies this counterintuitive outcome. & Shape is important for clustering of DLPs, and a degree of complexity invariance can be useful to ensure that clustering takes shape into account. \\[0.7cm]

            \bottomrule \hiderowcolors
        \end{tabular}
    \end{adjustbox}
    \caption{Common time series invariances and their relevance to the clustering of DLPs. Note that $X_t$ and $Y_t$ denote time series with $t\in \left \{ 1,2, \ldots \right \}$.}
    \label{Tab:Invariances}
\end{table}

\subsection{Design Strategy} \label{subsec:SyntheticData-Strategy}

Frequently synthetic SMTS data is produced in the form of multi-day load profiles \cite{Alonso2020,Zhang2018a}. These methods are not readily applicable to our DLP-focussed benchmark as it is not obvious how cluster labels should be meaningfully attributed to their constituent DLPs. As previously mentioned, it is important that cluster labels for a benchmark study correspond to natural, recoverable clusters \cite{Farber2010}. Furthermore, real DLPs have often been required as seeds in various synthetic generation processes \cite{Iftikhar2016,AlKhafaf2021AConsumers} --- particularly for Generative Adversarial Networks (GANs) which have become a popular method for producing synthetic load profiles \cite{Zhang2018a,Ravi2022,Chai2024}. When the goal is to generate synthetic instances representing different classes, a sufficient number of real DLPs would need to be accumulated for each desired class. Manual exploration of large datasets by domain experts would be prohibitively laborious, compounded by the challenging cognitive burden of defining and maintaining consistent classification criteria across the immense volume of DLPs. 
If clustering methods or similarity computations were used to automate or assist this process, then any benchmarking experiments would likely be biased towards the method used to segment the real data. 

\begin{figure}[!b]
    \centering
    \includegraphics[width=0.99\linewidth]{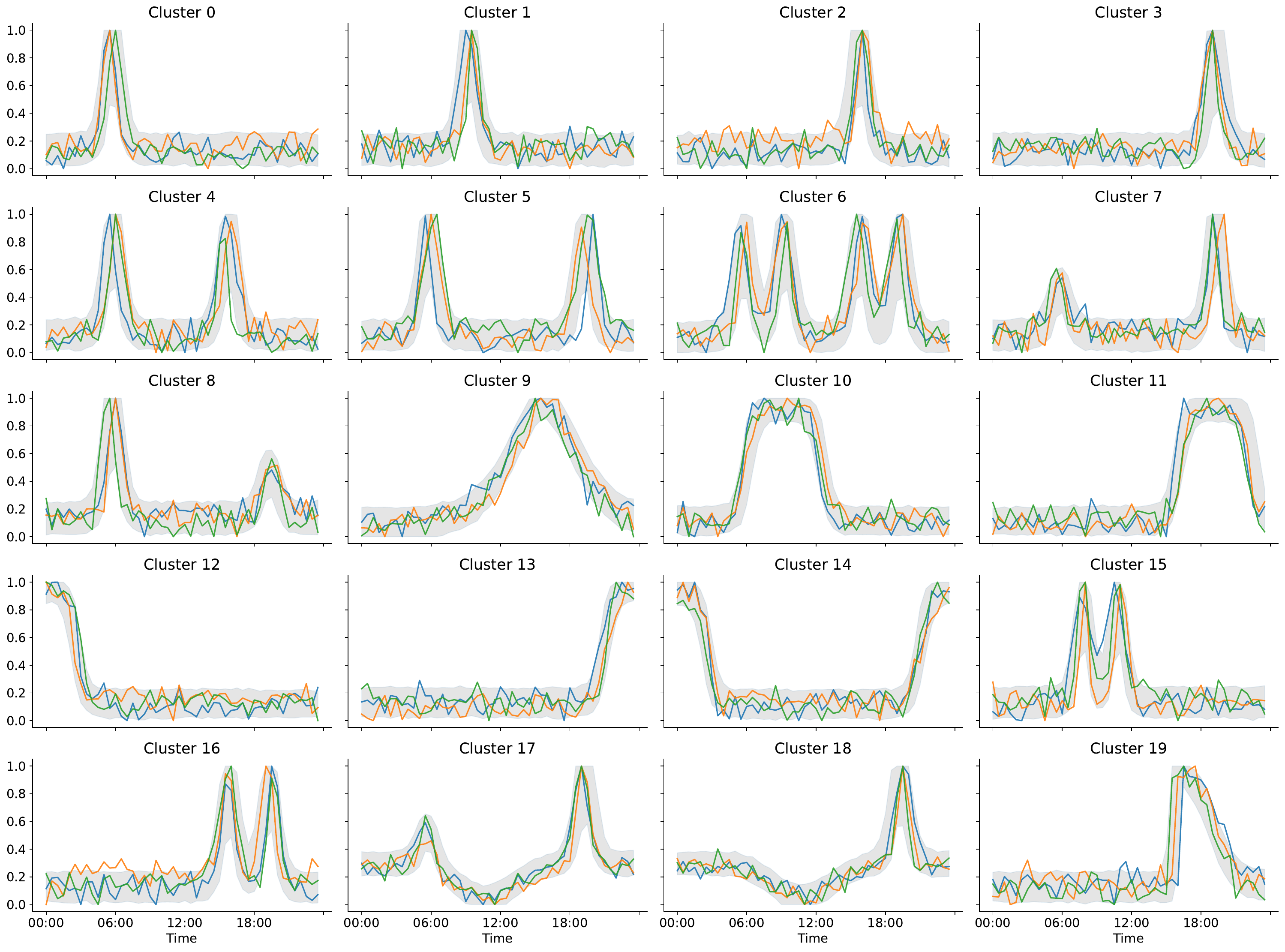}
    \caption{Three samples and pointwise 90 percentile interval for the suite of 20 synthetic clusters comprising our baseline synthetic data.}
    \label{Fig:ExampleSyntheticClusters}
\end{figure}

For these reasons, we have elected to design the collection of synthetic DLP clusters and their generators from scratch, balancing three key objectives. Firstly, the synthetic clusters should cover a range of DLP shapes which are encountered in real-world data and deemed fundamental by the engaged domain experts. Too few clusters would oversimplify the problem and fail to reflect real-world complexity, while too many clusters would be too prescriptive, reducing the study's applicability. Secondly, there should be potential for meaningful and informative conflict between the synthetic clusters when assessing the candidate methods, and that conflict should be in accordance with the aforementioned central organising principle, i.e. differences in the timing, number, shape and relative magnitudes of peak energy consumption events. Thirdly, the synthetic cluster labels must be recoverable so that external validation remains meaningful. Thus the within and between cluster variation should be coordinated to ensure that the accuracy of generated cluster labels can be relied upon.

Consideration of these requirements resulted in a suite of 20 distinct cluster shapes, which have been presented with random samples and a central 90 percentile interval in \Cref{Fig:ExampleSyntheticClusters}. The framework used to generate synthetic instances is discussed in \Cref{subsec:SyntheticData-GenerationFramework}, and validation of these instances with real-world data is discussed in \Cref{subsec:SyntheticData-Validation}. These shapes were selected as they represent fundamental diurnal consumption patterns that could occur across different seasons, rather than being tied to any specific season. While this selection of shapes may not be exhaustive, it provides a robust benchmark for candidate methods --- good performance with our synthetic data is a necessary condition for good performance on real-world data, as poorly performing methods cannot be expected to improve when faced with the further complexity present in real-world data, which includes additional and intermediary shapes. The 20 clusters strike a balance between comprehensiveness and generalisability, providing enough potential for informative and meaningful conflict between cluster shapes when comparing various candidate methods. These potential conflicts, which probe each aspect of our central organising principle, have been visualised and discussed in \Cref{sec:Appendix}

\subsection{Generation Framework} \label{subsec:SyntheticData-GenerationFramework}

\begin{figure}[!b]
    \centering
    \subfigure[Stage One: Characteristic Curve]{\includegraphics[width=0.32\textwidth]{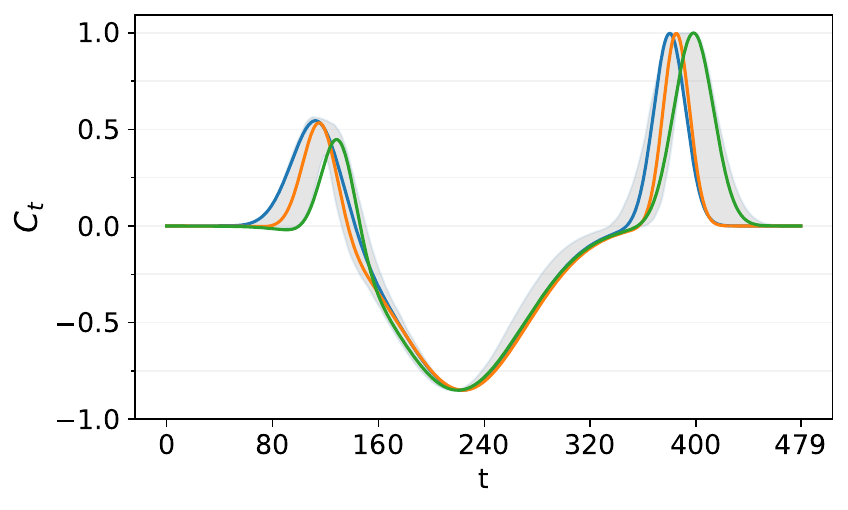}} 
    \subfigure[Stage Two: STAR Model Simulation]{\includegraphics[width=0.32\textwidth]{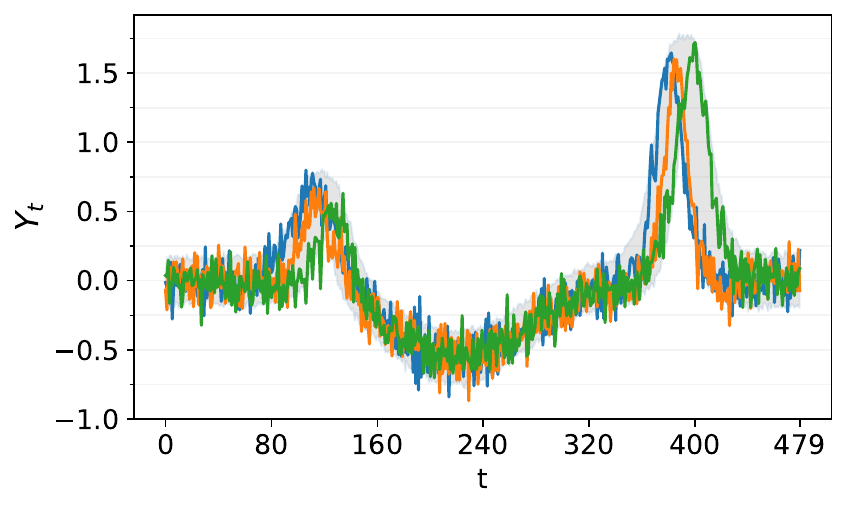}} 
    \subfigure[Stage Three: Downsampling and Normalisation]{\includegraphics[width=0.32\textwidth]{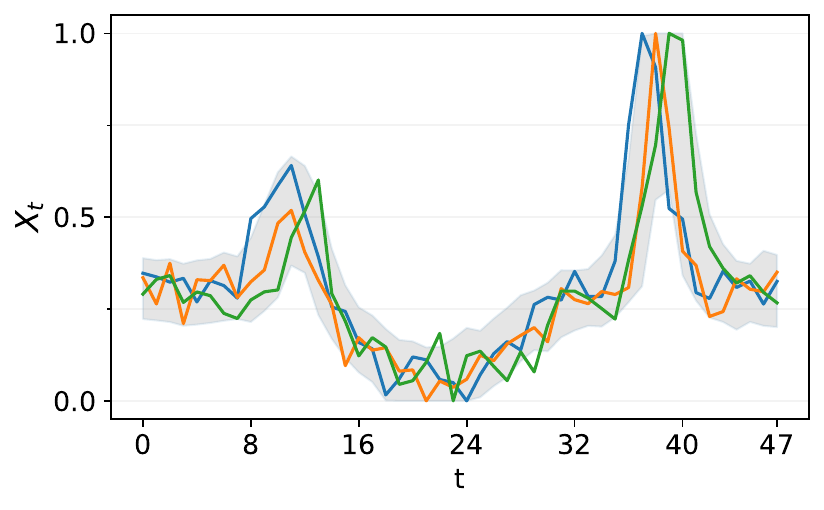}}
    \caption{The three stages of the synthetic data generation process for one particular cluster. A pointwise 90 percentile interval is presented for the time series from each stage with three random instances overlayed in orange, blue and green.}
    \label{Fig:SyntheticDataGenerator}
\end{figure}

Our synthetic data generator produces half-hourly diurnal DLPs, $X_t$ with $t \in \left\{0,1,\ldots,47\right\}$, through a three-staged process illustrated in \Cref{Fig:SyntheticDataGenerator}. Each cluster represents a distinct underlying usage pattern, characterised by the shape, number, timing and relative magnitude of peak energy consumption events, with controlled variation therein to maintain recoverable clusters. The peak consumption events are modelled independently, allowing for more realistic temporal variation within clusters. The first stage involves generating a random time series $C_t$ from the cluster's \textit{characteristic curve} generating function. These generating functions, developed in consultation with domain experts, combine one or more distributional functions with randomly varying parameters within predetermined constraints. All characteristic curves are long, bounded time series ($0 \leq | C_t | \leq 1$, $t \in \left\{0,1,\ldots,479\right\}$), with examples from one cluster shown in the left window of \Cref{Fig:SyntheticDataGenerator}. The specific implementation details for each cluster's generating function are not crucial to understanding the methodology, however they are accessible in the \href{https://github.com/yerbles/Smart-Meter-Time-Series-Clustering-Comparative-Study/tree/main}{supplementary materials}.

In the second stage, the characteristic curves are used like the transition function in a Smooth Transition Autoregressive (STAR) model simulation to produce a second long time series $Y_t$. In particular, 
\begin{equation}
    Y_t = \theta_{1} Y_{t-1} + \left ( 1 + \theta_{2} Y_{t-1} \right ) C_t + W_t, \text{  where  } 
    W_t \sim \begin{cases}
        N(0,\sigma_{L}) & \text{if }C_t < 0.5 \\
        N(0,\sigma_{H}) & \text{if }C_t \geq 0.5
    \end{cases},
    \label{Eqn:SyntheticDataGenerator}
\end{equation}
$\sigma_L=0.12$, $\sigma_H=0.1$, $\theta_1=-0.1$, $\theta_2=0.5$, and $Y_0 \sim N(0,\sigma_L)$ or $N(1.5, \sigma_H)$ depending on the cluster. These parameters were chosen to balance realism with label recoverability, drawing on domain expertise and validation against real-world smart meter data.

The simulated time series in the middle window of \Cref{Fig:SyntheticDataGenerator} have utilised the characteristic curves from the left window. The final DLPs ($X_t$) are then obtained from these simulated time series by subsampling
every 10\textsuperscript{th} point starting at a random index from 0 to 9 inclusive 
and subsequently applying Min-Max normalisation so that they range strictly between 0 and 1, i.e. 
\begin{equation*}
    X_t = \frac{Y_t - \min_{t} \left\{Y_t\right\}}{\max_{t} \left\{Y_t \right\} -  \min_{t} \left\{Y_t\right\}} \text{  for  }  t\in \left\{0+u,10+u,\ldots,470+u\right\} \text{  where  } u \sim \text{DU}\{0,9\},
\end{equation*}
where DU$\{a,b\}$ is the discrete uniform distribution from $a$ to $b$ inclusive. Datasets generated from these clusters with $\sigma_L=0.12$ and $\sigma_H=0.1$ are referred to as \textit{baseline} datasets. As previously mentioned, key characteristics of this baseline synthetic data will be varied in the course of this comparative study in an effort to compare the performance of clustering methods under conditions that more closely resemble those found in real-world datasets. These characteristics include the number of time series and clusters, amplitude noise, cluster balance, cluster separation, and the presence of outliers.

The code implementing this synthetic data generation framework has been made publicly available alongside other supplementary materials on our \href{https://github.com/yerbles/Smart-Meter-Time-Series-Clustering-Comparative-Study/tree/main}{GitHub repository}.

\subsection{Validation} \label{subsec:SyntheticData-Validation}

\begin{figure}[!b]
    \centering
    \includegraphics[width=0.99\linewidth]{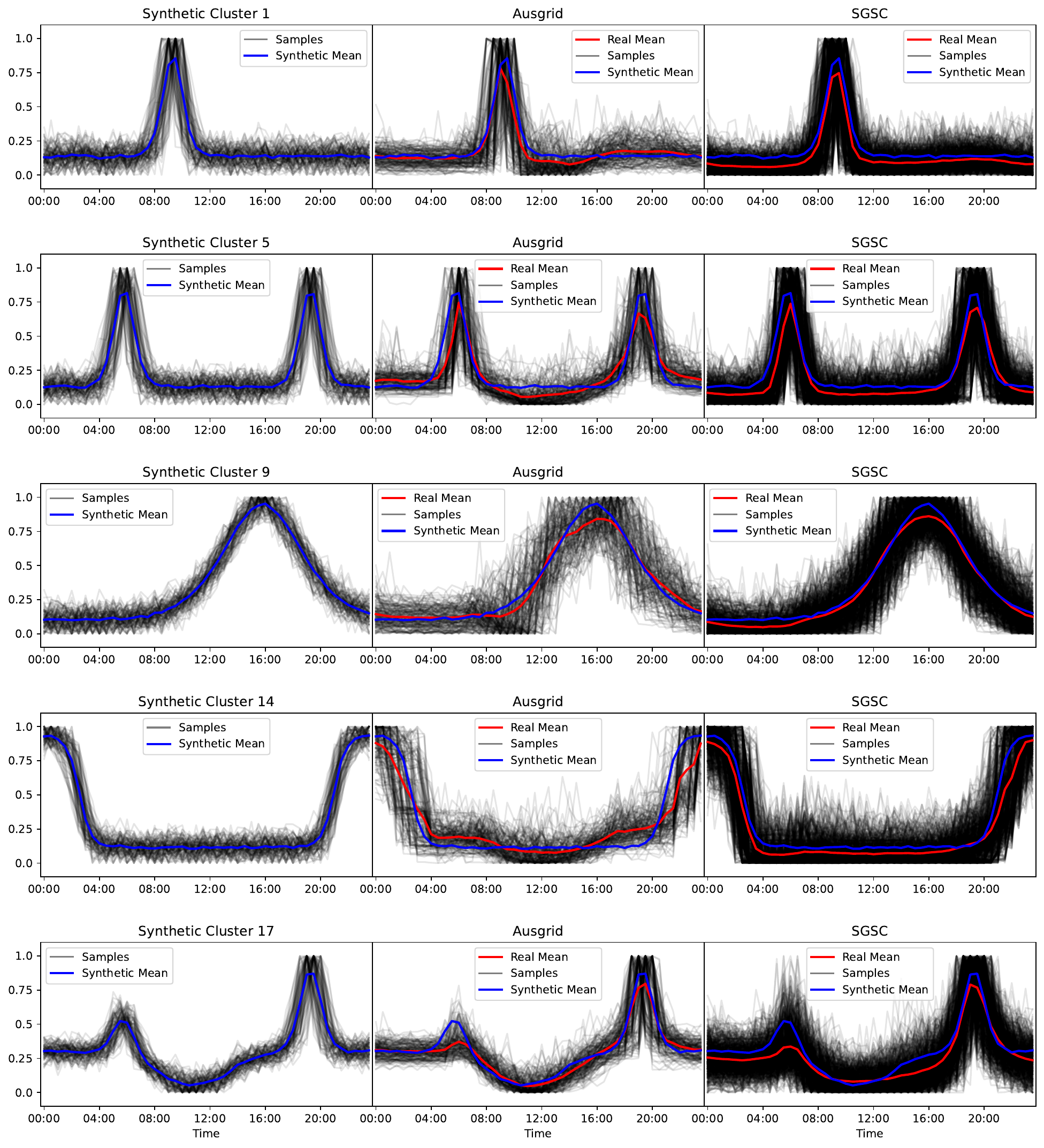}
    \caption{For each of the presented synthetic DLP clusters, real load curves have been collected which were in the closest $0.05\%$ of real DLPs to any of the synthetic profiles according to the Euclidean distance. There are 164 DLPs from the Ausgrid dataset and 721 from the SGSC dataset.}
    \label{Fig:DistancePlots}
\end{figure}

It should be emphasised that we are not claiming this synthetic data can stand in place of real-world data in terms of scope, utility or even fidelity. Indeed, our synthetic data was deliberately designed not to comprehensively imitate real-world data, which contains an abundance of DLP shapes due to various behavioural, geographical, cultural and meteorological factors. Were we able to capture a majority of these shapes within a synthetic dataset, clustering techniques would no longer be necessary --- we could consider classification techniques instead. Rather, we aimed to establish a controlled benchmark relying on fundamental consumption patterns that methods should be capable of identifying as a minimum requirement. We are claiming that this synthetic data contains fundamental shapes that are encountered in real-world data, making it useful for revealing key differences in the suitability of various methods for the clustering of residential DLP patterns. Our conclusions are therefore best interpreted as necessary (but not sufficient) conditions for good performance on real data --- methods that struggle with this simpler synthetic data are unlikely to yield better results when faced with the greater complexity of real-world data.

We now present results from two validation procedures used to refine the final load shapes and their parameters.
The first of these procedures establishes the presence of these synthetic DLP shapes in real-world datasets by comparing the similarity of the synthetic instances with real-world DLPs. The second procedure visualises the scope of shapes covered by our synthetic data within the scope of shapes from two real-world datasets.
Throughout this section we have utilised real residential DLPs from two publicly available Australian smart meter datasets: the Ausgrid Solar Home Electricity dataset \cite{Ausgrid2013} and the Smart-Grid Smart-City (SGSC) dataset \cite{Ausgrid2014}. We utilised all $328,800$ DLPs from the Ausgrid dataset, and used a subset of $1,442,845$ DLPs from the SGSC dataset which aligns with the pre-processing and filtering applied in \cite{Roberts2019}. These real DLPs have also been Min-Max normalised.

The first procedure confirms the relevance of the basic synthetic cluster shapes by collecting similarly shaped DLPs from the two real-world datasets. In \Cref{Fig:DistancePlots} the left plots show a sample of 100 baseline synthetic DLPs for a subset of five clusters, while to the right are the closest $0.05\%$ of DLPs from the entirety of the real-world datasets according to the Euclidean distance. We used the Euclidean distance to search for the most visually similar profiles due to its insistence on strict alignment of events in time. As noted above, the synthetic clusters are not exhaustive, but they contain shapes that are prevalent within real-world datasets. Similar plots for the remainder of the clusters are presented in the \href{https://github.com/yerbles/Smart-Meter-Time-Series-Clustering-Comparative-Study/tree/main}{supplementary materials}. There is less variability within the synthetic DLPs compared to the real DLPs as it is important that the baseline synthetic ground truth labels are guaranteed to be recoverable. Further experiments later in the paper explore how robust clustering methods are to increasing variability and other forms of additional real-world complexity.

\begin{figure}[!t]
    \centering
    \includegraphics[width=0.99\textwidth]{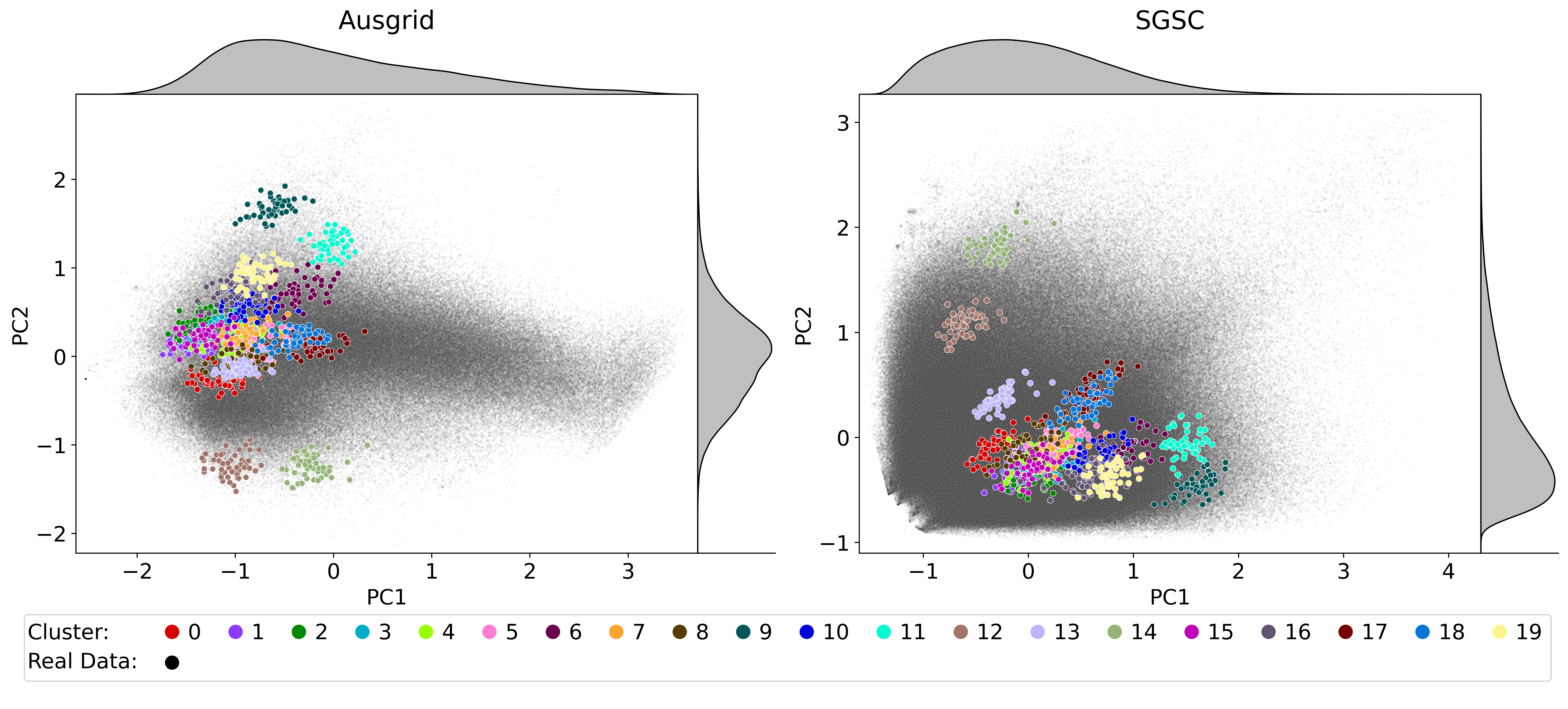}
    \caption{The first two principal components from a PCA decomposition of two real-world datasets (Ausgrid on the left, SGSC on the right) with subsequently transformed synthetic data overlayed. The first two principal components combined for the Ausgrid and SGSC datasets explain $49.6\%$ and $29.0\%$ of the variance respectively.}
    \label{Fig:PCA}
\end{figure}

The second of these procedures uses Principal Component Analysis (PCA) to provide an indication of the distribution of our synthetic data amongst real-world data \cite{Yoon2019Time-seriesNetworks}. A two-dimensional projection of a sample of the synthetic DLPs overlays the real DLPs in \Cref{Fig:PCA}. A sample of 50 synthetic DLPs from each cluster were transformed using the principal components computed for the entirety of the Ausgrid and SGSC datasets respectively, spanning multiple years of DLPs from multiple households. Most of the synthetic clusters are concentrated within the densest regions of the point cloud. Large regions in these projections are not covered by our cluster shapes due to the greater complexity of the real-world DLPs compared to the fundamental shapes in our synthetic data. This demonstrates that, while our synthetic data is not a comprehensive substitute for real-world data in terms of the scope of DLP shapes, it effectively captures a small range of core patterns, making it suitable for our comparative purposes.

\paragraph{} \noindent 
While the validation procedures presented above demonstrate that our synthetic cluster shapes are indeed represented in real-world datasets, we deliberately avoided attempting to quantify the ``coverage'' of real-world patterns by these fundamental shapes. Such attempts would contradict our methodological approach, as the value of these fundamental shapes lies in their ability to probe clustering methods’ capabilities rather than in comprehensive coverage of all possible load profiles. Moreover, quantification would also present an ill-posed problem requiring untenable assumptions. As noted in Section 3.2, if clustering methods or similarity computations were used to segment real data for synthetic data generation, it could bias the benchmarking experiments towards the chosen method. Similarly, if synthetic shapes were tailored to achieve some minimum coverage of real-world profiles according to an arbitrary distance threshold, this could bias our benchmark in favour of the distance measure used, such as the Euclidean Distance. The same concerns apply to the use of PCA, or any other representation methods or distance measures for that matter. Furthermore, we can infer that a method’s ability to effectively distinguish our synthetic data implies coverage of a much larger space of real-world patterns than just the 20 specific shapes themselves. A method capable of distinguishing between our fundamental shapes can be expected to effectively separate variations where peaks are shifted in time or magnitude within reasonable bounds, as well as profiles that combine multiple of these shapes. It is for this reason that our experts have described them as ``fundamental''.

        \section{Methods}
\label{sec:Methods}

This section begins by introducing the diverse range of both clustering components and approaches considered in this study. We then describe the phased experimental methodology used to compare their suitability for clustering DLPs.

\subsection{Clustering Methods}
This study focused on the three most pivotal components of a clustering approach: representation method, distance measure, and clustering algorithm. To provide a comprehensive comparison, we selected methods from a wide range of sources for each component. This included tools designed specifically for or frequently used in the SMTS clustering literature, but also considered suitable candidates from the broader time series and general clustering literature. It was also required for this study that any candidate methods be easily implemented, or have an implementation available already in Python. We will now describe these three components, and briefly introduce the methods we selected for each component.

\textbf{Representation method} is an umbrella term for any feature-extraction method or transformation of the \textit{raw} data objects which is intended to improve clustering outcomes or efficiency. An effective representation should emphasise characteristics that are most relevant or informative for grouping. Furthermore, representations are critical for dimensionality reduction in the context of large-scale data, where they can moderate the influence of noise and improve the computational efficiency of a clustering approach \cite{Aghabozorgi2015, Wang2013b}. Along with specialist distance measures, representations are often utilised specifically to make more complicated data types (like time-series) compatible with the wealth of existing feature-vector clustering tools.

The representation methods selected for the present study are described in \Cref{Tab:RepresentationMethods}. Half of these methods are commonly encountered in SMTS and general time series clustering and classification, including DWT, PAA, PCA and SAX. Both GAF and MTF were introduced in \cite{Wang2015} where they were paired with tiled convolutional neural networks for time series classification tasks. We have also considered features from LSTM autoencoders, using 16 distinct architectures (i.e. numbers of layers and nodes-per-layer --- see table in our \href{https://github.com/yerbles/Smart-Meter-Time-Series-Clustering-Comparative-Study/tree/main}{supplementary materials} for more details). Finally, we have proposed a feature bagging method which is a variant of those advanced in previous studies \cite{Wang2006,Rasanen2009}, where subsets of statistical features were curated to represent the time series. For the method we refer to as \textit{Bag of Features} (BOF), we have collected a diverse set of temporal, statistical, and spectral time series features, such as the mean, variance, longest streak below the mean, Hurst exponent, KPSS test statistic, spectral entropy and time of maximum value. In total, we extracted 76 unique features that capture various aspects of time series behaviour and structure (see table in our \href{https://github.com/yerbles/Smart-Meter-Time-Series-Clustering-Comparative-Study/tree/main}{supplementary materials} for a full list of included features). We then applied PCA to transform the features into orthogonal principal components ordered by explained variance, retaining the top $n_c$ components that explained the most variance.

\begin{table}[t]
    \scriptsize
    \rowcolors{1}{TableGray}{TableWhite}
    \begin{adjustbox}{center}
        \begin{tabular}{p{0.35cm} p{1.5cm} p{0.84cm} p{3.7cm} p{1.6cm} p{2.8cm} p{3.55cm}} \toprule \hiderowcolors 
            \textbf{Ref.} & \textbf{Name} & \textbf{Acronym} & \textbf{Parameters} & \makecell[tl]{\textbf{Python} \\ \textbf{Package}} & \textbf{Default Parameters}\textsuperscript{a} & \makecell[tl]{\textbf{Tested Parameter} \\ \textbf{Values}} \\ \midrule \showrowcolors
            
             --- & \makecell[tl]{Bag of \\ Features} & BOF & \makecell[tl]{Number principal components ($n_c$) \\ Feature Normalisation ($F$)} & tsfel, tsfresh, antropy, nolds, statsmodels & \makecell[tl]{$n_c=76$ (all features) \\ $F$: None} & \makecell[tl]{$n_c \in \left\{5,10,\ldots,75\right\}$ \\ $F$: Min-Max, None} \\[0.15cm]
             
             \cite{Turap} & Discrete Wavelet Transform & DWT & \makecell[tl]{Number interpolant points\textsuperscript{$\star$} ($n_{int}$) \\ Mother wavelet ($\Phi$) \\ Decomposition level ($l$) \\ Retained coefficients ($n_{c}$)} & scipy, pywt & \makecell[tl]{$n_{int} = 64$ (next power of 2) \\ $\Phi$: Daubechies 1 ($n_f^{\dagger}=2$) \\ $l = \lfloor \log{\left(\frac{n_{int}}{n_f-1}\right)}/\log(2)   \rfloor$ \\ $n_{c} = n_{int}$} & \makecell[tl]{$\Phi$: Daubechies 1-5 \\ $l$: All to maximum \\ $n_c$: All, approximation coefficients, \\ latest pair of approximation and \\ detail coefficients}\\[0.15cm]

            \cite{Wang2015} & \makecell[tl]{Grammian \\ Angular Field} & GAF & \makecell[tl]{Image Size ($n_i \times n_i$) \\ Type ($\tau$)} & pyts & \makecell[tl]{$n_i = n$ \\ $\tau$: Summation} & \makecell[tl]{$n_i \in \left\{2,3,\ldots,48 \right\}$ \\ $\tau$: Summation, \\ Difference} \\[0.15cm]

             \cite{6789445} & \makecell[tl]{LSTM \\ Autoencoder} & LSTM & \makecell[tl]{Architecture ($\alpha$) \\ Training batch size ($b$) \\ Number of training epochs ($n_e$)} & \makecell[tl]{pyts, \\ tensorflow} & \makecell[tl]{$\alpha$: A \\ $b$: 10 \\ $n_e$: 100} & \makecell[tl]{$\alpha$: $A,B,\ldots,O,P$ \\ $b \in \left\{ 10, 50, 100 \right\}$ \\ $n_e \in \left\{ 100, 500, 1000 \right\}$} \\[0.15cm]

             \cite{Wang2015} & \makecell[tl]{Markov \\ Transition Field} & MTF & \makecell[tl]{Image Size ($n_i \times n_i$) \\ Number of bins ($n_b$) \\ Bin boundary strategy ($b$)} & pyts & \makecell[tl]{$n_i = n$ \\ $n_b = 5$ \\ $b$: Quantile} & \makecell[tl]{$n_i \in \left\{2,3,\ldots,48 \right\}$ \\ $n_b \in \left\{2,3,\ldots,26\right\}$ \\ $b$: Quantile, Uniform, Normal} \\[0.15cm]

             \cite{Keogh2001DimensionalityDatabases} & \makecell[tl]{Piecewise \\Aggregate \\Approximation} & PAA & Window size ($w$) & pyts & \makecell[tl]{$w = 1$ \\ (identitcal to raw$+$ED)} & $w \in \left\{1,2,\ldots,24\right\}$ \\[0.15cm]
             
             \cite{Jollife2016} & \makecell[tl]{Principal \\ Component \\ Analysis} & PCA & Number principal components ($n_c$) & scikit-learn & \makecell[tl]{$n_c = n$ \\ (identitcal to raw$+$ED)} & $n_c \in \left\{1,2,\ldots,48\right\}$ \\[0.15cm]
             
             \cite{Lin2007ExperiencingSeries} & \makecell[tl]{Symbolic \\ Aggregate \\ Approximation} & SAX & \makecell[tl]{Distance measure ($d$) \\ Number of bins ($n_b$) \\ Bin boundary strategy ($b$)} & pyts & \makecell[tl]{$d$: MINDIST \\ $n_b=4$ \\ $b$: Quantile} & \makecell[tl]{$d$: MINDIST, LCSS, \\Levenshtein and variant \\ $n_b \in \left\{2,3,\ldots,26\right\}$ \\ $b$: Quantile, Uniform, Normal} \\[0.15cm]
             
            \bottomrule \hiderowcolors
            \multicolumn{7}{p{\dimexpr\textwidth-2\tabcolsep-2\arrayrulewidth-1.2cm}}{\textsuperscript{a} Defaults taken from package or proposing paper. If neither exists, a practical and fair default has been chosen.}  \\
            \multicolumn{7}{p{\dimexpr\textwidth-2\tabcolsep-2\arrayrulewidth-1.2cm}}{\textsuperscript{$\star$}These parameters have been fixed for our study.} \\
            \multicolumn{7}{p{\dimexpr\textwidth-2\tabcolsep-2\arrayrulewidth-1.2cm}}{\textsuperscript{$\dagger$}Wavelet filter length ($n_f$).} \\
        \end{tabular}
    \end{adjustbox}
    \caption{The 8 representation methods investigated in this study. The Euclidean distance has been used with all of these representations --- excepting SAX --- to establish DLP dissimilarities. For SAX, 4 different distance measures have been considered. Note that $n=48$ is the length of the half-hourly DLPs, and ``raw'' refers to the raw data representation.}
    \label{Tab:RepresentationMethods}
\end{table}

A \textbf{distance measure} is a function of two data objects which quantifies their ``sameness''. They are required for the vast majority of clustering algorithms, and representation methods typically specify a subset of compatible measures. The popular Euclidean Distance (ED) \cite{Aggarwal2014} can be applied across a swathe of different data representations, while others are specific to one representation, such as MINDIST for the Symbolic Aggregate Approximation (SAX) \cite{Lin2007ExperiencingSeries}. Some measures satisfy all of the properties of a mathematical metric including non-negativity, identity of indiscernibles, symmetry, and triangle inequality. These properties can be leveraged to increase computational efficiency \cite{CHEN2004}, but partial violations of them can introduce fruitful flexibility, granting measures those aforementioned invariances which have significant utility for more complex data types \cite{Dove2023, Paparrizos2015}. A distinction should be recognised between similarity and dissimilarity measures, with each term conveying how large values of the measure are to be interpreted. Similarity measures yield larger values for more similar objects (0 for minimally similar objects), while dissimilarity measures yield larger values for less similar objects (0 for maximally similar objects). All distance measures used in this paper are formulated as dissimilarity measures; thus from hereon the terms will be used interchangeably.

The distance measures selected for the present study are listed in \Cref{Tab:DistanceMeasures}. We included all distance measures identified in the previous comparative literature (\Cref{Tab:ComparativeStudies}), as well as compelling candidates from the time series classification and clustering literature, such as MSM, MPD, CID, TWED, and SBD. Two of the included distance measures were designed specifically for clustering SMTS data, namely Flexibility distance (FD) \cite{Yuan2023a} and $k$-sliding distance (KSD) \cite{Kang2015}. FD quantifies the amount of effort necessary to transform one time series into another by accounting for both variations in amplitude and shifts in time. KSD targets demand response applications of clustering by allowing flexible alignments of points within a specified window, providing local scaling invariance. Unlike DTW, both FD and KSD permit non-sequential point alignments. All of the distances in \Cref{Tab:DistanceMeasures} were applied to the DLPs using their raw representation, and ED was applied to the features obtained for all representations listed in \Cref{Tab:RepresentationMethods}, except for the string-based SAX representation. SAX was paired with four compatible distance measures: the original MINDIST measure proposed alongside it, the longest common subsequence, the standard Levenshtein distance, and a variant of Levenshtein distance. This variant penalises edits based on the magnitude of the difference between ordered letters in the SAX alphabet.

Representation methods and distance measures (alongside normalisation procedures) determine a unique spatial embedding of a set of data objects, dictating how the objects are positioned relative to one another. Changing either of these components \textit{or} their parameters fundamentally alters the pairwise relationships between objects. To recognise this fact, one combination (and parametrisation) of these components is referred to in this paper as a \textit{similarity paradigm}.

\begin{table}[t]
    \scriptsize
    \rowcolors{1}{TableGray}{TableWhite}
    \begin{adjustbox}{center}
        \begin{tabular}{p{0.35cm} p{3.45cm} p{0.8cm} p{3.2cm} p{1cm} p{2.9cm} p{2.75cm}} \toprule \hiderowcolors 
            \textbf{Ref.} & \textbf{Name} & \textbf{Acronym} & \textbf{Parameters} & \textbf{Python Package} & \textbf{Default Parameters\textsuperscript{a}} & \makecell[tl]{\textbf{Tested Parameter} \\ \textbf{Values}}\\ \midrule \showrowcolors

            \cite{Bray1957AnWisconsin} & Braycurtis Distance & BD & --- & scipy & --- & --- \\[0.15cm]

            \cite{Lance1967Mixed-DataSystems} & Canberra Distance & CaD & --- & scipy & --- & ---\\[0.15cm] 

            \cite{Aggarwal2014} & Chebyshev Distance & ChD & --- & scipy & --- & --- \\[0.15cm]

            \cite{Batista2014} & Complexity Invariant Distance & CID & --- & --- & --- & ---  \\[0.15cm]

            \cite{Aggarwal2014} & Cosine Distance & CoD & --- & scipy & --- & ---  \\[0.15cm]

            \cite{Sakoe1978} & Dynamic Time Warping & DTW & Window size ($w$) & aeon & $w = n$ (i.e. no window) & $w \in \left\{1,2,\ldots,48\right\}$   \\[0.15cm] 

            \cite{10.1145/1066157.1066213} & Edit distance for Real Sequences & ERS & \makecell[tl]{Window size ($w$)  \\ Matching threshold ($\varepsilon$)} & aeon & \makecell[tl]{$w = n$ \\ $\varepsilon=\max\left\{\sigma_{X}, \sigma_{Y}\right\}/4$} & \makecell[tl]{$w \in \left\{1,2,\ldots,48\right\}$ \\ $\varepsilon \in \left\{ 0,0.1,0.2,\ldots,1\right\}$}  \\[0.15cm]
            
            \cite{CHEN2004} & Edit distance with Real Penalty & ERP & \makecell[tl]{Window size ($w$)  \\ Gap penalty ($g$)} & aeon & \makecell[tl]{$w = n$ \\ $g=0$} & \makecell[tl]{$w \in \left\{1,2,\ldots,48\right\}$ \\ $g \in \left\{ 0,0.1,0.2,\ldots,1\right\}$} \\[0.15cm]

            \cite{Aggarwal2014} & Euclidean Distance & ED & --- & scipy & --- & ---  \\[0.15cm]

            \cite{Yuan2023a} & Flexibility Distance & FD & \makecell[tl]{Amplitude weights\textsuperscript{$\star$} ($A$) \\ Temporal weights\textsuperscript{$\star$} ($T$)} & --- & \makecell[tl]{$A_{i,j}=1$ \\ $T_{i,j} = \frac{\max_t\left\{X_t,Y_t\right\} - \min_t\left\{X_t, Y_t\right\}}{n}$} & ---   \\[0.15cm]

            \cite{Taha2015} & Hausdorff Distance & HD & --- & scipy & --- & ---  \\[0.15cm] 

            \cite{Howell2012StatisticalPsychology} & Kendall's Tau & KT & --- & scipy & --- & ---  \\[0.15cm]
            
            \cite{Kang2015} & $k$-Sliding Distance & KSD & Sliding window width ($w$) & --- & \makecell[tl]{$w = 9$ (15 min sampling) \\ we thus set $w = 5$ as default} & $w \in \left\{ 1, 2, \ldots, 48\right\}$ \\[0.15cm]

            \cite{Balasubramaniyan2005} & Local Shape-based Similarity & LSS & Min subsequence size ($min_k$) & --- & $min_k = n-2$ & $min_k \in \left\{ 44,45,46,47,48 \right\}$  \\[0.15cm]

            \cite{994784} & Longest Common Subsequence & LCSS & \makecell[tl]{Window size ($w$)  \\ Matching threshold ($\varepsilon$)} & aeon & \makecell[tl]{$w = n$ \\ $\varepsilon=1$} & \makecell[tl]{$w \in \left\{1,2,\ldots,48\right\}$ \\ $\varepsilon \in \left\{ 0,0.1,0.2,\ldots,1\right\}$}  \\[0.15cm]

            \cite{Mahalanobis1936OnStatistics} & Mahalanobis Distance & MAH & --- & scipy & --- & ---   \\[0.15cm]

            \cite{Aggarwal2014} & Manhattan Distance & MD & --- & scipy & --- & ---  \\[0.15cm]

            \cite{Gharghabi2018a} & Matrix Profile Distance & MPD & \makecell[tl]{Window size ($w$) \\ Threshold\textsuperscript{$\star$} ($\tau$)} & stumpy & \makecell[tl]{$w=n$ (identitcal to ED) \\ $\tau = 0.05$} & $w \in \left\{3,4,\ldots,48\right\}$ \\[0.15cm]

            \cite{Aggarwal2014} & Minkowski Metric & MM$_p$ & Norm order ($p=\frac12,3,4,5$) (note that $p=1$ gives MD, $p=2$ gives ED, and $p \to \infty $ gives ChD) & scipy & --- & ---    \\[0.15cm]

            \cite{6189346} & Move-Split-Merge & MSM & \makecell[tl]{Window size ($w$)  \\ Cost ($c$)} & aeon & \makecell[tl]{$w = n$ \\ $c=1$} & \makecell[tl]{$w \in \left\{1,2,\ldots,48\right\}$ \\ $c \in \left\{10^{i} : i=0,\pm 1, \pm 2\right\}$} \\[0.15cm]

            \cite{Howell2012StatisticalPsychology} & Pearson Correlation & PC & --- & scipy & --- & ---  \\[0.15cm]

            \cite{Paparrizos2015} & Shape-Based Distance & SBD & --- & aeon & --- & --- \\[0.15cm]

            \cite{Howell2012StatisticalPsychology} & Spearman Correlation & SC & --- & scipy & --- & ---  \\[0.15cm]

            \cite{Marteau2009} & Time-Warping Edit Distance & TWED & \makecell[tl]{Stiffness ($\nu$) \\ Penalty ($\lambda$)} & aeon & \makecell[tl]{$\nu=0.001$ \\ $\lambda = 1$} & \makecell[tl]{$\nu \in \left\{10^{-i}:i=0,1,\ldots,5\right\}$ \\ $\lambda \in \left\{ 0,0.25,0.5,0.75,1\right\}$}  \\[0.15cm]

            \bottomrule \hiderowcolors
            \multicolumn{7}{p{\dimexpr\textwidth-2\tabcolsep-2\arrayrulewidth-1.2cm}}{\textsuperscript{a} Defaults taken from package or proposing paper. If neither exists, a practical and fair default has been chosen.} \\
            \multicolumn{7}{p{\dimexpr\textwidth-2\tabcolsep-2\arrayrulewidth-1.2cm}}{\textsuperscript{$\star$}These parameters have been fixed for our study.} \\
        \end{tabular}
    \end{adjustbox}
    \caption{The 27 distance measures investigated with the raw DLPs in this study. Note that $X_t$ and $Y_t$ denote DLPs for $t\in\left\{0,1,\ldots,n-1\right\}$ where $n=48$, and $\sigma_X$ denotes the standard deviation of $X_t$.}
    \label{Tab:DistanceMeasures}
\end{table}

A \textbf{clustering algorithm} is a procedure which produces a partition between a set of objects, where objects placed in the same group are more closely related than objects separated into different groups. The obtained partitions are typically hard or crisp, indicating that each object belongs to a single cluster, though partial cluster membership can be obtained by calling upon fuzzy or probabilistic clustering algorithms \cite{Aggarwal2014}.
The clustering algorithms selected for the present study are detailed in \Cref{Tab:ClusteringAlgorithms}. Regarding each hierarchical linkage type as a unique clustering algorithm, we have included 9 algorithms that are compatible with any given similarity paradigm. That is they are modular components that can be used to form distinct clustering approaches. In contrast, the last two rows of \Cref{Tab:ClusteringAlgorithms} are referred to as \textit{indivisible} clustering approaches as the algorithms and objective criteria are inherently tied to particular similarity paradigms or prototypes.

In this study, we focused exclusively on clustering algorithms and approaches that generate crisp partitions and allow for the pre-specification of the number of clusters --- a criterion that excludes many density-based algorithms. Furthermore, in \cite{Jin}, the popular density-based algorithm DBSCAN \cite{Ester1996ANoise} was observed to identify a significant portion of DLPs as noise, suggesting more work may be necessary to make density-based methods effective with time series. Hidden Markov Models \cite{NIPS1996_6a61d423} fitted using the Expectation Maximisation algorithm \cite{Duda1974PatternCA}, alongside $k$-means with DTW and Barycenter averaging, were also initially considered as candidate indivisible clustering approaches for this study. However, due to unpromising preliminary results and significant computational demands for both methods, they were ultimately excluded.

\begin{table}[t]
    \scriptsize
    \rowcolors{1}{TableGray}{TableWhite}
    \begin{adjustbox}{center}
        \begin{tabular}{p{0.35cm} p{2.6cm} p{0.84cm} p{3.1cm} p{1.9cm} p{5.05cm}} \toprule \hiderowcolors 
            \textbf{Ref.} & \textbf{Name} & \textbf{Acronym} & \textbf{Parameters} & \makecell[tl]{\textbf{Python Package}} & \makecell[tl]{\textbf{Parameter Settings}} \\ \midrule \showrowcolors
            
            \multicolumn{6}{c}{Clustering Algorithms} \\ \hline \showrowcolors
            
            \cite{Zhang1996} & \makecell[tl]{Balanced Iterative \\ Reducing and Clustering \\ using Hierarchies} & BIRCH & \makecell[tl]{Threshold ($\tau$) \\ Branching factor\textsuperscript{$\star$} ($f$)} & scikit-learn & \makecell[tl]{$\tau = 0.1$ and reduced by factors of $10$ until the \\ requested number of clusters returned\textsuperscript{$\dagger$} \\ $f=50$} \\[0.15cm]
            
            \cite{Gagolewski2021GenieclustClustering} & Genieclust & GC & Threshold\textsuperscript{$\star$} ($\tau$) & genieclust & $\tau = 0.3$ \\[0.15cm]
            
            \cite{Aggarwal2014} & \makecell[tl]{Hierarchical \\ Agglomerative Clustering} & HAC & Linkage ($l$) & \makecell[tl]{scipy, fastcluster} & \makecell[tl]{$l$: Each of Single (S), Complete (C), \\ Average (A), Weighted (We) and Ward (Wa)} \\[0.15cm]
            
            \cite{Aggarwal2014} & $k$-Medoids & KMd & \makecell[tl]{Maximum iterations\textsuperscript{$\star$} ($m_i$) \\ Initialisation\textsuperscript{$\star$} ($i$) \\ Number of initialisations\textsuperscript{$\star$} ($n_i$) \\ Method\textsuperscript{$\star$} ($M$)} & scikit-learn-extra & \makecell[tl]{$m_i = 200$ \\ i: $k$-medoids$++$ \\ $n_i=30$ \\ $M$: Partitioning Around Medoids (PAM)} \\[0.15cm]
            
            \cite{Aggarwal2014} & Spectral Clustering & SC & \makecell[tl]{Kernel\textsuperscript{$\star$} ($\kappa$) \\ Kernel width ($\delta$)} & scikit-learn & \makecell[tl]{$\kappa$: Gaussian (RBF) \\ $\delta = 20$ and incremented by $20$ until the \\ requested number of clusters returned\textsuperscript{$\dagger$} } \\[0.15cm] \midrule \hiderowcolors

            \multicolumn{6}{c}{Indivisible Clustering Approaches} \\ \hline \showrowcolors

            \cite{Aggarwal2014} & $k$-Means & KMn & \makecell[tl]{Maximum iterations\textsuperscript{$\star$} ($m_i$) \\ Initialisation\textsuperscript{$\star$} ($i$) \\ Number of initialisations\textsuperscript{$\star$} ($n_i$) \\ Tolerance\textsuperscript{$\star$} ($tol$)} & scikit-learn & \makecell[tl]{$m_i = 200$ \\ $i$: $k$-means$++$ \\ $n_i = 30$ \\ $tol=10^{-5}$} \\[0.15cm]
            
            \cite{Paparrizos2015} & $k$-Shape & KS & \makecell[tl]{Maximum iterations\textsuperscript{$\star$} ($m_i$) \\ Number of initialisations\textsuperscript{$\star$} ($n_i$) \\ Tolerance\textsuperscript{$\star$} ($tol$)} & tslearn & \makecell[tl]{$m_i = 200$ \\ $n_i=30$ \\ $tol = 10^{-5}$} \\[0.15cm] \midrule
             
            \bottomrule \hiderowcolors
            \multicolumn{6}{p{\dimexpr\textwidth-2\tabcolsep-2\arrayrulewidth-1.2cm}}{\textsuperscript{$\star$}These parameters have been fixed for our study.} \\
            \multicolumn{6}{p{\dimexpr\textwidth-2\tabcolsep-2\arrayrulewidth-1.2cm}}{\textsuperscript{$\dagger$}As in \cite{Yerbury2024}} \\
        \end{tabular}
    \end{adjustbox}
    \caption{The 11 clustering methods investigated in the current study, separated into 9 clustering algorithms (including 5 different linkage criterion for HAC) and 2 indivisible clustering approaches.}
    \label{Tab:ClusteringAlgorithms}
\end{table}

Whilst it is known that the choice of normalisation procedure can also significantly affect pairwise similarities and clustering outcomes \cite{Yerbury2024}, a comparison of normalisation procedures was determined to be beyond the scope of the current study, as it would require an alternative synthetic data generation process. It should also be noted that prototype definitions are not expressly necessary for most clustering approaches, only being required within the subroutines of a subset of ``prototype-based'' clustering algorithms. Prototypes are also commonly utilised post-clustering for visualisations, summary purposes or as exemplars for downstream applications. Comparisons of prototype methods would also require a different approach, and thus fall outside the scope of the current study.

\subsection{Experimental Methodology}
To conduct a comprehensive comparative study of such a wide range of clustering methods, the experiments were divided into two main stages. In the first stage, the aim was to identify the better-performing distance measures and representation methods. The most suitable techniques were then selected for further evaluation in the second stage, which focused on identifying effective pairings with clustering algorithms and assessing their robustness to changes in key dataset properties.

For the first stage we elected to use leave-one-out One-Nearest Neighbour (1NN) classification accuracy to compare the suitability of each distance measure and representation method for determining the dissimilarity of DLPs. This method temporarily removes each time series from a dataset, identifies its nearest neighbour among the remaining series (using the chosen distance measure or representation method), and compares their class labels. The 1NN accuracy is calculated as the proportion of cases where the nearest neighbour's class label matches that of the removed time series. An accuracy of 1 indicates that the nearest neighbour of every time series had the same class label (perfect classification), while 0 indicates complete misclassification.

This methodology has been used extensively for such comparisons \cite{Paparrizos2020,Alonso2020,Paparrizos2015,Montero2014TSclust:Clustering,Wang2013b,CHEN2004,Keogh2002} as it offers several key advantages. Firstly, the underlying representation and distance measure are known to critically influence 1NN accuracy \cite{Wang2013b}. The performance evaluation is also kept independent from interactions with any particular clustering algorithms, allowing the methods to be compared on their own merits. Secondly, the task parallels the similarity search problem solved by distances and representations within a clustering approach. Thirdly, the methodology is parameter free and straightforward to implement. Finally, and importantly for our application, 1NN accuracy is an intuitive measure which indicates the extent to which each method's notion of similarity conforms with that provided by the domain experts within the recoverable synthetic DLP cluster labels.

In the second stage, we compared clustering approaches using both the baseline synthetic data and modified versions covering multiple levels of various dataset characteristics. While a full factorial study could have captured any interactions between these characteristics, computational constraints would have required limiting the scope of our investigation. We therefore elected to examine a wider array of characteristics one-dimensionally, leaving interaction investigations to future work. The comparisons were obtained by computing various External Validity Indices (EVIs). We primarily report results obtained using the Adjusted Rand Index (ARI); however, this index is not invariant to changes in the number of clusters and lacks sensitivity to errors in small clusters \cite{Rezaei2016}. Therefore, when varying the number of clusters we use the Normalised Van Dongen (NVD) index \cite{vanDongen2000PerformanceExperiments}, which is invariant to changes in the number of clusters. Throughout this paper however, we report 1 minus the NVD value, so that higher values also indicate better performance. Furthermore, when manipulating the balance of cluster sizes we  report the Pair Sets Index (PSI) \cite{Rezaei2016}, which is equally affected by errors in small and large clusters. Results obtained using additional indices for each experiment --- including the Adjusted Mutual Information (AMI) \cite{Vinh2010} --- are available in the \href{https://github.com/yerbles/Smart-Meter-Time-Series-Clustering-Comparative-Study/tree/main}{supplementary materials}. 
ARI, AMI and PSI/NVD represent the three main categories of EVIs, namely pair-counting, information theoretic and set-matching EVIs respectively \cite{Warrens2022,Arinik2021}. These categories distinguish EVIs according to the distinct techniques they use to assess similarity between partitions. Thus robust conclusions can be drawn where these EVIs are in agreeance.
Further experimental details are provided prior to the presentation and discussion of the respective results in \Cref{sec:Results}.

        \section{Results and Discussion}
\label{sec:Results}

\subsection{Stage One: Comparison of Distance Measures and Representation Methods} \label{Subsec:StageOne}
\label{sec:Results---StageOne}

We evaluated 38 distinct combinations of distance measures and representation methods using 1NN accuracy. These combinations were derived from seven representations using Euclidean distance, SAX with its four specific distance measures (from \Cref{Tab:RepresentationMethods}), and the 27 additional distance measures applied to the raw time series representation and detailed in \Cref{Tab:DistanceMeasures}. For the Minkowski Metric, we counted different values of the parameter $p$ ($p=0.5, 3, 4, 5$) as separate measures, while noting that the cases $p=1$ and $p=2$ correspond to the Manhattan and Euclidean distances respectively.

The evaluation was performed on 100 baseline synthetic datasets. The 20 clusters were all evenly balanced with 10 DLPs, resulting in datasets with 200 time series. For those methods that require the setting of parameters, we have reported results for two cases. The first case used default parameter settings, which were primarily obtained by consulting the documentation of the python implementations or otherwise were found in the proposing paper. In the absence of such options, we selected the most natural or practical defaults. The default parameters for each method can be found listed in the second-last column of \Cref{Tab:RepresentationMethods,Tab:DistanceMeasures}. The second case used the optimal accuracy obtained across all tested parameter combinations. These combinations were drawn exhaustively from the respective parameter lists found in the final column of \Cref{Tab:RepresentationMethods,Tab:DistanceMeasures}. Efforts were made to thoroughly explore a wide range of parameter settings in order to gauge the optimal performance theoretically possible for each method on these datasets.

\afterpage{%
    \clearpage
    \begin{landscape}
        \begin{table}[p]
            \tiny
            \begin{adjustbox}{center}
                \begin{tabular}{r|c|cccccccccccccccccccc|} \toprule
                    \multicolumn{2}{l}{} & \multicolumn{20}{c}{\scriptsize{\textbf{Accuracy by Cluster}}} \\ \cmidrule(l){3-22}
                    \scriptsize{\textbf{Method}} & \scriptsize{\textbf{Mean}} & \scriptsize{\textbf{0}} & \scriptsize{\textbf{1}} & \scriptsize{\textbf{2}} & \scriptsize{\textbf{3}} & \scriptsize{\textbf{4}} & \scriptsize{\textbf{5}} & \scriptsize{\textbf{6}} & \scriptsize{\textbf{7}} & \scriptsize{\textbf{8}} & \scriptsize{\textbf{9}} & \scriptsize{\textbf{10}} & \scriptsize{\textbf{11}} & \scriptsize{\textbf{12}} & \scriptsize{\textbf{13}} & \scriptsize{\textbf{14}} & \scriptsize{\textbf{15}} & \scriptsize{\textbf{16}} & \scriptsize{\textbf{17}} & \scriptsize{\textbf{18}} & \scriptsize{\textbf{19}} \\ \midrule
                    
                    SBD & \mc{0.970} & \mc{0.974} & \mc{0.992} & \mc{0.990} & \mc{0.929} & \mc{1.000} & \mc{0.926} & \mc{0.999} & \mc{0.841} & \mc{0.902} & \mc{1.000} & \mc{1.000} & \mc{1.000} & \mc{1.000} & \mc{1.000} & \mc{1.000} & \mc{0.996} & \mc{0.993} & \mc{0.921} & \mc{0.936} & \mc{1.000} \\
                    MM$_3$ & \mc{0.966} & \mc{0.978} & \mc{1.000} & \mc{1.000} & \mc{0.929} & \mc{0.999} & \mc{0.838} & \mc{1.000} & \mc{0.856} & \mc{0.912} & \mc{1.000} & \mc{1.000} & \mc{1.000} & \mc{1.000} & \mc{1.000} & \mc{1.000} & \mc{1.000} & \mc{0.996} & \mc{0.879} & \mc{0.944} & \mc{0.997} \\
                    MM$_4$ & \mc{0.965} & \mc{0.977} & \mc{1.000} & \mc{1.000} & \mc{0.902} & \mc{0.999} & \mc{0.845} & \mc{1.000} & \mc{0.854} & \mc{0.919} & \mc{1.000} & \mc{1.000} & \mc{1.000} & \mc{1.000} & \mc{1.000} & \mc{1.000} & \mc{1.000} & \mc{0.997} & \mc{0.888} & \mc{0.938} & \mc{0.987} \\
                    MM$_5$ & \mc{0.963} & \mc{0.980} & \mc{1.000} & \mc{1.000} & \mc{0.892} & \mc{0.999} & \mc{0.845} & \mc{1.000} & \mc{0.835} & \mc{0.912} & \mc{1.000} & \mc{1.000} & \mc{1.000} & \mc{1.000} & \mc{1.000} & \mc{1.000} & \mc{0.999} & \mc{0.997} & \mc{0.898} & \mc{0.930} & \mc{0.979} \\
                    ED & \mc{0.962} & \mc{0.973} & \mc{1.000} & \mc{0.999} & \mc{0.937} & \mc{0.997} & \mc{0.800} & \mc{0.998} & \mc{0.855} & \mc{0.887} & \mc{1.000} & \mc{1.000} & \mc{1.000} & \mc{1.000} & \mc{1.000} & \mc{1.000} & \mc{1.000} & \mc{0.996} & \mc{0.861} & \mc{0.942} & \mc{1.000} \\
                    FD & \mc{0.962} & \mc{0.976} & \mc{0.943} & \mc{0.986} & \mc{0.965} & \mc{0.984} & \mc{0.886} & \mc{0.997} & \mc{0.907} & \mc{0.928} & \mc{1.000} & \mc{1.000} & \mc{1.000} & \mc{1.000} & \mc{1.000} & \mc{1.000} & \mc{0.895} & \mc{0.920} & \mc{0.911} & \mc{0.954} & \mc{0.991} \\
                    CID & \mc{0.962} & \mc{0.960} & \mc{1.000} & \mc{0.999} & \mc{0.933} & \mc{1.000} & \mc{0.857} & \mc{1.000} & \mc{0.849} & \mc{0.865} & \mc{1.000} & \mc{1.000} & \mc{1.000} & \mc{1.000} & \mc{1.000} & \mc{1.000} & \mc{1.000} & \mc{0.997} & \mc{0.858} & \mc{0.922} & \mc{0.999} \\
                    PC & \mc{0.961} & \mc{0.967} & \mc{1.000} & \mc{0.998} & \mc{0.925} & \mc{1.000} & \mc{0.880} & \mc{1.000} & \mc{0.779} & \mc{0.821} & \mc{1.000} & \mc{1.000} & \mc{1.000} & \mc{1.000} & \mc{1.000} & \mc{1.000} & \mc{0.991} & \mc{0.997} & \mc{0.909} & \mc{0.953} & \mc{1.000} \\
                    CoD & \mc{0.955} & \mc{0.948} & \mc{1.000} & \mc{0.998} & \mc{0.867} & \mc{1.000} & \mc{0.856} & \mc{0.999} & \mc{0.767} & \mc{0.848} & \mc{1.000} & \mc{1.000} & \mc{1.000} & \mc{1.000} & \mc{1.000} & \mc{1.000} & \mc{0.988} & \mc{0.997} & \mc{0.907} & \mc{0.928} & \mc{0.999} \\
                    ChD & \mc{0.943} & \mc{0.965} & \mc{0.999} & \mc{0.999} & \mc{0.805} & \mc{0.999} & \mc{0.811} & \mc{0.999} & \mc{0.691} & \mc{0.901} & \mc{1.000} & \mc{1.000} & \mc{1.000} & \mc{1.000} & \mc{1.000} & \mc{1.000} & \mc{0.999} & \mc{0.993} & \mc{0.880} & \mc{0.875} & \mc{0.954} \\
                    MD & \mc{0.938} & \mc{0.951} & \mc{1.000} & \mc{0.998} & \mc{0.919} & \mc{0.955} & \mc{0.693} & \mc{0.982} & \mc{0.762} & \mc{0.811} & \mc{1.000} & \mc{1.000} & \mc{1.000} & \mc{1.000} & \mc{1.000} & \mc{1.000} & \mc{0.999} & \mc{0.971} & \mc{0.805} & \mc{0.920} & \mc{1.000} \\
                    BD & \mc{0.937} & \mc{0.883} & \mc{1.000} & \mc{0.988} & \mc{0.850} & \mc{0.992} & \mc{0.764} & \mc{0.999} & \mc{0.744} & \mc{0.799} & \mc{1.000} & \mc{1.000} & \mc{1.000} & \mc{1.000} & \mc{1.000} & \mc{1.000} & \mc{0.998} & \mc{0.989} & \mc{0.848} & \mc{0.886} & \mc{1.000} \\
                    MM$_{0.5}$ & \mc{0.898} & \mc{0.900} & \mc{0.999} & \mc{0.974} & \mc{0.879} & \mc{0.842} & \mc{0.543} & \mc{0.924} & \mc{0.690} & \mc{0.696} & \mc{1.000} & \mc{1.000} & \mc{1.000} & \mc{1.000} & \mc{1.000} & \mc{1.000} & \mc{0.997} & \mc{0.880} & \mc{0.751} & \mc{0.876} & \mc{1.000} \\
                    CaD & \mc{0.830} & \mc{0.719} & \mc{0.955} & \mc{0.826} & \mc{0.685} & \mc{0.843} & \mc{0.448} & \mc{0.962} & \mc{0.497} & \mc{0.514} & \mc{1.000} & \mc{0.998} & \mc{0.996} & \mc{0.996} & \mc{0.995} & \mc{1.000} & \mc{0.943} & \mc{0.804} & \mc{0.692} & \mc{0.752} & \mc{0.971} \\
                    KT & \mc{0.667} & \mc{0.413} & \mc{0.455} & \mc{0.371} & \mc{0.235} & \mc{0.908} & \mc{0.334} & \mc{1.000} & \mc{0.280} & \mc{0.267} & \mc{1.000} & \mc{0.971} & \mc{0.982} & \mc{0.830} & \mc{0.761} & \mc{1.000} & \mc{0.475} & \mc{0.626} & \mc{0.962} & \mc{0.833} & \mc{0.629} \\
                    SC & \mc{0.642} & \mc{0.405} & \mc{0.413} & \mc{0.354} & \mc{0.219} & \mc{0.920} & \mc{0.325} & \mc{1.000} & \mc{0.248} & \mc{0.235} & \mc{1.000} & \mc{0.961} & \mc{0.963} & \mc{0.822} & \mc{0.716} & \mc{0.998} & \mc{0.440} & \mc{0.555} & \mc{0.964} & \mc{0.853} & \mc{0.441} \\
                    MAH & \mc{0.505} & \mc{0.288} & \mc{0.670} & \mc{0.433} & \mc{0.209} & \mc{0.305} & \mc{0.153} & \mc{0.229} & \mc{0.116} & \mc{0.166} & \mc{0.937} & \mc{0.903} & \mc{0.878} & \mc{0.793} & \mc{0.805} & \mc{0.641} & \mc{0.579} & \mc{0.244} & \mc{0.460} & \mc{0.583} & \mc{0.699} \\
                    HD & \mc{0.166} & \mc{0.115} & \mc{0.117} & \mc{0.126} & \mc{0.118} & \mc{0.098} & \mc{0.069} & \mc{0.265} & \mc{0.152} & \mc{0.161} & \mc{0.489} & \mc{0.170} & \mc{0.163} & \mc{0.212} & \mc{0.194} & \mc{0.122} & \mc{0.067} & \mc{0.085} & \mc{0.327} & \mc{0.167} & \mc{0.108} \\ \midrule
                    DTW$_{\text{opt}}$ & \mc{0.994} & \mc{0.992} & \mc{1.000} & \mc{1.000} & \mc{0.990} & \mc{1.000} & \mc{0.992} & \mc{1.000} & \mc{0.979} & \mc{0.988} & \mc{1.000} & \mc{1.000} & \mc{1.000} & \mc{1.000} & \mc{1.000} & \mc{1.000} & \mc{1.000} & \mc{0.999} & \mc{0.969} & \mc{0.980} & \mc{1.000} \\
                    DTW$_{\text{def}}$ & \mc{0.824} & \mc{0.552} & \mc{0.460} & \mc{0.505} & \mc{0.603} & \mc{0.569} & \mc{0.591} & \mc{0.998} & \mc{0.933} & \mc{0.963} & \mc{0.971} & \mc{0.917} & \mc{0.936} & \mc{1.000} & \mc{1.000} & \mc{1.000} & \mc{0.764} & \mc{0.844} & \mc{0.955} & \mc{0.973} & \mc{0.949} \\
                    KSD$_{\text{opt}}$ & \mc{0.993} & \mc{0.990} & \mc{0.996} & \mc{1.000} & \mc{0.991} & \mc{1.000} & \mc{0.991} & \mc{1.000} & \mc{0.978} & \mc{0.995} & \mc{1.000} & \mc{1.000} & \mc{1.000} & \mc{1.000} & \mc{1.000} & \mc{1.000} & \mc{0.966} & \mc{0.997} & \mc{0.974} & \mc{0.986} & \mc{0.999} \\
                    KSD$_{\text{def}}$ & \mc{0.916} & \mc{0.982} & \mc{0.680} & \mc{0.971} & \mc{0.916} & \mc{0.990} & \mc{0.981} & \mc{0.980} & \mc{0.980} & \mc{0.994} & \mc{1.000} & \mc{0.982} & \mc{0.968} & \mc{1.000} & \mc{1.000} & \mc{1.000} & \mc{0.318} & \mc{0.686} & \mc{0.982} & \mc{0.994} & \mc{0.921} \\
                    MPD$_{\text{opt}}$ & \mc{0.981} & \mc{0.988} & \mc{1.000} & \mc{1.000} & \mc{0.972} & \mc{0.999} & \mc{0.885} & \mc{0.999} & \mc{0.920} & \mc{0.958} & \mc{1.000} & \mc{1.000} & \mc{1.000} & \mc{1.000} & \mc{1.000} & \mc{1.000} & \mc{1.000} & \mc{0.998} & \mc{0.918} & \mc{0.977} & \mc{1.000} \\
                    MPD$_{\text{def}}$ & \mc{0.962} & \mc{0.973} & \mc{1.000} & \mc{0.999} & \mc{0.937} & \mc{0.997} & \mc{0.800} & \mc{0.998} & \mc{0.855} & \mc{0.887} & \mc{1.000} & \mc{1.000} & \mc{1.000} & \mc{1.000} & \mc{1.000} & \mc{1.000} & \mc{1.000} & \mc{0.996} & \mc{0.861} & \mc{0.942} & \mc{1.000} \\
                    PCA$_{\text{opt}}$ & \mc{0.978} & \mc{0.992} & \mc{1.000} & \mc{1.000} & \mc{0.981} & \mc{1.000} & \mc{0.839} & \mc{1.000} & \mc{0.883} & \mc{0.946} & \mc{1.000} & \mc{1.000} & \mc{1.000} & \mc{1.000} & \mc{1.000} & \mc{1.000} & \mc{1.000} & \mc{0.996} & \mc{0.957} & \mc{0.972} & \mc{1.000} \\
                    PCA$_{\text{def}}$ & \mc{0.962} & \mc{0.973} & \mc{1.000} & \mc{0.999} & \mc{0.937} & \mc{0.997} & \mc{0.800} & \mc{0.998} & \mc{0.855} & \mc{0.887} & \mc{1.000} & \mc{1.000} & \mc{1.000} & \mc{1.000} & \mc{1.000} & \mc{1.000} & \mc{1.000} & \mc{0.996} & \mc{0.861} & \mc{0.942} & \mc{1.000} \\
                    ERP$_{\text{opt}}$ & \mc{0.976} & \mc{0.983} & \mc{0.995} & \mc{1.000} & \mc{0.979} & \mc{0.999} & \mc{0.890} & \mc{1.000} & \mc{0.911} & \mc{0.938} & \mc{1.000} & \mc{1.000} & \mc{1.000} & \mc{1.000} & \mc{1.000} & \mc{1.000} & \mc{0.961} & \mc{0.993} & \mc{0.903} & \mc{0.970} & \mc{0.999} \\
                    ERP$_{\text{def}}$ & \mc{0.957} & \mc{0.921} & \mc{0.939} & \mc{0.940} & \mc{0.916} & \mc{0.933} & \mc{0.842} & \mc{0.999} & \mc{0.897} & \mc{0.924} & \mc{1.000} & \mc{1.000} & \mc{1.000} & \mc{1.000} & \mc{1.000} & \mc{1.000} & \mc{0.981} & \mc{0.994} & \mc{0.897} & \mc{0.957} & \mc{0.998} \\
                    LSTM$_{\text{opt}}$ & \mc{0.970} & \mc{0.974} & \mc{0.996} & \mc{0.999} & \mc{0.972} & \mc{0.999} & \mc{0.847} & \mc{0.999} & \mc{0.899} & \mc{0.921} & \mc{1.000} & \mc{1.000} & \mc{0.999} & \mc{1.000} & \mc{1.000} & \mc{1.000} & \mc{0.987} & \mc{0.993} & \mc{0.896} & \mc{0.929} & \mc{1.000} \\
                    LSTM$_{\text{def}}$ & \mc{0.897} & \mc{0.897} & \mc{0.792} & \mc{0.988} & \mc{0.906} & \mc{0.925} & \mc{0.687} & \mc{0.979} & \mc{0.714} & \mc{0.720} & \mc{1.000} & \mc{1.000} & \mc{1.000} & \mc{1.000} & \mc{1.000} & \mc{1.000} & \mc{0.764} & \mc{0.865} & \mc{0.897} & \mc{0.887} & \mc{0.928} \\
                    DWT$_{\text{opt}}$ & \mc{0.968} & \mc{0.983} & \mc{0.999} & \mc{1.000} & \mc{0.966} & \mc{0.998} & \mc{0.795} & \mc{0.999} & \mc{0.884} & \mc{0.903} & \mc{1.000} & \mc{1.000} & \mc{1.000} & \mc{1.000} & \mc{1.000} & \mc{1.000} & \mc{0.990} & \mc{0.993} & \mc{0.898} & \mc{0.953} & \mc{1.000} \\
                    DWT$_{\text{def}}$ & \mc{0.967} & \mc{0.982} & \mc{1.000} & \mc{0.999} & \mc{0.957} & \mc{0.998} & \mc{0.810} & \mc{0.999} & \mc{0.882} & \mc{0.896} & \mc{1.000} & \mc{1.000} & \mc{1.000} & \mc{1.000} & \mc{1.000} & \mc{1.000} & \mc{1.000} & \mc{0.996} & \mc{0.872} & \mc{0.946} & \mc{1.000} \\
                    MSM$_{\text{opt}}$ & \mc{0.968} & \mc{0.971} & \mc{0.997} & \mc{0.999} & \mc{0.961} & \mc{0.991} & \mc{0.826} & \mc{1.000} & \mc{0.880} & \mc{0.903} & \mc{1.000} & \mc{1.000} & \mc{1.000} & \mc{1.000} & \mc{1.000} & \mc{1.000} & \mc{0.982} & \mc{0.986} & \mc{0.905} & \mc{0.957} & \mc{1.000} \\
                    MSM$_{\text{def}}$ & \mc{0.939} & \mc{0.951} & \mc{1.000} & \mc{0.998} & \mc{0.919} & \mc{0.955} & \mc{0.693} & \mc{0.989} & \mc{0.762} & \mc{0.811} & \mc{1.000} & \mc{1.000} & \mc{1.000} & \mc{1.000} & \mc{1.000} & \mc{1.000} & \mc{0.999} & \mc{0.971} & \mc{0.805} & \mc{0.920} & \mc{1.000} \\
                    TWED$_{\text{opt}}$ & \mc{0.968} & \mc{0.974} & \mc{0.994} & \mc{0.997} & \mc{0.963} & \mc{0.990} & \mc{0.829} & \mc{0.999} & \mc{0.883} & \mc{0.906} & \mc{1.000} & \mc{1.000} & \mc{1.000} & \mc{1.000} & \mc{1.000} & \mc{1.000} & \mc{0.978} & \mc{0.987} & \mc{0.906} & \mc{0.952} & \mc{0.999} \\
                    TWED$_{\text{def}}$ & \mc{0.948} & \mc{0.959} & \mc{1.000} & \mc{0.998} & \mc{0.925} & \mc{0.985} & \mc{0.752} & \mc{1.000} & \mc{0.789} & \mc{0.837} & \mc{1.000} & \mc{1.000} & \mc{1.000} & \mc{1.000} & \mc{1.000} & \mc{1.000} & \mc{0.997} & \mc{0.986} & \mc{0.814} & \mc{0.923} & \mc{1.000} \\
                    PAA$_{\text{opt}}$ & \mc{0.966} & \mc{0.985} & \mc{1.000} & \mc{1.000} & \mc{0.955} & \mc{0.998} & \mc{0.791} & \mc{0.999} & \mc{0.883} & \mc{0.893} & \mc{1.000} & \mc{1.000} & \mc{1.000} & \mc{1.000} & \mc{1.000} & \mc{1.000} & \mc{1.000} & \mc{0.997} & \mc{0.878} & \mc{0.943} & \mc{1.000} \\
                    PAA$_{\text{def}}$ & \mc{0.962} & \mc{0.973} & \mc{1.000} & \mc{0.999} & \mc{0.937} & \mc{0.997} & \mc{0.800} & \mc{0.998} & \mc{0.855} & \mc{0.887} & \mc{1.000} & \mc{1.000} & \mc{1.000} & \mc{1.000} & \mc{1.000} & \mc{1.000} & \mc{1.000} & \mc{0.996} & \mc{0.861} & \mc{0.942} & \mc{1.000} \\
                    SAX-MINDIST$_{\text{opt}}$ & \mc{0.960} & \mc{0.973} & \mc{1.000} & \mc{1.000} & \mc{0.911} & \mc{1.000} & \mc{0.792} & \mc{1.000} & \mc{0.823} & \mc{0.891} & \mc{1.000} & \mc{1.000} & \mc{1.000} & \mc{1.000} & \mc{1.000} & \mc{1.000} & \mc{1.000} & \mc{0.997} & \mc{0.871} & \mc{0.936} & \mc{0.999} \\
                    SAX-MINDIST$_{\text{def}}$ & \mc{0.594} & \mc{0.352} & \mc{0.318} & \mc{0.286} & \mc{0.178} & \mc{0.896} & \mc{0.287} & \mc{0.998} & \mc{0.215} & \mc{0.211} & \mc{1.000} & \mc{0.934} & \mc{0.885} & \mc{0.775} & \mc{0.622} & \mc{0.996} & \mc{0.349} & \mc{0.442} & \mc{0.955} & \mc{0.814} & \mc{0.360} \\
                    LCSS$_{\text{opt}}$ & \mc{0.947} & \mc{0.992} & \mc{1.000} & \mc{0.998} & \mc{0.978} & \mc{0.929} & \mc{0.875} & \mc{0.987} & \mc{0.731} & \mc{0.747} & \mc{1.000} & \mc{1.000} & \mc{1.000} & \mc{1.000} & \mc{1.000} & \mc{1.000} & \mc{0.991} & \mc{0.959} & \mc{0.941} & \mc{0.817} & \mc{0.999} \\
                    LCSS$_{\text{def}}$ & \mc{0.050} & \mc{1.000} & \mc{0.000} & \mc{0.000} & \mc{0.000} & \mc{0.000} & \mc{0.000} & \mc{0.000} & \mc{0.000} & \mc{0.000} & \mc{0.000} & \mc{0.000} & \mc{0.000} & \mc{0.000} & \mc{0.000} & \mc{0.000} & \mc{0.000} & \mc{0.000} & \mc{0.000} & \mc{0.000} & \mc{0.000} \\
                    ERS$_{\text{opt}}$ & \mc{0.947} & \mc{0.991} & \mc{1.000} & \mc{0.999} & \mc{0.960} & \mc{0.900} & \mc{0.834} & \mc{0.965} & \mc{0.746} & \mc{0.803} & \mc{1.000} & \mc{1.000} & \mc{1.000} & \mc{1.000} & \mc{1.000} & \mc{1.000} & \mc{0.992} & \mc{0.927} & \mc{0.946} & \mc{0.874} & \mc{1.000} \\
                    ERS$_{\text{def}}$ & \mc{0.482} & \mc{0.050} & \mc{0.099} & \mc{0.166} & \mc{0.098} & \mc{0.387} & \mc{0.317} & \mc{0.601} & \mc{0.104} & \mc{0.115} & \mc{1.000} & \mc{1.000} & \mc{0.997} & \mc{0.810} & \mc{0.363} & \mc{0.631} & \mc{0.131} & \mc{0.339} & \mc{0.764} & \mc{0.732} & \mc{0.936} \\
                    SAX-vLev$_{\text{opt}}$ & \mc{0.939} & \mc{0.952} & \mc{1.000} & \mc{0.998} & \mc{0.925} & \mc{0.955} & \mc{0.711} & \mc{0.985} & \mc{0.761} & \mc{0.794} & \mc{1.000} & \mc{1.000} & \mc{1.000} & \mc{1.000} & \mc{1.000} & \mc{1.000} & \mc{1.000} & \mc{0.969} & \mc{0.817} & \mc{0.908} & \mc{1.000} \\
                    SAX-vLev$_{\text{def}}$ & \mc{0.622} & \mc{0.372} & \mc{0.376} & \mc{0.358} & \mc{0.200} & \mc{0.907} & \mc{0.383} & \mc{0.999} & \mc{0.219} & \mc{0.185} & \mc{1.000} & \mc{0.968} & \mc{0.980} & \mc{0.857} & \mc{0.672} & \mc{0.999} & \mc{0.317} & \mc{0.535} & \mc{0.966} & \mc{0.728} & \mc{0.424} \\
                    GAF$_{\text{opt}}$ & \mc{0.925} & \mc{0.932} & \mc{0.981} & \mc{0.976} & \mc{0.931} & \mc{0.994} & \mc{0.640} & \mc{1.000} & \mc{0.699} & \mc{0.665} & \mc{1.000} & \mc{0.994} & \mc{1.000} & \mc{1.000} & \mc{1.000} & \mc{1.000} & \mc{0.937} & \mc{0.980} & \mc{0.880} & \mc{0.893} & \mc{0.998} \\
                    GAF$_{\text{def}}$ & \mc{0.890} & \mc{0.916} & \mc{0.984} & \mc{0.981} & \mc{0.884} & \mc{0.964} & \mc{0.491} & \mc{0.980} & \mc{0.576} & \mc{0.566} & \mc{1.000} & \mc{1.000} & \mc{1.000} & \mc{0.999} & \mc{1.000} & \mc{1.000} & \mc{0.989} & \mc{0.954} & \mc{0.730} & \mc{0.788} & \mc{0.992} \\
                    SAX-LCSS$_{\text{opt}}$ & \mc{0.901} & \mc{0.993} & \mc{0.999} & \mc{1.000} & \mc{0.963} & \mc{0.928} & \mc{0.890} & \mc{0.981} & \mc{0.706} & \mc{0.714} & \mc{1.000} & \mc{1.000} & \mc{1.000} & \mc{1.000} & \mc{1.000} & \mc{1.000} & \mc{0.901} & \mc{0.945} & \mc{0.596} & \mc{0.404} & \mc{0.999} \\
                    SAX-LCSS$_{\text{def}}$ & \mc{0.602} & \mc{0.467} & \mc{0.328} & \mc{0.408} & \mc{0.304} & \mc{0.585} & \mc{0.383} & \mc{0.980} & \mc{0.219} & \mc{0.184} & \mc{1.000} & \mc{0.960} & \mc{0.833} & \mc{0.938} & \mc{0.672} & \mc{0.998} & \mc{0.341} & \mc{0.392} & \mc{0.962} & \mc{0.710} & \mc{0.380} \\
                    SAX-Lev$_{\text{opt}}$ & \mc{0.895} & \mc{0.993} & \mc{0.999} & \mc{0.998} & \mc{0.975} & \mc{0.852} & \mc{0.798} & \mc{0.929} & \mc{0.692} & \mc{0.722} & \mc{1.000} & \mc{1.000} & \mc{1.000} & \mc{1.000} & \mc{1.000} & \mc{0.999} & \mc{0.960} & \mc{0.877} & \mc{0.619} & \mc{0.480} & \mc{0.998} \\
                    SAX-Lev$_{\text{def}}$ & \mc{0.655} & \mc{0.472} & \mc{0.438} & \mc{0.458} & \mc{0.295} & \mc{0.829} & \mc{0.426} & \mc{0.995} & \mc{0.245} & \mc{0.227} & \mc{1.000} & \mc{0.963} & \mc{0.959} & \mc{0.947} & \mc{0.815} & \mc{0.998} & \mc{0.363} & \mc{0.517} & \mc{0.962} & \mc{0.711} & \mc{0.473} \\
                    MTF$_{\text{opt}}$ & \mc{0.887} & \mc{0.885} & \mc{0.979} & \mc{0.986} & \mc{0.909} & \mc{0.951} & \mc{0.529} & \mc{0.979} & \mc{0.497} & \mc{0.466} & \mc{1.000} & \mc{0.996} & \mc{1.000} & \mc{1.000} & \mc{1.000} & \mc{1.000} & \mc{0.962} & \mc{0.917} & \mc{0.846} & \mc{0.857} & \mc{0.976} \\
                    MTF$_{\text{def}}$ & \mc{0.642} & \mc{0.403} & \mc{0.343} & \mc{0.465} & \mc{0.381} & \mc{0.836} & \mc{0.378} & \mc{0.953} & \mc{0.294} & \mc{0.267} & \mc{0.996} & \mc{0.777} & \mc{0.951} & \mc{0.654} & \mc{0.549} & \mc{0.999} & \mc{0.507} & \mc{0.751} & \mc{0.749} & \mc{0.724} & \mc{0.867} \\
                    BOF$_{\text{opt}}$ & \mc{0.850} & \mc{0.743} & \mc{0.700} & \mc{0.747} & \mc{0.746} & \mc{0.718} & \mc{0.609} & \mc{0.995} & \mc{0.771} & \mc{0.733} & \mc{0.998} & \mc{1.000} & \mc{1.000} & \mc{1.000} & \mc{1.000} & \mc{0.997} & \mc{0.915} & \mc{0.881} & \mc{0.766} & \mc{0.739} & \mc{0.948} \\
                    BOF$_{\text{def}}$ & \mc{0.725} & \mc{0.747} & \mc{0.675} & \mc{0.746} & \mc{0.695} & \mc{0.484} & \mc{0.478} & \mc{0.758} & \mc{0.580} & \mc{0.538} & \mc{0.946} & \mc{0.983} & \mc{0.945} & \mc{0.977} & \mc{0.920} & \mc{0.855} & \mc{0.493} & \mc{0.476} & \mc{0.707} & \mc{0.701} & \mc{0.791} \\
                    LSS$_{\text{opt}}$ & \mc{0.665} & \mc{0.471} & \mc{0.478} & \mc{0.404} & \mc{0.257} & \mc{0.941} & \mc{0.345} & \mc{0.999} & \mc{0.262} & \mc{0.274} & \mc{1.000} & \mc{0.981} & \mc{0.971} & \mc{0.835} & \mc{0.733} & \mc{0.999} & \mc{0.435} & \mc{0.587} & \mc{0.975} & \mc{0.866} & \mc{0.477} \\
                    LSS$_{\text{def}}$ & \mc{0.655} & \mc{0.471} & \mc{0.467} & \mc{0.394} & \mc{0.282} & \mc{0.936} & \mc{0.350} & \mc{0.999} & \mc{0.270} & \mc{0.288} & \mc{1.000} & \mc{0.983} & \mc{0.949} & \mc{0.800} & \mc{0.686} & \mc{0.996} & \mc{0.418} & \mc{0.523} & \mc{0.980} & \mc{0.871} & \mc{0.432} \\ \bottomrule
                \end{tabular}
            \end{adjustbox}
            \caption{Mean 1NN accuracies for 38 distance measures and representation methods, shown both overall and by individual cluster. Results are visualised using the Viridis colourmap (\Cref{Fig:viridis_colorbar}), where lighter colours indicate higher accuracies. Methods are divided into two groups: parameterless methods and parametrised methods. For parametrised methods, results are shown for both optimal parameter settings (denoted as X$_{\text{opt}}$) and default parameter settings (X$_{\text{def}}$). Within each group, methods are arranged in descending order of overall mean accuracy (using the optimal parametrisation for the parametrised methods). For SAX, the only representation not paired solely with ED, results are shown as SAX-\textit{distance} to indicate the specific distance measure used.}
            \label{Tab:ExperimentOneResults}
        \end{table}
    \end{landscape}
\clearpage
}
\clearpage

\begin{figure}[!htb]
    \centering
    \includegraphics[width=0.6\textwidth]{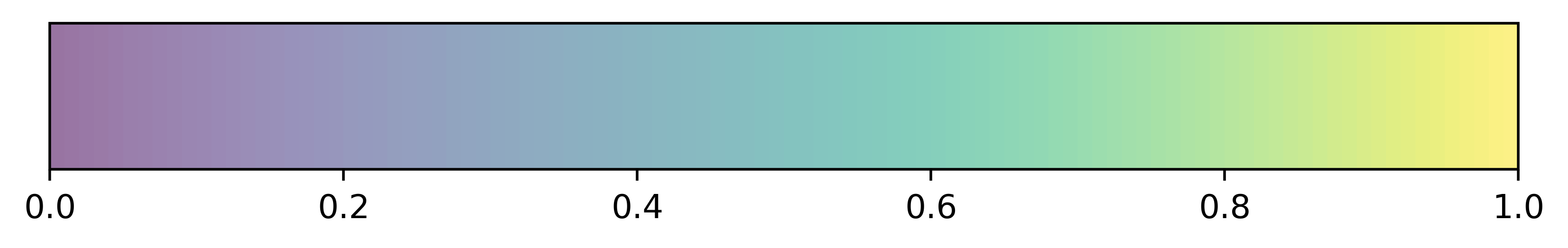}
    \caption{Reference scale for the Viridis colourmap.}
    \label{Fig:viridis_colorbar}
\end{figure}

\Cref{Tab:ExperimentOneResults} reports the average 1NN accuracy for each distance and representation combination overall (``Mean'' column) and by cluster. The table makes use of the Viridis colourmap (\Cref{Fig:viridis_colorbar}) to assist in interpretation. The parameterless methods are listed first in descending order of mean accuracy. These are followed by the optimal and default parametrised methods, denoted by X$_{\text{opt}}$ and X$_{\text{def}}$ respectively for any method X. The parametrised methods are similarly arranged in descending order based on the mean accuracy of their optimal parameter setting. This arrangement makes it easy to identify and compare the best methods in each category.

\begin{figure}[b]
    \centering
    \includegraphics[width=0.99\linewidth]{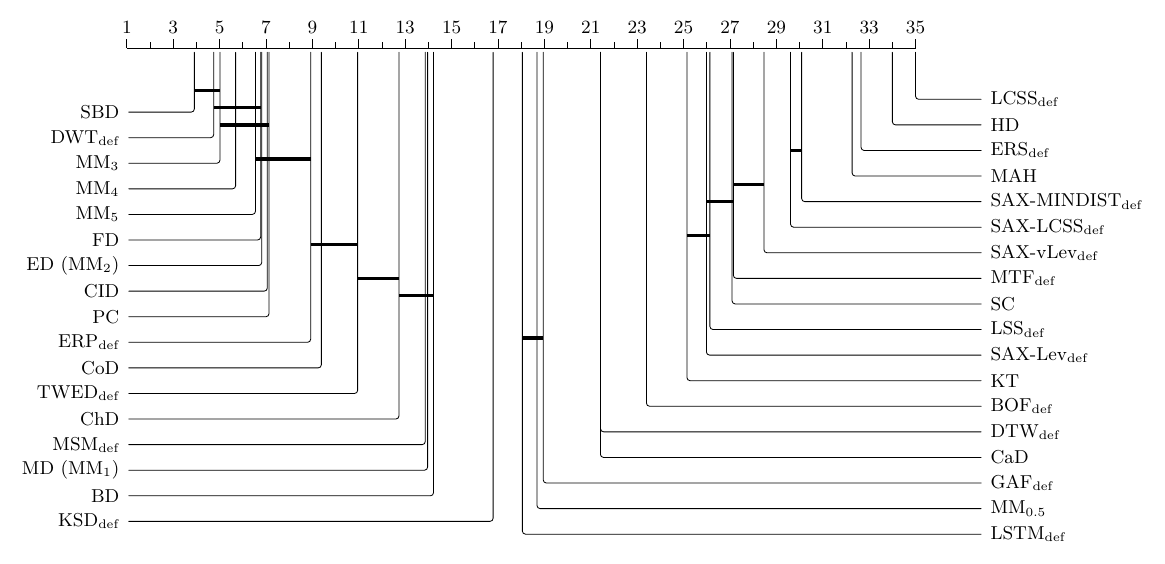}
    \caption{The mean ranks of methods computed across all datasets, including both parametrised methods using \textbf{default} settings and parameterless methods. Methods are ordered from best (left) to worst (right), with lower ranks indicating better performance. Methods connected by a thick line do not perform statistically significantly different according to a Bonferroni corrected Wilcoxon signed-rank post-hoc test. Note that PCA$_\text{def}$, PAA$_\text{def}$ and MPD$_\text{def}$ are equivalent to ED, so have been left out of the figure.}
    \label{Fig:ExpOne-CDD-NoTuning}
\end{figure}

Additionally, \Cref{Fig:ExpOne-CDD-NoTuning} plots the average rank of the default parametrised and parameterless methods across all datasets. Lower ranks (to the left) indicate better performance. The Friedman test resulted in a rejection of the null hypothesis that all methods performed similarly at a significance level of $0.05$. A Wilcoxon signed-rank post-hoc test with Bonferroni correction was used to test the significance of all pairwise differences in ranks. The solid lines in \Cref{Fig:ExpOne-CDD-NoTuning} indicate cliques of methods that did not perform significantly differently according to this test. 

\begin{figure}[!htb]
    \centering
    \includegraphics[width=0.99\linewidth]{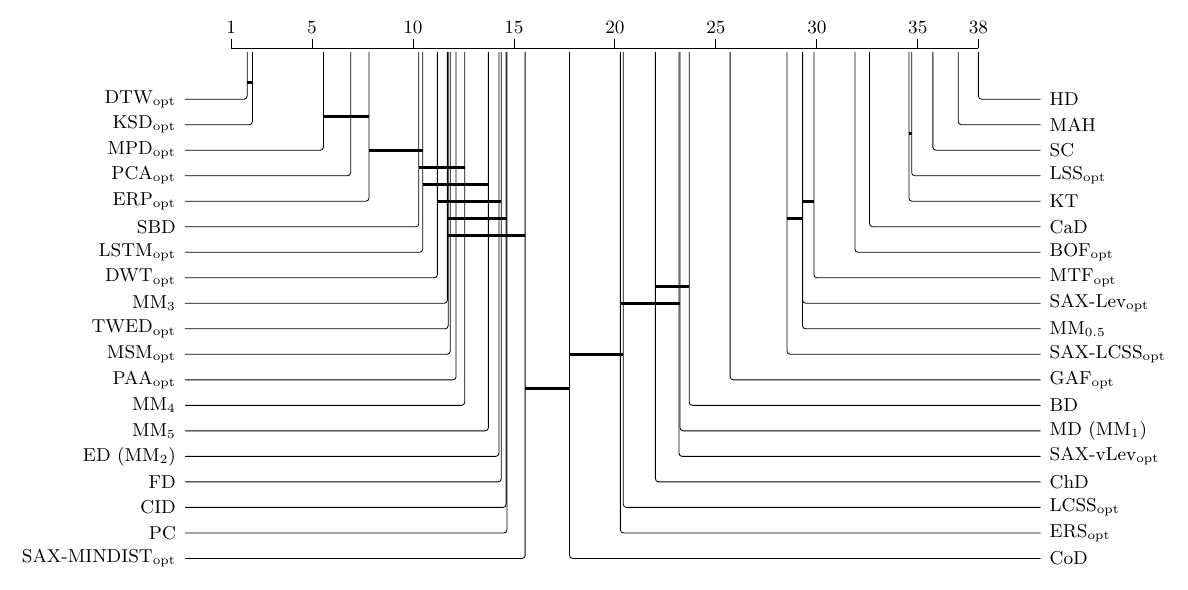}
    \caption{The mean ranks of methods computed across all datasets, including both parametrised methods using \textbf{optimal} settings and parameterless methods. Methods are ordered from best (left) to worst (right), with lower ranks indicating better performance. Methods connected by a thick line do not perform statistically significantly different according to a Bonferroni corrected Wilcoxon signed-rank post-hoc test.}
    \label{Fig:ExpOne-CDD-All}
\end{figure}

We conducted a similar statistical analysis for methods using optimal parameters, presented in \Cref{Fig:ExpOne-CDD-All}. While comparing parameterless methods with optimally-tuned parametrised methods may not be methodologically sound, this analysis provides valuable insights into the theoretical performance ceiling of parametrised methods relative to their parameterless counterparts. When parametrised methods are outperformed by parameterless methods even after thorough parameter tuning, this strongly indicates their fundamental unsuitability for DLP clustering. Finally, the distributions of 1NN accuracies for each method are depicted with boxplots in the \href{https://github.com/yerbles/Smart-Meter-Time-Series-Clustering-Comparative-Study/tree/main}{supplementary materials}.

SBD had the highest average accuracy out of the parameterless methods, followed closely by the higher order Minkowski metrics, which narrowly outperformed ED. The lower order Minkowski metrics ($p\in \left\{0.5, 1\right\}$) performed significantly worse, despite suggestions they are more appropriate for high dimensional data \cite{Aggarwal2001}. The domain specific FD and noise robust CID did not perform differently from ED with statistical significance. 
Out of the correlation measures, PC vastly outperformed KT and SC. For its appeal as an extension of ED that accounts for data correlation, MAH has frequently been considered as a candidate for load profiling \cite{Li2022,Leprince2020AForecasting,Iglesias2013}, yet it was outperformed by nearly all of the other parameterless methods.

DTW$_{\text{opt}}$ achieved the highest average accuracy out of all methods, followed very closely by the domain specific KSD$_{\text{opt}}$ which did not perform differently with statistical significance. This pair ranked higher than the next clique containing MPD$_{\text{opt}}$, PCA$_{\text{opt}}$ and ERP$_{\text{opt}}$ with statistical significance. Notably for all parametrised methods, the defaults were never selected as optimal. All defaults produced lower mean accuracies than SBD, and only DWT$_{\text{def}}$ exceeded the mean accuracies of the next best parameterless methods --- MM$_{p}$ for $p\in\left\{2,3,4,5\right\}$, FD, CID and PC. This suggests that when practitioners clustering DLPs are uncertain about parameter tuning for distances and representations, it is likely better to opt for parameter-free methods, rather than relying on the default parameter settings. With this being said, we observed that in most cases there was a moderately sized subset of parameter combinations around the optimal which could provide results comparable to those of the optimal setting. This observation has ramifications for the next stage of our experiments, and those ramifications will be discussed later in this section.

The proposed Levenshtein variant distance outperformed the normal Levenshtein and LCSS distances in combination with SAX, though it was still inferior to MINDIST. Despite the large range of features incorporated into BOF, its optimal parametrisation performed second-worst out of the optimally parametrised methods. The optimal number of principal components was 60 with an accuracy of 0.850, though limited improvement was observed above 10 components, where an accuracy of 0.834 was achieved. This suggests that many of the 76 features contained limited information capable of contributing to cluster identification. Note that the interesting result for LCSS$_{\text{def}}$ is due to the default value of the matching threshold, $\varepsilon$, being 1. This means that for our normalised synthetic time series ranging from 0 to 1, all of the points have a distance less than or equal to the threshold, so all of the time series are equal and the pairwise distance matrix is the zero matrix. Then, during 1NN classification, the sklearn Nearest Neighbour function splits these ``ties'' by assigning all points to the lowest class (first when sorted). Thus the 1NN accuracy is 1 for cluster 0, and exactly 0.05 overall with zero variance. 

Looking at each column of accuracies by cluster in \Cref{Tab:ExperimentOneResults}, we can see that none of the individual clusters were perfectly identified by all methods. Furthermore, numerous methods equipped with default parameters struggled to classify clusters that were easily classified by their optimally parametrised counterparts. This suggests that whilst many clusters appear as though they were easily identifiable in \Cref{Tab:ExperimentOneResults}, those clusters were important for discriminating appropriate method parameters. The best performing pair (DTW$_{\text{opt}}$ and KSD$_{\text{opt}}$) tended to show consistent accuracies across all of the 20 clusters (DTW, KSD), whilst a majority of the other methods tended to struggle with specific clusters. In particular, it appears that differentiating between two time series where one of the peaks differs in terms of its relative magnitude is the most difficult task for the majority of investigated methods. This is revealed by generally poorer accuracies for clusters 5, 7 and 8. Despite this, both DTW$_{\text{opt}}$ and KSD$_{\text{opt}}$ demonstrated proficiency in this scenario. Clusters 3, 17 and 18 also displayed lower accuracies overall. 
This suggests that evidence of PV generation in DLPs may tend to be dominated in dissimilarity assessments by peak energy consumption events. If identification of PV generation within DLPs is a particular outcome sought from an application of clustering, practitioners may achieve greater success by restricting dissimilarity computations to the relevant subsequences where PV is expected. There is insufficient space to explore the typical failure modes of each individual method in \Cref{Tab:ExperimentOneResults}, however we have made average confusion matrix plots available in the \href{https://github.com/yerbles/Smart-Meter-Time-Series-Clustering-Comparative-Study/tree/main}{supplementary materials}. 

\begin{wraptable}{r}{0.3\textwidth}
    \scriptsize
    \rowcolors{1}{TableGray}{TableWhite}
    \begin{adjustbox}{center}
        \begin{tabular}{p{0.84cm} p{2.5cm}} \toprule \hiderowcolors 
            \makecell[tl]{\textbf{Retained} \\ \textbf{Methods}} & \makecell[tl]{\textbf{Retained Parameter} \\ \textbf{Values}} \\ \midrule \showrowcolors

             CID & ---  \\[0.15cm]

             DTW & $w \in \left\{1,2,3,4,5,6\right\}$   \\[0.15cm] 

             ERP & \makecell[tl]{$w \in \left\{1,2,3,4,5\right\}$ \\ $g \in \left\{0,0.1,0.2,0.3\right\}$} \\[0.15cm]

             ED & ---  \\[0.15cm]

             FD & ---   \\[0.15cm]

             KSD & $w \in \left\{1,2,3\right\}$ \\[0.15cm]
             
             LSTM & \makecell[tl]{$\alpha$: $A,B,\ldots,K,L$ \\ $b=100$ \\ $n_e=1000$} \\[0.15cm]

             MPD & $w \in \left\{41,42,\ldots,47\right\}$ \\[0.15cm] 

             MM$_3$ & ---    \\[0.15cm]
             
             MSM & \makecell[tl]{$w \in \left\{1,2,3,4,5,6\right\}$ \\ $c=0.1$} \\[0.15cm]
             
             PCA & $n_c \in \left\{5,6,\ldots,14\right\}$ \\[0.15cm]

             SBD & --- \\[0.15cm]

             TWED & \makecell[tl]{$\nu \in \left\{0.0001,0.001,0.01\right\}$ \\ $\lambda \in \left\{ 0.25,0.5,0.75\right\}$}  \\[0.15cm]

            \bottomrule \hiderowcolors
        \end{tabular}
    \end{adjustbox}
    \caption{The 13 methods retained for the second stage of experiments, including the subset of parameters used to establish their expected performance (X$_{\text{exp}}$).}
    \label{Tab:RetainedMethods}
\end{wraptable}

The \textit{difference} in mean accuracy between default and optimal parameter settings was smallest for DWT (0.001), PAA (0.004) and TWED (0.02)\footnote{The differences in mean accuracy between default and optimal parameters should not be interpreted as indicating parameter sensitivity. In fact, a small difference could indicate either low sensitivity to parameter choice OR that the default parameters were already close to optimal. Conversely, a large difference might suggest that the default parameters were far from optimal, rather than high parameter sensitivity.}. Apart from the unique LCSS case, this same difference was most pronounced for ERS (0.465) and the SAX methods (0.366, 0.317, 0.299 and 0.240). Of particular note is the accuracy difference from DTW$_{\text{def}}$ (0.824) to DTW$_{\text{opt}}$ (0.994). It appears that many practitioners fail to recognise the importance of enforcing a maximum allowed warping window when applying DTW. Indeed, a commonly encountered critique for DTW$_{\text{def}}$ is that it is too computationally inefficient \cite{Hino2013,Iglesias2013}. Yet the improvement from DTW$_{\text{def}}$ to DTW$_{\text{opt}}$ reinforces an important observation made in multiple studies \cite{Dau2018,Ratanamahatana2004a,Holder2023AClustering}: the application of a window to constrain the warping in DTW significantly improves markers of performance \textit{and} has the added benefit of improving efficiency. In \cite{Ratanamahatana2004a} the authors suggest that ``a little warping is a good thing, but too much warping is a bad thing''. Where constrained DTW has been used for time series data mining, the window parameter is commonly set to 5\%, 10\% or 20\% of the length of the time series \cite{Paparrizos2015,Cai2021a,Wu2020,Holder2023AClustering}, including when clustering DLPs \cite{Li2022}. However, in \cite{Dau2018} they show that the best warping window is quite sensitive to various dataset characteristics, including the size of the dataset and unsurprisingly, the kinds of shapes present in the dataset. Accordingly they suggest a semi-supervised approach for tuning the window parameter. 

In the next stage of experiments, we examine how a subset of the better-performing distances and representations examined here perform, in combination with clustering algorithms, when dataset characteristics are varied to better reflect the complexity of real-world data. A minimum threshold of 0.8 was required for all of the by-cluster 1NN accuracies in order to retain a modest subset of the best performing methods to use in the second stage of experiments. This value struck a balance between the number and variety of methods, and the capacity of those methods to identify and separate all of the synthetic clusters. To provide more practical insights, we have elected to indicate the performance that could be expected if a practitioner randomly selected parameters for the parametrised distance measures and representation methods from within a reasonable subset. Here, reasonable was taken to be a subset of parameters that produced mean 1NN accuracies of at least 0.95. If all values were found to be above this threshold, a cut-off was selected after the 1NN accuracy stopped changing significantly. This principle was applied to all methods carried into the next stage of experiments. \Cref{Tab:RetainedMethods} shows the retained methods and the parameter subsets used to calculate their \textit{expected} performance (X$_{\text{exp}}$), which was determined by averaging the relevant metric's values across these parameter settings. This reflects a similar approach used in \cite{Rodriguez2019}, where a reasonable subset of parameters were randomly sampled to provide an expected performance.

\subsection{Stage Two: Clustering Approaches and Varying Dataset Characteristics}
\label{sec:Results---StageTwo}

In this stage of experiments we compare clustering approaches in a variety of scenarios. These approaches are either the indivisible clustering approaches listed in \Cref{Tab:ClusteringAlgorithms}, or combinations of the similarity paradigms retained from stage one and the clustering algorithms listed in \Cref{Tab:ClusteringAlgorithms}. All of these clustering approaches require the number of clusters to be specified a priori. In order to provide an accurate comparison, we have provided the true number of clusters to all approaches. The estimation of this parameter is outside the scope of this current study, usually relying on appeals to practitioner knowledge or RVIs, and is a difficult problem in its own right. 

\subsubsection{Combinations with Clustering Algorithms}
\label{sec:Results---StageTwo---Combinations}
To begin the second stage of experiments, we examined the recovery of ground truth labels when the 13 retained distances and representations were paired with the 9 clustering algorithms listed in \Cref{Tab:ClusteringAlgorithms} in order to identify the most effective combinations. These combinations are subsequently denoted using the naming convention of a `$+$' between method acronyms (e.g. DTW$_{\text{exp}}+$HAC-Wa) throughout the remainder of the paper. These combinations are also compared with the two indivisible clustering approaches (KS and KMn) from \Cref{Tab:ClusteringAlgorithms}, resulting in a total of 119 clustering approaches. We computed the ARI for each clustering approach with 100 baseline synthetic datasets. We first sampled 1,000 cluster labels uniformly at random, then generated the corresponding 1,000 time series from their respective clusters. This resulted in roughly balanced clusters, with 50 expected DLPs per cluster.

\begin{figure}[!htp]
    \centering
    \subfigure[Heatmap of mean ARI]{\includegraphics[width=0.49\textwidth]{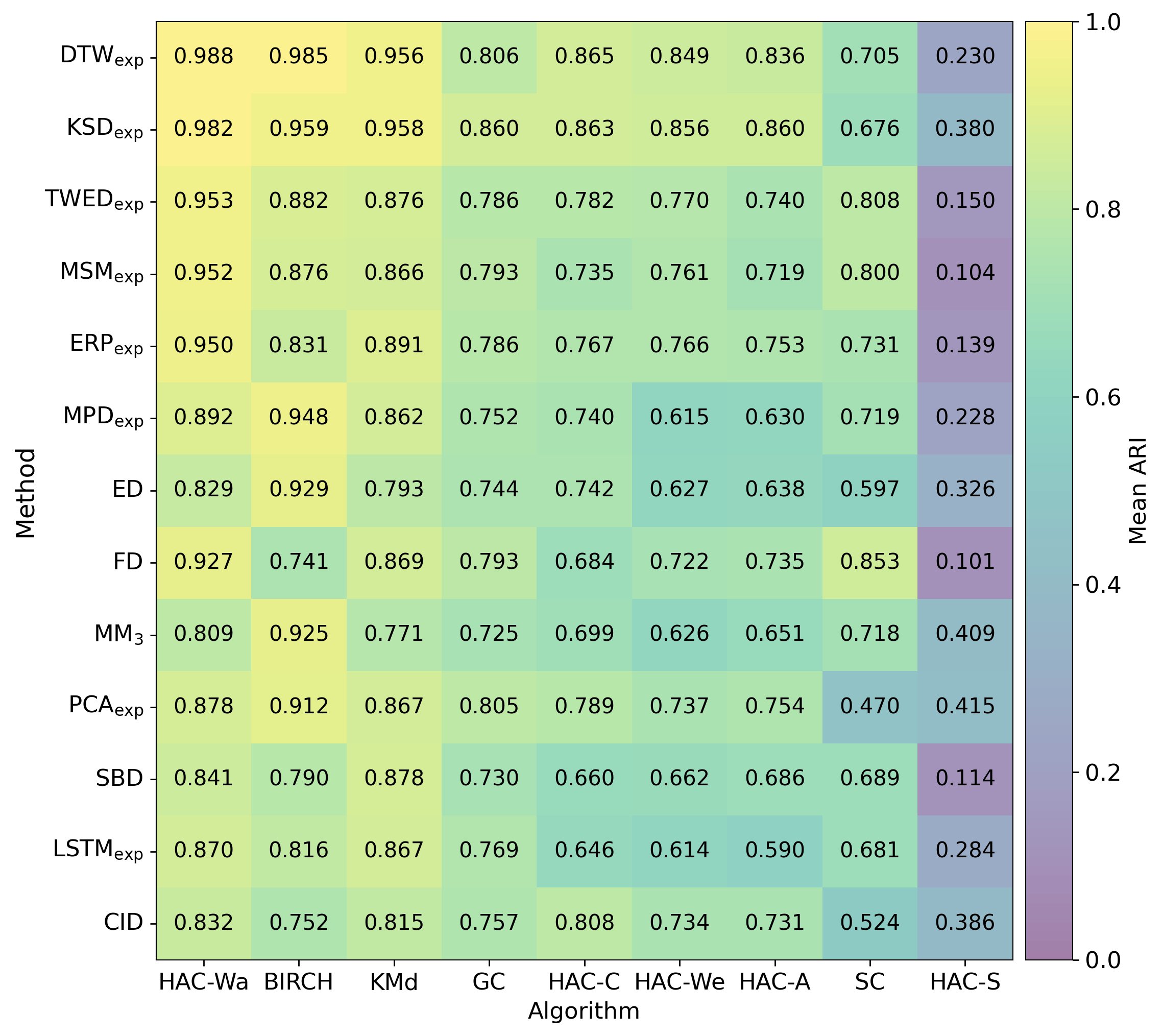}} 
    \subfigure[Heatmap of mean ranks (coloured markers indicate the leading sets of statistically indistinguishable methods)]{\includegraphics[width=0.49\textwidth]{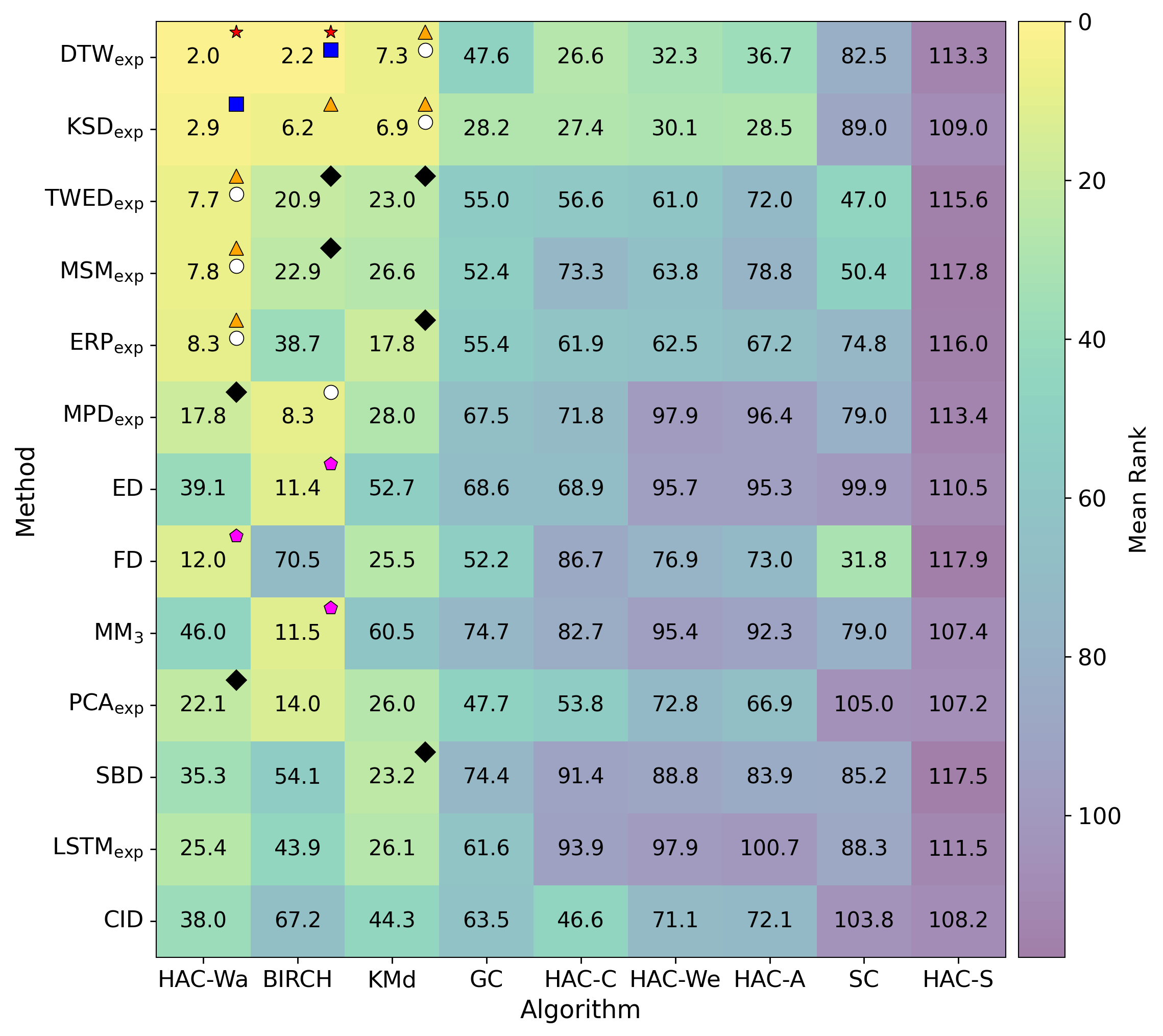}}     
    \caption{Results for combinations of clustering algorithms and similarity paradigms over 100 baseline datasets with 1000 time series. Heatmaps of the mean ARI and mean ranks are shown in a) and b) respectively. Combinations in b) containing the same coloured markers in the top right are not ranked statistically significantly different according to a Bonferroni corrected Wilcoxon signed-rank post-hoc test. We have limited the inclusion of such shapes to cliques containing ranks less than or equal to 20.}
    \label{Fig:ExpTwo-ARI}
\end{figure}

\Cref{Fig:ExpTwo-ARI} summarises the results from this experiment, excluding KS and KMn which are discussed in the text. On the left, a heatmap displays the mean ARI over all datasets for each combination of clustering algorithm and similarity paradigm. On the right, another heatmap shows the mean ranks of the combinations according to the ARI. Again, a Friedman test was consulted to test whether all combinations performed similarly, and a Wilcoxon signed-rank test with Bonferroni correction was used to test for significant pairwise differences in ranks at a 0.05 significance level. Instead of the solid lines used to connect cliques in \Cref{Fig:ExpOne-CDD-All,Fig:ExpOne-CDD-NoTuning}, this time cliques with indistinguishable performance are indicated using coloured markers in the top right of the relevant squares. For example, the red stars in the squares for DTW$_{\text{exp}}+$HAC-Wa and DTW$_{\text{exp}}+$BIRCH indicate that these two methods did not perform differently with statistical significance. 
We only labelled the cliques with at least one mean rank less than or equal to 20 to assist in readability. Similar results can be observed for AMI and PSI, plots for which can be found in the \href{https://github.com/yerbles/Smart-Meter-Time-Series-Clustering-Comparative-Study/tree/main}{supplementary materials}.

The dominance of DTW and KSD identified in stage one was unaffected by interactions with clustering algorithms. The highest mean ARIs overall were obtained by DTW$_{\text{exp}}+$HAC-Wa $(0.988)$ and DTW$_{\text{exp}}+$BIRCH $(0.985)$, whose mean ranks were found not to differ significantly. KSD$_{\text{exp}}+$HAC-Wa $(0.982)$ was a close third, found not to differ from DTW$_{\text{exp}}+$BIRCH with significance. 

Interestingly, KS was vastly outperformed by all combinations using SBD (the distance measure proposed within the KS clustering approach) except when paired with HAC-S. The mean ARI for KS was 0.355, with an average rank of 109.5. This suggests that the novel distance used within the KS clustering approach is better suited to the clustering of DLPs than the novel prototype and procedure are. Despite SBD outperforming the Minkowski metrics in stage one, the opposite was true in stage two. Both ED and MM$_3$ achieved better (lower) mean ranks with BIRCH than SBD with KMd, its most effective combination. 

Meanwhile among combinations using ED, only BIRCH and HAC-Wa performed better than KMn with statistical significance. The mean ARI for KMn was 0.813, with an average rank of 44.9. This clustering approach is by far the most popular for clustering DLPs \cite{Trittenbach2019,Lavin2015ClusteringMeters,Rajabi2020}, yet our experiments suggest that a host of potentially more suitable approaches can be obtained without the need for careful parameter tuning.

Consistently, HAC-Wa, BIRCH and KMd accounted for the top 3 highest ranked algorithms across all similarity paradigms, except for FD and CID where the third ranked algorithms were SC and HAC-C respectively rather than KMd. Unsurprisingly, single linkage was the worst performing algorithm, ranking last in all cases. This aligns with single linkage's tendency towards chaining, which often results in highly imbalanced partitions consisting of a few very large clusters alongside multiple singleton or small clusters \cite{Almeida2007ImprovingClustering}.

\subsubsection{Varying Dataset Characteristics}
\label{sec:Results---StageTwo---VaryingCharacteristics}
Given the consistent superiority of HAC-Wa, BIRCH and KMd across all 13 distances and representations, we report results exclusively for combinations with these clustering algorithms in the remainder of this experimental stage. However, results for all combinations are available in the \href{https://github.com/yerbles/Smart-Meter-Time-Series-Clustering-Comparative-Study/tree/main}{supplementary materials}, alongside generally consistent results according to other EVIs.

\subsubsection*{Amplitude Noise}

\begin{figure}[!htbp]
    \centering
    \includegraphics[width=0.99\linewidth]{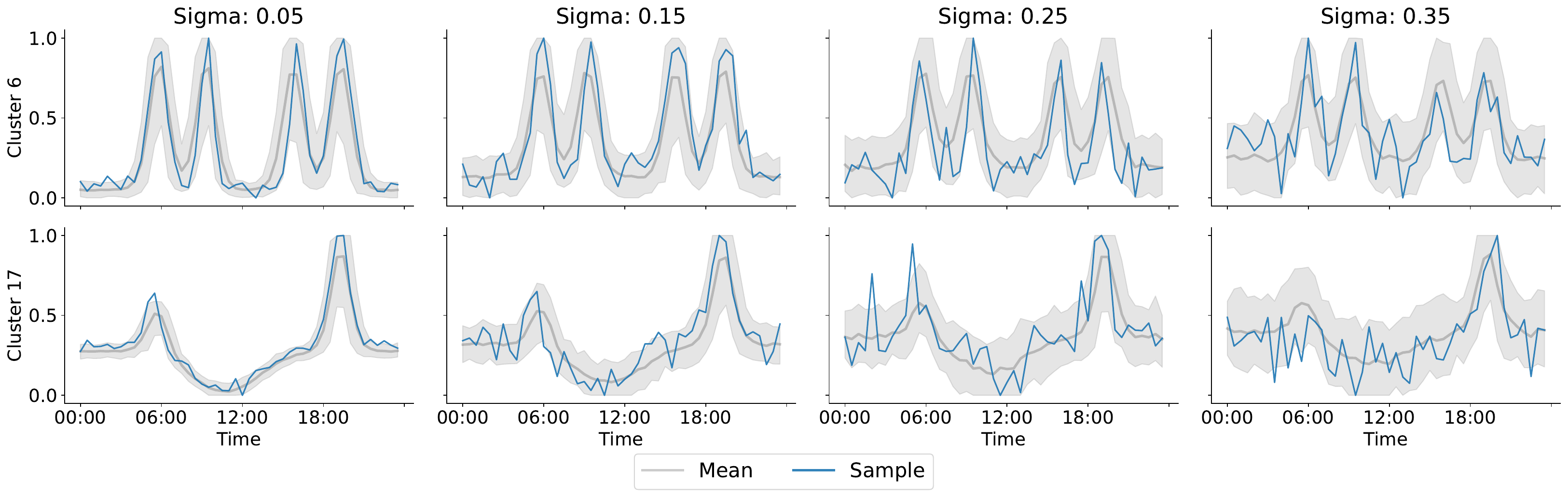}
    \caption{A single sample, the mean and a pointwise 90 percentile interval are plotted for clusters 6 and 17 with different levels of white noise variance ($\sigma$).}
    \label{Fig:NoiseProgression}
\end{figure}

\begin{figure}[!htbp]
    \centering
    \includegraphics[width=0.99\linewidth]{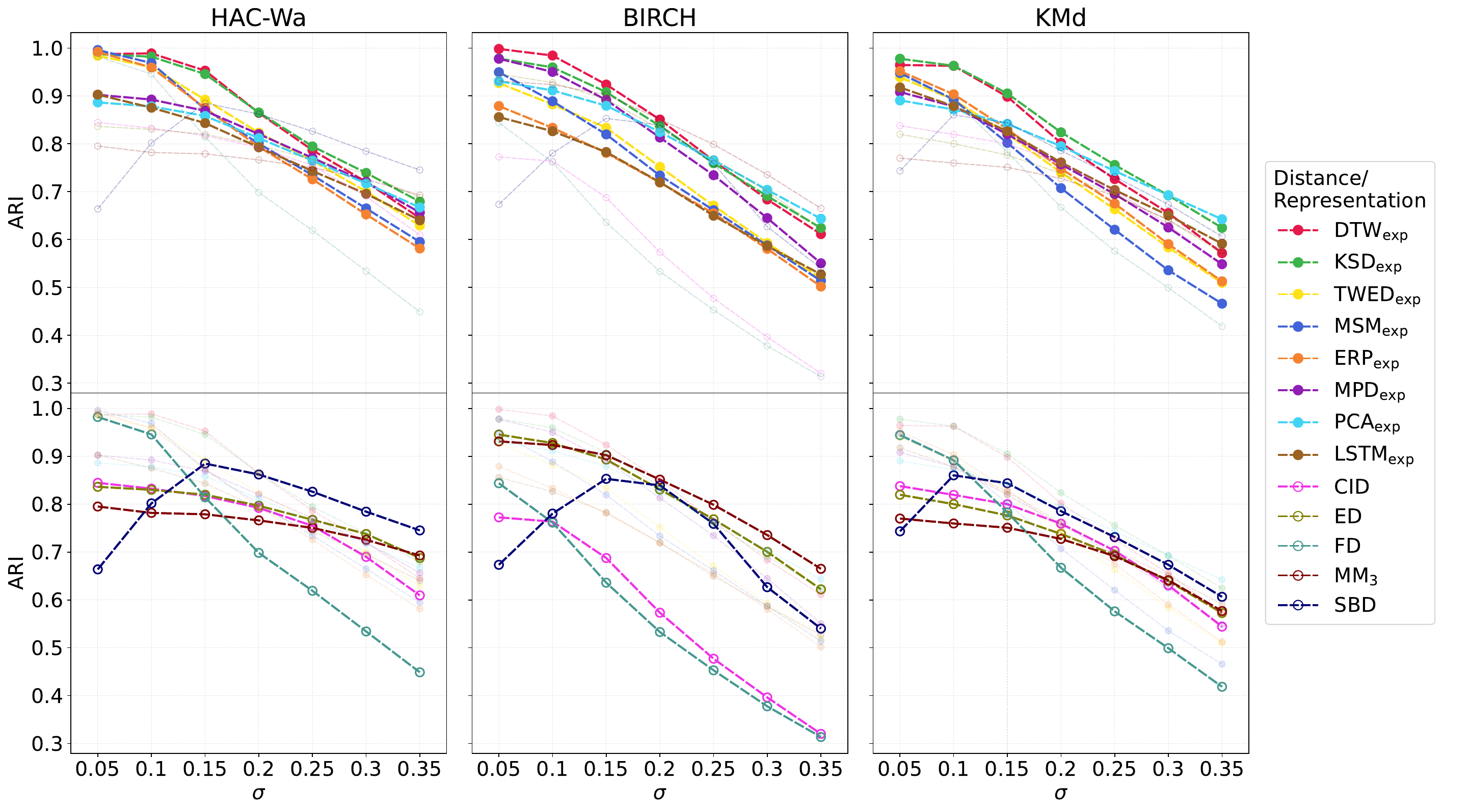}
    \caption{Mean ARI values for clustering approaches applied to datasets with different levels of white noise variance ($\sigma$). The two rows show identical results, with parametrised methods (solid markers) highlighted in the top row and parameterless methods (open markers) highlighted in the bottom row for clarity.}
    \label{Fig:AmplitudeNoise}
\end{figure}

To evaluate the influence of amplitude noise on the performance of clustering approaches, we computed the ARI for datasets where the variance of the white noise in \Cref{Eqn:SyntheticDataGenerator} was set to different levels. For simplicity we set $\sigma_L = \sigma_H = \sigma$ and allowed $\sigma$ to vary from 0.05 to 0.35 in steps of 0.05. \Cref{Fig:NoiseProgression} provides an indication of what these different levels of noise look like for two particular synthetic clusters. We generated 100 datasets for each $\sigma$. For each dataset, we first sampled 1,000 cluster labels uniformly at random, then generated the corresponding time series from their respective clusters. The average ARI values are presented in \Cref{Fig:AmplitudeNoise}. Note that for this figure, the parametrised methods (with solid markers) were separated from the parameterless methods (open markers) to improve legibility. 

An increase in the level of amplitude noise reduced label recovery monotonically for all methods other than SBD. Label recovery initially improved for SBD across each clustering algorithm, reaching a maximum before deteriorating. This trend was consistent across all three EVIs and for all clustering algorithms (other than HAC-S), though the location of the peak varied. Moreover, SBD$+$HAC-Wa outperformed all other combinations in the presence of significant noise ($\sigma>0.2$). This suggests that dissimilarities based on cross-correlation are more robust when cluster shapes are less pronounced due to the influence of noise. 

KSD$_{\text{exp}}$ and DTW$_{\text{exp}}$ demonstrated the most consistent robustness across the different levels of $\sigma$ and clustering algorithms, though the latter tended to underperform for significant levels of noise. PCA$_{\text{exp}}$ in particular performed competitively for significant levels of noise. Despite its design premise, the Complexity Invariant Distance (CID) did not demonstrate a noteworthy aptitude for clustering DLPs with greater complexity (i.e. noise) in combination with any of the 9 clustering algorithms. Also FD appeared to be particularly affected by the presence of amplitude noise with a steeper decline in performance, likely because this distance quantifies transformation effort by taking amplitude variation into account as well as time shifts.

\subsubsection*{Number of Time Series}
\begin{figure}[!htbp]
    \centering
    \includegraphics[width=0.99\linewidth]{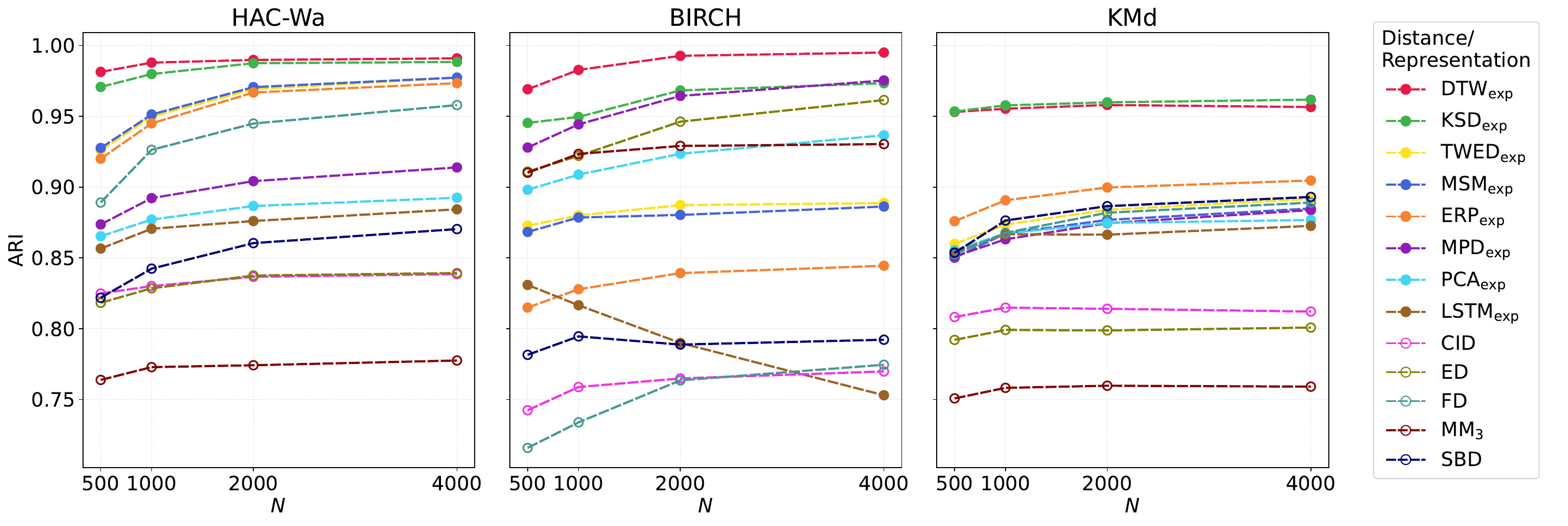}
    \caption{Mean ARI values for clustering approaches applied to datasets with different numbers of time series ($N$).}
    \label{Fig:NumberTimeSeries}
\end{figure}

Hennig argued that for clustering simulation studies, varying the size of the dataset ($N$) ``is very often the least interesting factor'' \cite{Hennig2018}. To verify this statement, we computed the ARI for 100 datasets for each $N \in \left\{ 500, 1000, 2000, 4000 \right\}$. For each dataset, we first sampled $N$ cluster labels uniformly at random, again generating the corresponding time series from their respective clusters. The average ARI values are presented in \Cref{Fig:NumberTimeSeries}. 

Coinciding with Hennig's claim, the relative performances of each combination appeared little changed from those in \Cref{Fig:ExpTwo-ARI}. The examined methods overwhelmingly showed a tendency toward better ARI values for larger $N$. While small errors in label recovery may have a more pronounced effect on EVI values when $N$ is small, the improvement in clustering with larger $N$ is more likely due to increased cluster resolution, i.e. a greater number of time series in each cluster makes the cluster patterns more well-defined. Peculiarly, LSTM$_{\text{exp}}$ appears to fare quite poorly with BIRCH and some of the other hierarchical algorithms as $N$ increases. If the same were true with HAC-Wa and KMd this could be attributed to the fixed number of training epochs or batches. As such, the mechanism is uncertain. ED was the best performing parameterless method across levels when combined with BIRCH, though it performed much worse with the other algorithms.

\subsubsection*{Number of Clusters}
\begin{figure}[!htbp]
    \centering
    \includegraphics[width=0.99\linewidth]{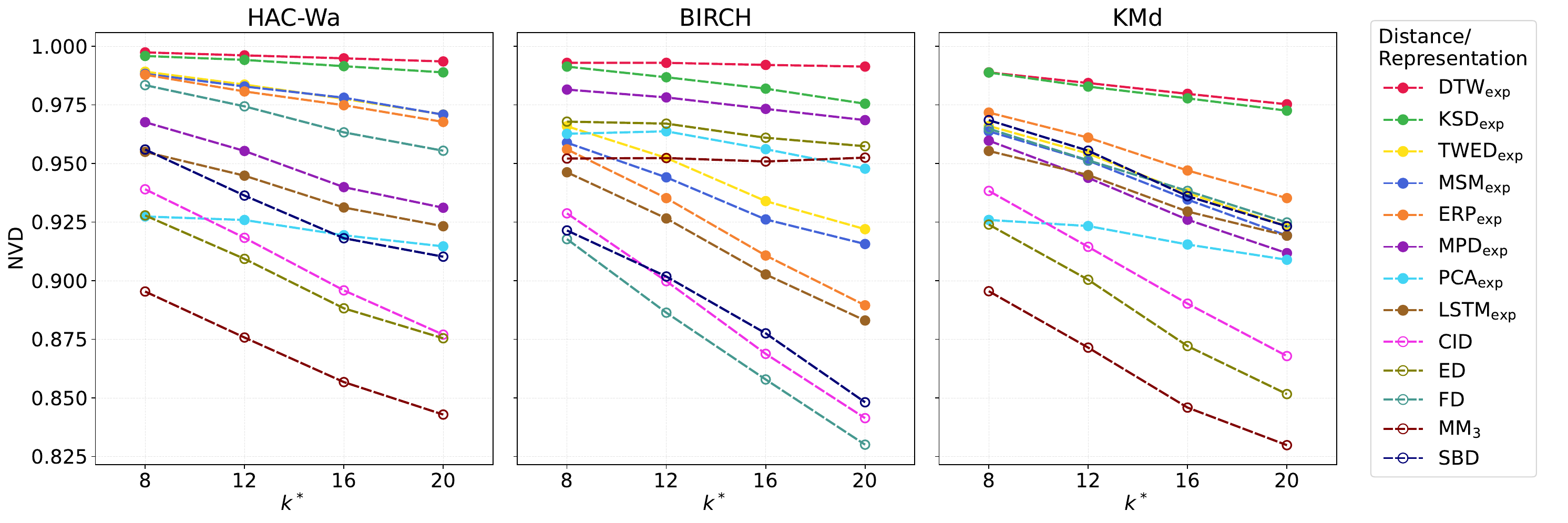}
    \caption{Mean NVD values for clustering approaches applied to datasets with different numbers of clusters ($k^*$).}
    \label{Fig:NumberClusters}
\end{figure}

To investigate the impact of the number of clusters ($k^*$) present within a dataset on the performance of clustering approaches, we computed the NVD for 500 datasets with $50k^*$ time series for each $k^* \in \left\{ 8, 12, 16, 20 \right\}$. We first randomly selected $k^*$ clusters without replacement from the full set of 20 clusters. Then, for each dataset, we sampled $50 \times k^*$ cluster labels uniformly at random from these selected clusters, resulting in an expected 50 time series per cluster, similar to \cite{Rodriguez2019}. We then generated time series according to these sampled labels. Results in \Cref{Fig:NumberTimeSeries} suggest that the resulting variation in dataset sizes ($N = 50 \times k^*$) will have negligible impact on performance. The average NVD values are presented in \Cref{Fig:NumberClusters}.

DTW$_{\text{exp}}$ with BIRCH and HAC-Wa, the two best performing clustering approaches, were essentially unaffected by increasing $k^*$. In contrast, most other methods showed a noticeable decline in performance, although over this particular range it was much smaller in magnitude than the decrease observed with increasing $\sigma$. MM$_3+$BIRCH also showed a distinct robustness to changes in $k^*$, as did PCA$_{\text{exp}}$ with HAC-Wa and KMd, though they were still outperformed by many other combinations. Though performing worst in combination with BIRCH, FD$+$HAC-Wa was the best parameterless method across the different numbers of clusters.

\subsubsection*{Balance of Cluster Sizes}
\begin{figure}[!htbp]
    \centering
    \includegraphics[width=0.99\linewidth]{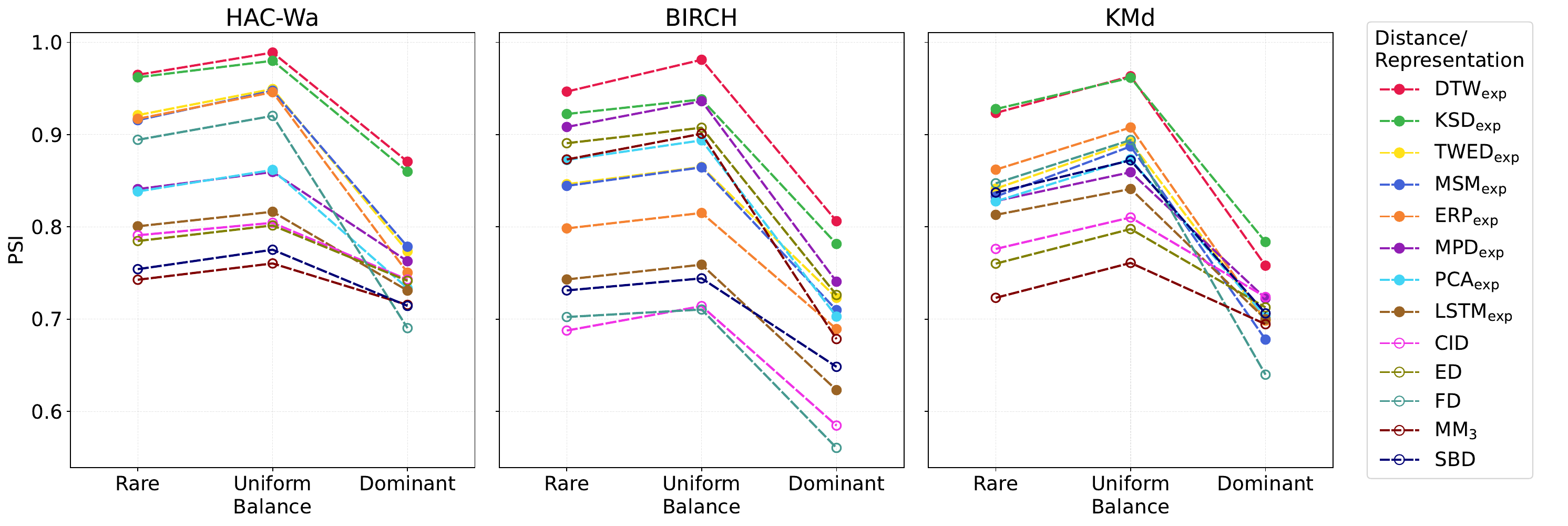}
    \caption{Mean PSI values for clustering approaches applied to datasets with different cluster size distributions.}
    \label{Fig:ClusterBalance}
\end{figure}

Real-world datasets often contain clusters of varying sizes, with both dominant and rare patterns, particularly in DLP clustering. To evaluate robustness to cluster imbalance, we computed the PSI for 500 datasets with three types of cluster balance. In ``Rare'' datasets, one randomly selected cluster had exactly 5 elements, while in ``Dominant'' datasets, one cluster had 500 elements. The remaining cluster labels were sampled uniformly at random, with a minimum of 10 elements per cluster, until a total of 1,000 were obtained. ``Balanced'' datasets were generated as before by uniformly sampling cluster labels. Time series were then generated according to the obtained cluster labels. The average PSI values are presented in \Cref{Fig:ClusterBalance}.

The DTW$_{\text{exp}}+$HAC-Wa and KSD$_{\text{exp}}+$HAC-Wa combinations delivered the highest PSIs for all three balance categories. Furthermore, DTW$_{\text{exp}}$ and KSD$_{\text{exp}}$ outperformed all other methods in combination with the remaining clustering algorithms. However all methods combined with HAC-Wa, BIRCH and KMd were found to be sensitive to the levels of cluster imbalance examined in this study, demonstrating a reduction in performance compared to the balanced case. Regarding the other clustering algorithms (figures for which can be found in the \href{https://github.com/yerbles/Smart-Meter-Time-Series-Clustering-Comparative-Study/tree/main}{supplementary materials}), whilst their PSI values were lower overall, combinations with HAC-C, HAC-A and HAC-We were largely indifferent to imbalances. Additionally, GC performed better when a dominant cluster was present, but BIRCH appeared to be more significantly affected in this particular scenario. Overall, the performance of FD also appeared particularly sensitive to the presence of dominant clusters.

\subsubsection*{Outliers}
Real-world DLP datasets frequently contain outlier time series that represent anomalous behaviour, as human activities will sporadically deviate from regular patterns. Ideally, something like a density-based clustering algorithm could be used to distinguish noise from pattern; however, as previously mentioned, some such methods have been observed to over-allocate objects to the noise group \cite{Jin}. Therefore, it is important to understand the capacity of different clustering approaches to recover the main patterns in a dataset amid ``noisy objects'', also known as outliers.

\begin{figure}[hp]
    \centering
    \includegraphics[width=0.99\linewidth]{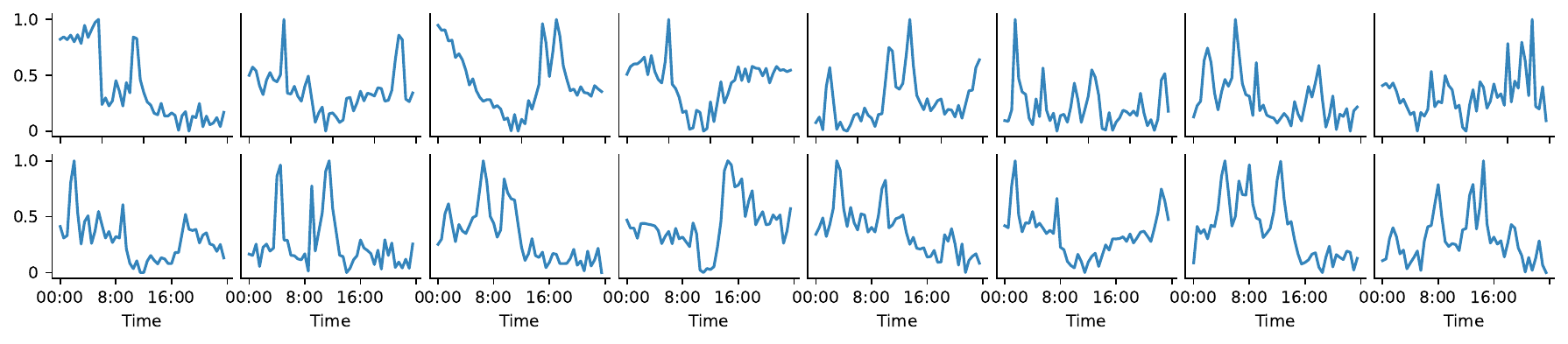}
    \caption{A random sample of 16 synthetic outliers.}
    \label{Fig:Outliers-TS}
\end{figure}

A method was developed to generate outlier time series with a high probability of not fitting into any of the 20 existing clusters, and not forming clusters of their own.
Briefly, it generates characteristic curves with a random number of randomly located normal probability distribution function peaks, along with additional features (such as a PV generation dip, or a logistic step up/down) added probabilistically. Some example outliers are presented in \Cref{Fig:Outliers-TS}. While this approach does not guarantee that a generated outlier will indeed be an outlier, it was used for two reasons. Firstly, we are only interested in assessing the recovery of ground-truth labels for the 20 baseline synthetic clusters, not identifying outliers. Accordingly, the methods have been scored by ignoring the outlier labels. Candidate outliers that fall within or on the boundary of a cluster should not negatively impact the recovery of that cluster's labels. Secondly, most methods we could utilise to filter the generated outliers rely upon a specific distance measure or representation method. For instance, filtering generated outliers via their $k$-Nearest Neighbours outlier score would confer an advantage to the specific method used to infer the nearest neighbours.

\begin{figure}[!htbp]
    \centering
    \includegraphics[width=0.99\linewidth]{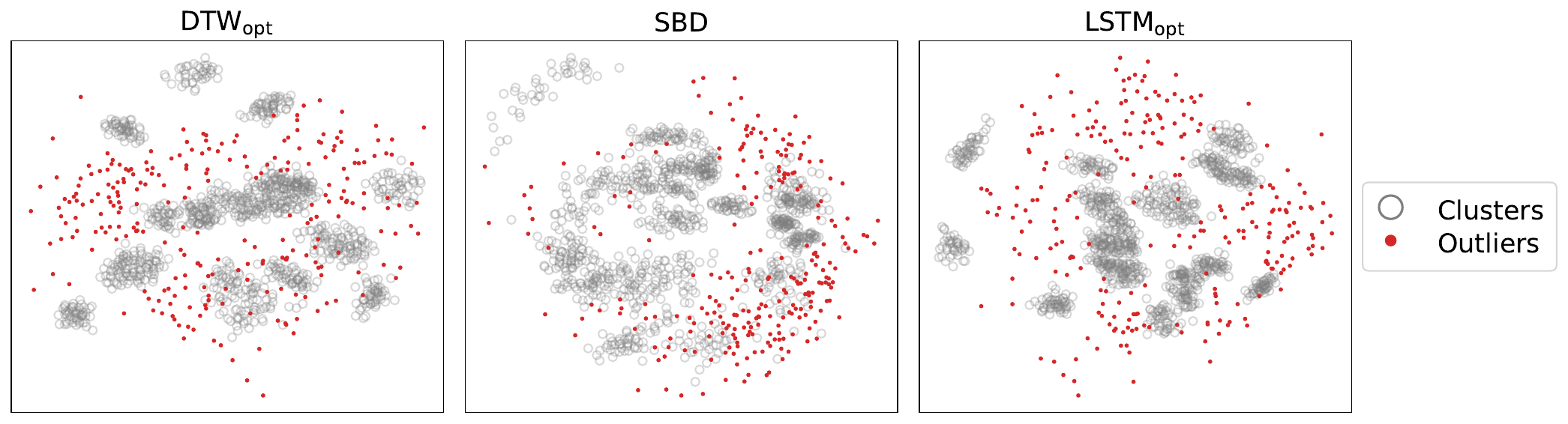}
    \caption{MDS embeddings for three of the thirteen similarity paradigms showing the locations of generated outliers relative to the baseline synthetic clusters. The optimal parameters obtained from stage one experiments have been used for the parametrised methods.}
    \label{Fig:Outliers-MDS-Subset}
\end{figure}

To show how each similarity paradigm regards the outliers relative to the actual clusters, we produced 2D Multi-Dimensional Scaling (MDS) embeddings \cite{Borg2005} for a dataset with 250 of our outliers and 1,000 time series sampled uniformly from the baseline synthetic clusters. MDS attempts to find a set of points in a low-dimensional space which optimally preserve the pairwise dissimilarities in a given distance matrix. A subset of three such embeddings are shown in \Cref{Fig:Outliers-MDS-Subset}, and the remainder can be found in the \href{https://github.com/yerbles/Smart-Meter-Time-Series-Clustering-Comparative-Study/tree/main}{supplementary materials}.

\begin{figure}[!htbp]
    \centering
    \includegraphics[width=0.99\linewidth]{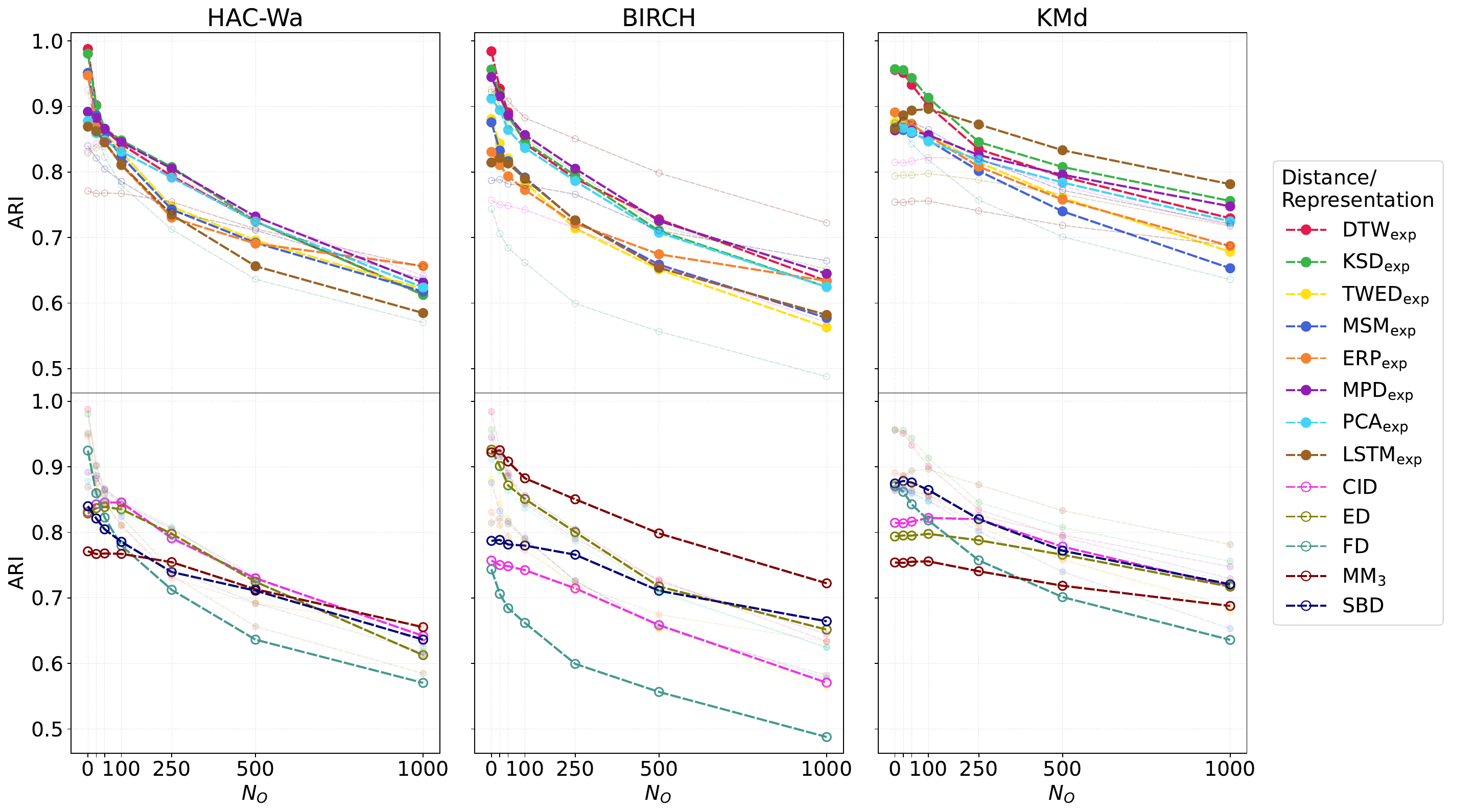}
    \caption{Mean ARI values for clustering approaches applied to datasets with different quantities of additional outliers ($N_O$). The two rows show identical results, with parametrised methods (solid markers) highlighted in the top row and parameterless methods (open markers) highlighted in the bottom row for clarity.}
    \label{Fig:Outliers-Results}
\end{figure}

To evaluate robustness in the presence of outliers, we computed the ARI for 500 datasets for each $N_O \in \left\{ 0,25,50,100,250,500,1000 \right\}$. Here, $N_O$ denotes the number of outliers added to 1,000 time series generated according to uniformly sampled cluster labels. The average ARI values (computed disregarding outliers) are presented in \Cref{Fig:Outliers-Results}.

Unsurprisingly, increasing the prevalence of outliers negatively impacts recovery of the cluster labels. Overall, KMd is the algorithm that appears to be the most robust to this effect, while HAC-Wa and BIRCH show steeper deterioration in performance. This observation is even more pronounced for the remaining hierarchical methods (see \href{https://github.com/yerbles/Smart-Meter-Time-Series-Clustering-Comparative-Study/tree/main}{supplementary materials}). The greedy, irreversible merging decisions in agglomerative hierarchical clustering allow outliers to be absorbed into growing clusters early in the process. These suboptimal merges compound over time, progressively steering subsequent merges further from more globally optimal solutions.

Focusing now on combinations, DTW$_{\text{exp}}$ and KSD$_{\text{exp}}$ with KMd demonstrate a significant loss in performance up to the addition of 100 outliers, while LSTM$_{\text{exp}}+$KMd initially improves over this same range of $N_O$. This initial improvement in LSTM$_{\text{exp}}+$KMd could be a side effect of access to additional training data. This combination then outperforms DTW$_{\text{exp}}$ and KSD$_{\text{exp}}$ for more extreme numbers of outliers ($N_O \geq 250)$. This is likely due to the fact that LSTM learns a global data representation, where outliers may be identified as such and have less influence than the main patterns. On the other hand, FD is the worst performing method, appearing to be particularly sensitive to the presence of outliers regardless of the clustering algorithm, but most notably when in combination with BIRCH.

\subsubsection*{Cluster Separation}
The separation of clusters, or cluster overlap, is another crucial dataset characteristic for consideration \cite{Costa2023,Rodriguez2019}. Real-world DLP datasets rarely exhibit well-separated clusters with clear outliers. Instead, large datasets often show continuous spectrums of peak behaviours across many time series, making cluster boundaries uncertain. Given how significantly their parametrisations can affect similarity assessments, we focus this analysis on distance measures and representations, utilising 1NN accuracy as in \Cref{Subsec:StageOne}. The flexibility of each method in handling overlapping clusters provides insight into their invariances and how their extent is influenced by parameter choices. Assessments of ``better'' or ``worse'' made when varying the previous dataset characteristics do not so much apply for this characteristic --- different clustering aims may require differing degrees of flexibility when discriminating cluster boundaries. 

\begin{figure}[!htbp]
    \centering
    \includegraphics[width=0.99\linewidth]{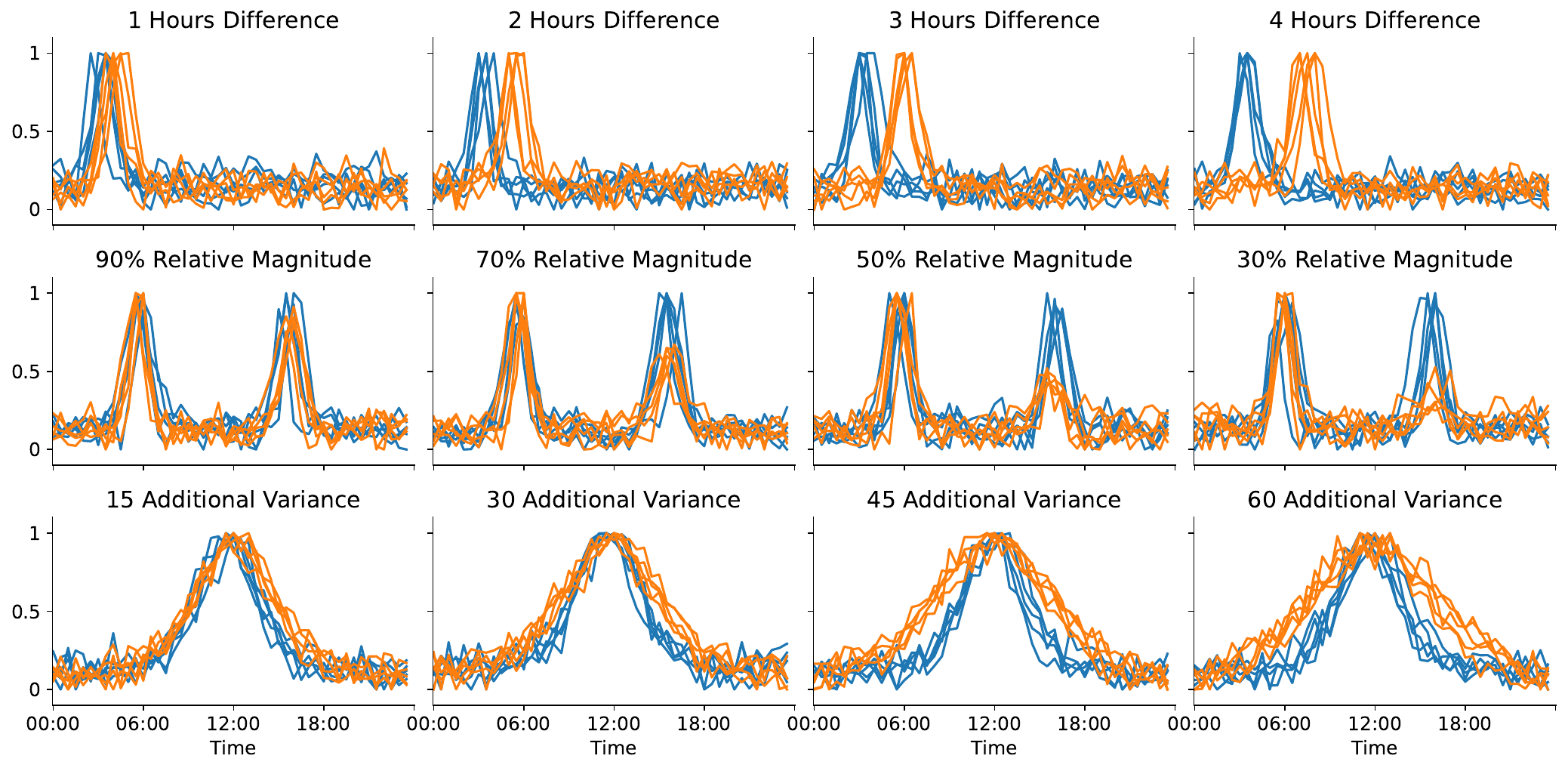}
    \caption{This figure show five samples from each cluster for the three different cluster separation scenarios: timing, relative magnitude and width. Four indicative levels of separation are shown for each scenario.}
    \label{Fig:ClusterMorphing-Demonstration}
\end{figure}

We assessed each method's flexibility using 1NN accuracies, considering three two-cluster scenarios separately. In each scenario, one cluster remained fixed while the second cluster was gradually changed from being a copy of the first cluster to being more and more distinct from it. We generated 100 datasets with 100 time series respectively for each level of separation, with 50 time series in each cluster. This allowed us to compare how both the methods \textit{and} their retained parameters handled different levels of separation in peak \textit{timing}, relative peak \textit{magnitude} and peak \textit{width}. Example time series from each cluster at different levels of separation are shown in \Cref{Fig:ClusterMorphing-Demonstration}. 

\begin{figure}[p]
    \centering
    \includegraphics[width=0.96\linewidth]{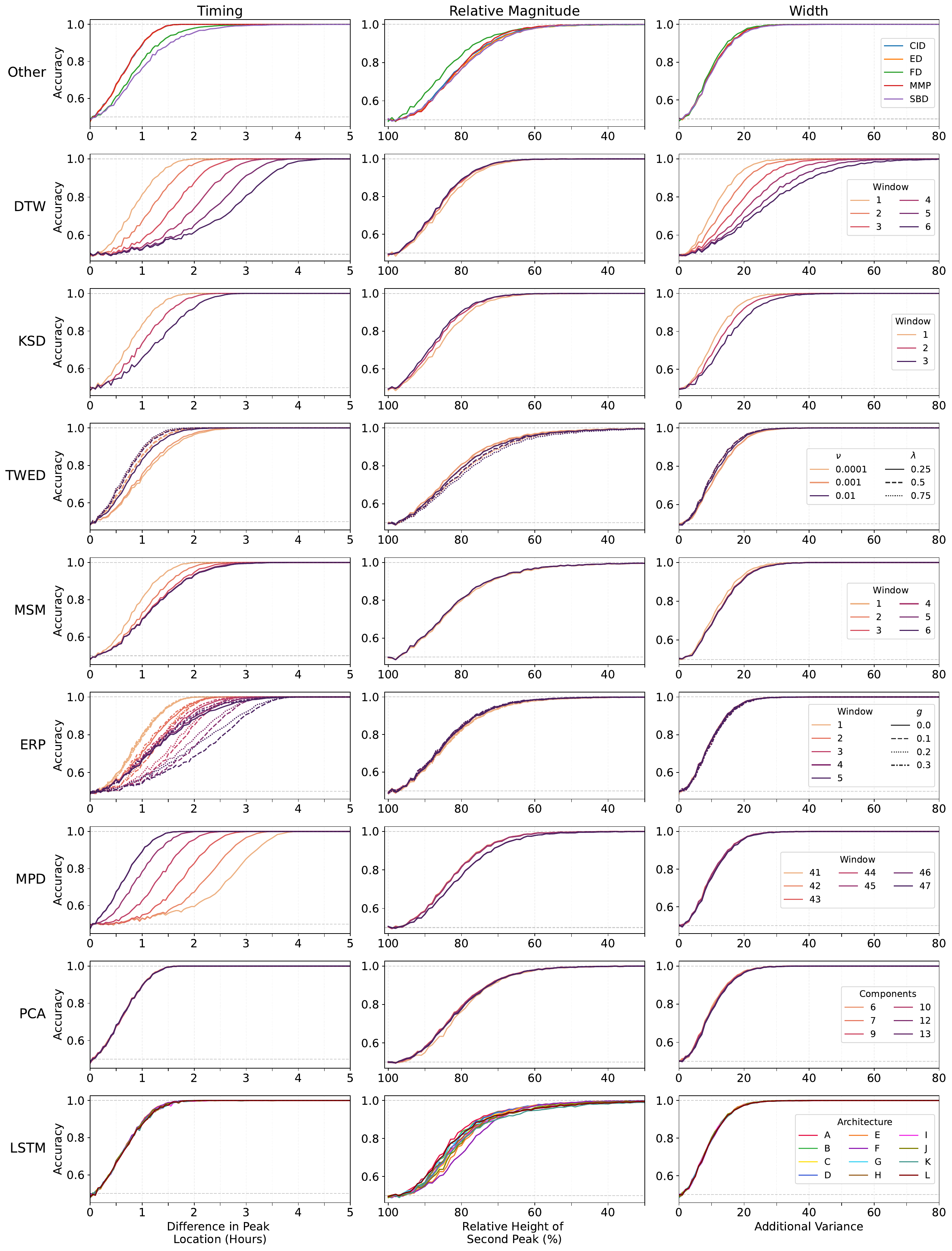}
    \caption{Mean 1NN accuracies for different levels of cluster separation by method and parameters. Each column corresponds to one of three scenarios where two initially identical clusters are slowly separated by changing the \textit{timing}, \textit{relative magnitude} or \textit{width} of one of the clusters, as shown in \Cref{Fig:ClusterMorphing-Demonstration}.}
    \label{Fig:ClusterMorphing-Results}
\end{figure}

We now describe how the clusters were systematically generated for each of the three scenarios. For timing, the first cluster was a single early peak similar to cluster 0. The downsampling step within our synthetic data generator then allowed for fine-grained variation in the timing of the second cluster's peak, with an additional $\left\{0,0.05,\ldots,1.95,2,2.1,\ldots,3.9,4,4.5,5\right\}$ hours added to the location of the normal probability distribution function in the characteristic curve. For relative magnitude, the first cluster was the same as baseline cluster 4. To create the second cluster, the relative magnitude of the second peak in cluster 4's characteristic curve was reduced from $100\%$ to $0\%$ in steps of $1\%$. Finally for width, the first cluster was a modified version of baseline cluster 9 where the peak location was centralised. The second cluster was produced by adding an additional $\left\{0,1,\ldots,80 \right\}$ units of variance to the characteristic curve's normal probability distribution function. The average 1NN accuracies for each method, parametrisation and separation scenario are presented in \Cref{Fig:ClusterMorphing-Results}. For such a binary classification problem, an accuracy of 0.5 implies a method is completely unable to differentiate the two clusters, whilst an accuracy of 1 implies perfect separation. Note that in the following discussion we define ``convergence'' as the first moment the average accuracy exceeds 0.99.

The peak timing experiment revealed the greatest variation among methods and their parameters. Naturally, FD and SBD demonstrated greater flexibility than the other parameterless methods (ED, MM$_3$, and CID). While FD and SBD required 2.4 and 2.8 hours of peak separation respectively to achieve convergence, the others converged after just 1.5 hours. Elastic distance measures with window parameters demonstrated the greatest sensitivity regarding peak timing. For instance, the largest convergence range was observed for DTW, which converged after 1.75 or 4.5 hours with window sizes of 1 and 6\footnote{This convergence point may appear unusual for a window size of 6, given that this ``should'' directly correspond to a 3 hour warping window. However, just as for baseline clusters 0 through 4, the peaks in these clusters are generated with a degree of location variance. This accounts for the extra 1.5 hours required for convergence --- which happens to coincide with the convergence of DTW with no window parameter (i.e. ED).} respectively. A similar convergence trend was observed for KSD over its 3 window sizes. The lack of local warping in MPD meant that over effectively an equivalent range (window sizes 47 to 42\footnote{Note that for MPD, a window size of 42 allows dissimilarities to be computed using the ED between contiguous subsequences offset by a maximum of 6 timepoints.}), the MPD converged after only 1.5 to 3.3 hours. MSM with $c=0.1$ was much less flexible with changes to its window parameter, with convergence achieved at 1.8 to 2.8 hours over the same window sizes as DTW and MPD. For ERP, increasing either parameter resulted in a more flexible measure, whilst for TWED increasing either parameter reduced flexibility.

For the relative magnitude experiment, most parameters had very little impact on convergence, with most variation being found amongst the methods themselves. Overall, KSD and DTW achieved convergence the fastest, at relative magnitudes around 65\%. This strong discriminatory power explains their superior ability to classify clusters 5, 7 and 8 as demonstrated in the stage one experiments. 
In contrast, MSM, TWED and LSTM converged around 42\% on average --- meaning the second peak needed to be \textit{smaller} than half the height of the first before the two clusters were separable. A convergence point less than 50\% is an issue for practitioners hoping to use these methods to discriminate between clusters with peaks of different relative magnitudes. This explains observations from the stage one experiments where MSM, TWED and LSTM were less capable than KSD and DTW at discriminating between clusters 0, 3, 5, 7 and 8. Methods converging at values less than 50\% will tend to incorrectly allocate time series from clusters 7 or 8 which contain smaller secondary peaks. MPD, ERP, SBD, FD, CID and PCA all demonstrated convergence points above this key value of 50\%. 

For the width experiment, most methods performed very similarly, generally converging around a variance level of 25. DTW and KSD were the only methods whose parameters had a significant impact on convergence. DTW converged at variance levels of 27, 33, 40, 49, 56 and 64 for each of the windows 1 through 6. KSD followed a similar trend, though mildly more discriminative. Such flexibility may be undesirable if the quantity of energy consumed is equally important a consideration as the timing of that consumption. Convergence for MPD is unaffected by the window parameter, further demonstrating the lack of local warping for this method. Whilst convergence for LSTM was unaffected by the architecture in both timing and width experiments, the global nature of this representation means this observation may not hold if time series outside of these two clusters were also present in the dataset.

        \section{Validation with Real World Data}
\label{sec:RealData}

While our comprehensive synthetic data experiments offer insights into method performance under controlled conditions, an important question remains: Do these findings translate to real data? To bridge this gap between theory and practice, we conducted a focused validation study comparing method performance on synthetic data against results from a real smart meter dataset manually classified by domain experts. By aligning the characteristics of our synthetic datasets with those of the real-world data --- matching dataset properties such as the number of time series and clusters, amplitude noise, cluster balance, and number of outliers --- we can directly assess whether the relative effectiveness of methods observed in our simulation environment persists when confronted with the greater complexity of natural residential consumption patterns. This validation exercise tangibly demonstrates how our synthetic data findings translate to real-world applications of SMTS clustering.

The selected real dataset consisted of 365 half-hourly DLPs from a single residential consumer from the Ausgrid dataset \cite{Ausgrid2013}, taken from 1 July 2012 to 30 June 2013. This consumer was chosen as they demonstrated a good clustering tendency --- that is they displayed a good variety of recurring DLP shapes throughout that period.
Consumers were screened by domain experts for clustering tendency using various clustering approaches taken from those introduced in \cref{sec:Results---StageTwo}. Once selected, this consumer's DLPs were manually clustered according to the organising principle described in \cref{subsec:SyntheticData-Philosophy}. The resulting ground truth consisted of 16 clusters containing a combined 268 DLPs, with the remaining 97 profiles labelled as outliers.
The complete dataset with assigned labels is available through our \href{https://github.com/yerbles/Smart-Meter-Time-Series-Clustering-Comparative-Study/tree/main}{supplementary materials}, alongside a visualisation of the clusters.

To enable direct comparison with this real dataset, we generated 500 synthetic datasets with matching properties. Each synthetic dataset contained exactly 365 time series, with 97 outliers and 268 clustered time series distributed across 16 clusters. For each dataset, we randomly selected 16 clusters from our 20 synthetic clusters, then randomly assigned the real-world cluster counts to these selected synthetic clusters to emulate the cluster balance observed in the real data. In order to imitate the amplitude noise we analysed the average pointwise standard deviation within each real cluster, finding values ranging from 0.077 to 0.124, with an average of 0.105 across all 16 clusters. Thus $\sigma$ was set to 0.105 for the synthetic data.

We applied the same 119 clustering approaches utilised in \cref{sec:Results---StageTwo---Combinations,sec:Results---StageTwo---VaryingCharacteristics} to both the synthetic and real datasets. To equally reflect errors in the smaller and larger clusters, their performance was assessed using PSI. For parameterised methods, we again present the expected performance over all parametrisations from \cref{Tab:RetainedMethods}. Outliers were ignored during EVI evaluations, consistent with our analysis in \cref{sec:Results---StageTwo---VaryingCharacteristics}. The outliers include both traditional outliers --- series distinct from any cluster without sufficient similar profiles to constitute their own cluster --- and time series that sat squarely between multiple clusters. The latter were labelled as such so that these time series would not affect evaluation. 

\begin{figure}[!htbp]
    \centering
    \includegraphics[width=0.99\linewidth]{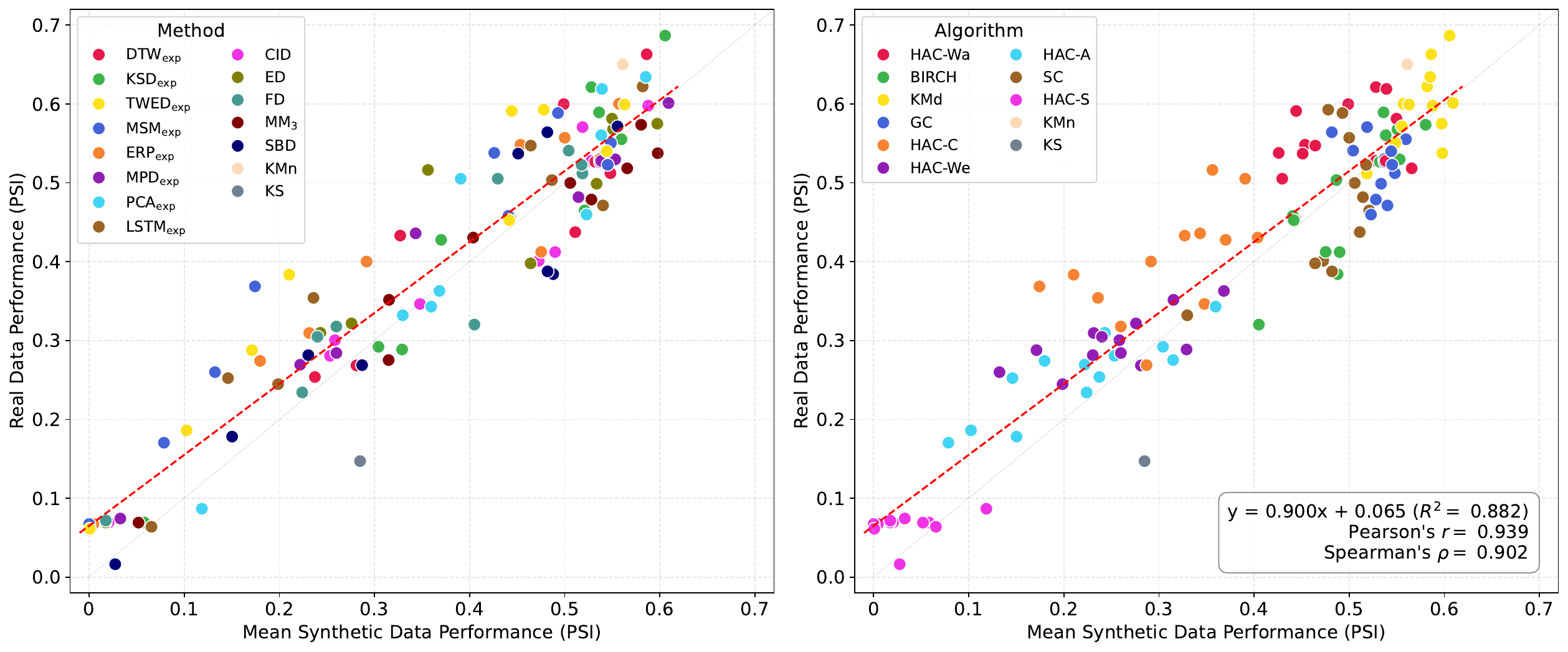}
    \caption{Scatterplot comparing mean clustering performance on synthetic datasets against performance on the real dataset, where each point represents one of 119 clustering approaches. Performance was evaluated using PSI. The left panel shows points coloured by method, while the right panel shows the same points coloured by clustering algorithm.}
    \label{Fig:RealDataValidationResults---PSI}
\end{figure}


\Cref{Fig:RealDataValidationResults---PSI} presents a scatter plot comparing the mean PSI values achieved by the clustering approaches on the synthetic datasets against their PSI values on the real dataset. Statistical analysis confirms a significant relationship between synthetic and real-world performance, with both Pearson's correlation ($r_{\text{PSI}} = 0.939$, $p<10^{-56}$)
and Spearman's rank correlation ($\rho_\text{PSI} = 0.902$, $p < 10^{-44}$) 
indicating strong positive associations. This strong correlation provides compelling evidence that method performance on our synthetic data is highly predictive of performance on real-world data. This observation is supported by similar results for ARI ($r_{\text{ARI}} = 0.935$, $\rho_{\text{ARI}} = 0.899$) and AMI ($r_{\text{AMI}} = 0.950$, $\rho_{\text{AMI}} = 0.882$). Analagous plots are available for these EVIs in the \href{https://github.com/yerbles/Smart-Meter-Time-Series-Clustering-Comparative-Study/tree/main}{supplementary materials}. The regression line for PSI, which explains $88.2\%$ of the variance, suggests performance scores between real and synthetic datasets are close to parity ($y=0.900x+0.065$). However, performance tended to be slightly higher on the real dataset than on synthetic data, particularly for those methods with lower PSI values.


Several findings from our synthetic data experiments were reinforced by this real-world validation. Despite not being the outright best performers on these emulated synthetic datasets, DTW and KSD did perform best on the real-world data when paired with KMd. This aligns with our identification of these methods as the most robust across our previous experimental stages. Interestingly, KMn was not far behind the expected values for these two parametrised combinations, performing more competitively on the real-world data than the emulated synthetic data.

This investigation has also corroborated our earlier observation that KMd can robustly recover cluster structure in the presence of outliers. While BIRCH and HAC-Wa excelled in many previous experiments, their performance degraded in the presence of outliers. Additionally, BIRCH's vulnerability to cluster imbalance explains why KMd outperformed them in this real-world context. We previously theorised that a vulnerability of agglomerative hierarchical methods is found in their greedy, irreversible merging decisions, which can lead to outliers being absorbed into growing clusters early in the process, with these suboptimal merges compounding over time. It should then be unsurprising that another partition-based clustering algorithm, KMn, also outperformed hierarchical methods on a real-world dataset featuring outliers. 

Other patterns from our synthetic experiments repeated on the real-world data include: SBD with various algorithms (BIRCH, SC, HAC-Wa, GC, KMd) significantly outperforming KS; GC performing similarly to BIRCH and HAC-Wa, which is likely due to its superior robustness to cluster imbalance and outliers; also LSTM$_{\text{exp}}+$KMd demonstrating a strong performance in the presence of outliers.

The results of this investigation should be interpreted with appropriate caution due to its limitation to a single real dataset. The generalisability of these findings would ideally be confirmed through additional validations with more consumers, larger datasets and different SMTS clustering scenarios. However as previously discussed, manual classification of SMTS data is extremely labour-intensive, requiring domain expertise and considerable time investment, making extensive real-world validation studies practically challenging. Nevertheless, the strong correlation observed here provides compelling evidence that the insights gained from our synthetic benchmark can reasonably inform method selection for real-world SMTS clustering tasks. This validation exercise thus strengthens the practical utility of our comprehensive synthetic analysis, offering practitioners greater assurance that our recommendations will translate to improved clustering outcomes in real-world smart meter analytics.

        \section{Conclusion}
\label{sec:Conclusion}

When designing a clustering solution for smart meter time series analyses, practitioners are faced with a daunting array of methods and unreliable tools for their direct comparison \cite{Yerbury2024}. Whilst previous comparative studies have recognised the capacity of benchmarking to provide valuable insights, their findings have been constrained by their narrow scope and methodological limitations. In this study, we utilised expert-curated synthetic datasets designed to incorporate fundamental cluster concepts to evaluate the performance of 31 distance measures, 8 representations, and 11 clustering methods. Our staged methodology facilitated this expansive comparison, enabling a deeper analysis of robustness in the best-performing combinations to variations in dataset characteristics relevant to real-world applications. 

Several clear patterns emerged from our analysis regarding the relative strengths and limitations of different distances and representations for clustering approaches. Most notably, methods that could accommodate local temporal shifts while maintaining sensitivity to amplitude differences, such as Dynamic Time Warping (DTW) and $k$-Sliding Distance (KSD), provided the most consistent strong performances across both experimental stages. However, this superior performance was highly dependent on parameter selection --- default parametrisations of all parametrised methods routinely underperformed relative to even parameterless methods when evaluated in terms of their discriminatory capacity independent of any specific clustering strategy. 
Importantly, our findings suggest that practitioners don’t need to pinpoint precise optimal values for parameters. As identified in previous studies, the application of a small warping constraint to DTW was essential to achieving strong results --- with results from our second stage of experiments and real data validation suggesting that any justifiable window size from 0.5 to 3 hours can be expected to yield stronger results than default settings or parameterless alternatives. With only a single such intuitive window parameter to select, DTW and KSD are particularly strong candidates compared to other more complex or highly parametrised methods.

When combined with $k$-medoids or hierarchical clustering using Ward's linkage (HAC-Wa), DTW and KSD demonstrated remarkable robustness to changes across the dataset characteristics examined in our second stage of experiments. Even without precise parameter tuning, these particular combinations outperformed the widely-used $k$-means with Euclidean Distance (ED), which has historically dominated the smart meter clustering literature. Our systematic variation of dataset properties revealed, unsurprisingly, that clustering performance was most significantly affected by cluster imbalances, amplitude noise, and the presence of outliers. In particular, Shape-Based Distance (SBD) with HAC-Wa emerged as the most robust approach for handling amplitude noise, suggesting that correlation-based dissimilarity measures may be more appropriate for datasets where underlying patterns are obscured by noise. Meanwhile, clustering approaches involving $k$-medoids proved to be the most effective at recovering the main dataset patterns in the presence of outliers for both real and synthetic datasets.

A key challenge revealed in this study was the difficulty most methods have in distinguishing between time series with peaks of different relative magnitude. Relatedly, subtle features like PV generation patterns also tended to be overshadowed by larger peak consumption events. These findings emphasise that practitioners should not expect clustering methods to capture subtle features of interest in daily load profiles when applied off-the-shelf. For instance, if the identification of specific characteristics like PV generation or EV charging is a desired clustering outcome, practitioners should design their clustering approach by either applying weights to or focusing specifically on the relevant periods in the day (e.g., daylight hours for PV, overnight for EV charging) to prevent peak consumption events from dominating the clustering.

While our synthetic data approach enabled a principled comparison of clustering methods, it is important to acknowledge its intentional simplification of real-world complexity. Our synthetic clusters captured fundamental consumption shapes but did not attempt to replicate the full array of complex shapes encountered in real SMTS data. This controlled design ensured that good performance on our synthetic data represented a necessary but not sufficient condition for success with real-world data. Following Hennig's argument that simulation studies offer insights unique from those available from theory or analyses with real-data, our study provided, for the first time, complementary guidance for practitioners, despite the inherent limitation that EVIs cannot be applied to unlabelled real-world datasets --- where practitioners still face the significant challenges posed by RVI limitations. Our validation exercise with a labelled real-world dataset offers greater assurance for practitioners that our recommendations will translate to improved clustering outcomes in scenarios involving real data.

Future research could build upon the observations from our study in several directions. Our work did not evaluate different normalisation procedures or prototype definitions, both of which warrant dedicated investigation. Furthermore, while our study assumed the number of clusters was known, evaluating automatic cluster number determination methods would address another challenging aspect of real-world applications. Additionally, as noted by \cite{Jin}, outlier identification in smart meter time series specifically represents a difficult problem not actively addressed by clustering methods in our study, meriting focused research. The comparative framework we established could be extended to methods focused on clustering consumers rather than daily load profiles, including approaches for long time series or representative load profile clustering. Other modern deep learning approaches such as transformer-based architectures and end-to-end clustering frameworks \cite{Lafabregue2022End-to-endStudy}, deserve investigation specifically for SMTS clustering to determine whether their benefits outweigh their increased implementation complexity and hyperparameter tuning requirements. Finally, the insights established herein regarding method strengths and limitations could inform the development of new approaches better adapted to the specific challenges of clustering daily smart meter load profiles. To facilitate further work, all datasets and code used in our experiments have been made available through our GitHub \href{https://github.com/yerbles/Smart-Meter-Time-Series-Clustering-Comparative-Study/tree/main}{repository}, enabling direct comparison of novel or overlooked methods without needing to replicate our entire experimental framework.

        \begin{appendices}
\renewcommand{\thesection}{A.}
\renewcommand{\thefigure}{A.\arabic{figure}}  
\setcounter{figure}{0}  
\section{Designing Challenge into the Synthetic Data} \label[appendix]{sec:Appendix}

When designing the synthetic data it was important to ensure that the collection of cluster shapes would provide opportunities for meaningful conflict so that candidate methods could be appropriately challenged. Recall from \Cref{subsec:SyntheticData-Philosophy} that the synthetic data was designed around the following central organising principle: \textit{Differences in the timing, number, shape and relative magnitudes of peak energy consumption events establish suitable grounds for the separation of residential daily load profiles into clusters.} We identified six main opportunities for conflict between pairs of clusters which probe each aspect of our organising principle:

\begin{enumerate}
    \item[] \textbf{Timing of Peaks} conflicts occur when consumption patterns share identical or similar peak shapes, but differ in their temporal positioning within the daily profile. These conflicts test a method's sensitivity to temporal alignment, which is a basic requirement when clustering DLPs.
    \item[] \textbf{Number of Peaks} conflicts arise between profiles that share identical or similarly-shaped peaks but differ in the number of such peaks. This tests a method's ability to distinguish between single-peak and multi-peak consumption patterns.
    \item[] \textbf{Shape of Peak/Profile} conflicts emerge when events occur at the same time but exhibit different characteristics in terms of duration (support) or rate of change (gradient). These conflicts evaluate a method's sensitivity to the detailed structure of consumption events.
    \item[] \textbf{Relative Magnitude of Peaks} conflicts occur between profiles sharing identical temporal positioning and number of peaks, but with differing peak amplitudes. This tests a method's ability to respect scale variations.
    \item[] \textbf{Temporal Symmetry} conflicts appear between profiles that are mirror images of each other along the time axis. Given the importance of temporal ordering in load profiles, methods should maintain sensitivity to these differences.
    \item[] \textbf{Feature Dominance} conflicts arise when prominent peaks could potentially overshadow other significant but subtler features, such as smaller consumption events or PV generation dips. This tests a method's ability to consider multiple features with varying prominence.
\end{enumerate}

\begin{figure}[!t]
    \centering
    \includegraphics[width=0.65\linewidth]{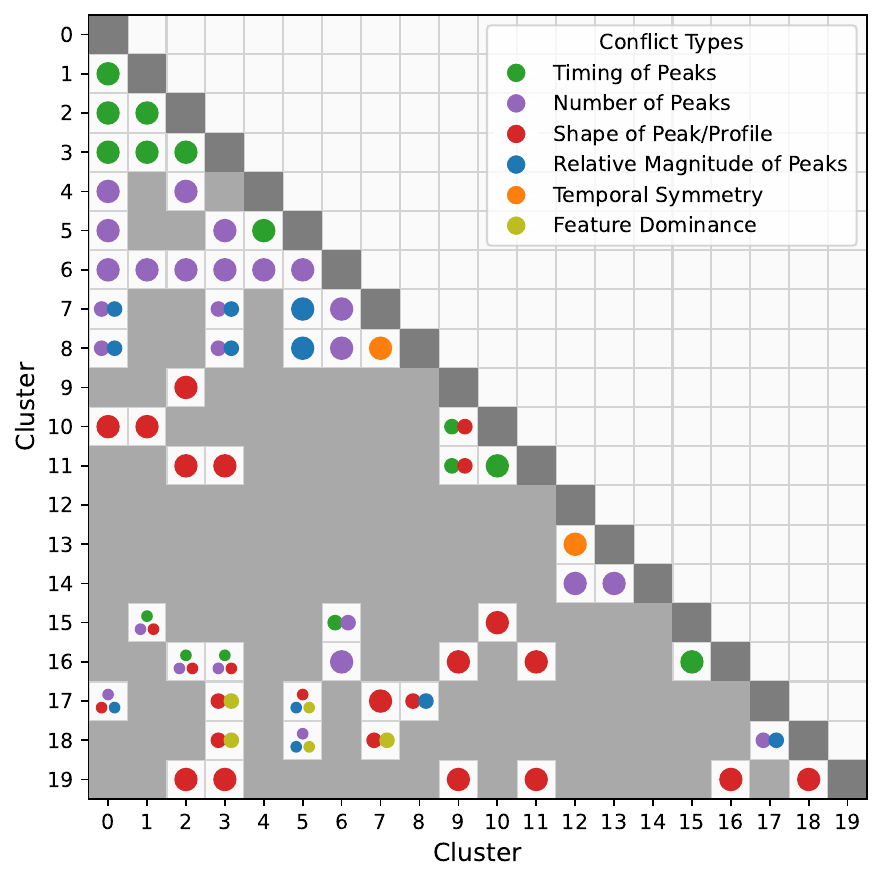}
    \caption{This matrix visualises the potential conflict types between pairs of cluster shapes that may challenge clustering methods. The $(i,j)\textsuperscript{th}$ cell is marked with coloured dots to represent specific types of conflicts that could cause methods to incorrectly classify members of clusters $i$ and $j$ as similar. These conflict types are described in the text. Superior clustering methods will demonstrate greater resilience to these conflicts, while less effective methods may fail to distinguish between clusters that show potential conflicts. Multiple markers in a single cell indicates the presence of multiple potential conflict types between those clusters.}
    \label{Fig:ConflictMap}
\end{figure}

The conflict map in \Cref{Fig:ConflictMap} reveals which of these six types of conflict occur between each pair of the 20 cluster shapes. For example, clusters 5 and 7 share a ``Relative Magnitude of Peaks'' conflict because cluster 7 contains two peaks with the same timing and shape as in cluster 5, but the earlier peak has a smaller relative magnitude in cluster 7. Meanwhile, some of these conflict types directly relate to time series invariances that different clustering methods may exhibit. For example, clusters 0 through 3 share peaks modelled from the same Gaussian probability distribution function, differing only in temporal position. This design choice specifically challenges methods with varying degrees of phase invariance, such as the Matrix Profile Distance (MPD) \cite{Gharghabi2018a}. The MPD's window parameter determines its sensitivity to phase shifts - with a small window parameter, these four clusters would become indistinguishable, highlighting how method parameters can critically affect cluster delineation. Similarly, Relative Magnitude conflicts test amplitude invariance, while Shape conflicts probe a method's invariance to warping in the time dimension. Meanwhile sensitivity to local scaling invariance is assessed by the variation allowed within each cluster.

Many cluster pairs exhibit multiple conflict types simultaneously, creating more complex evaluation scenarios that better reflect the nuanced differences found in real-world load profiles. Methods that competently distinguish these 20 synthetic clusters despite their various forms of potential conflict satisfy a necessary (but not sufficient) condition for suitability in real-world DLP clustering tasks.

\end{appendices}       

	\bibliographystyle{plain}
	\bibliography{references}

\end{document}